\documentclass[a4paper,aps,prd,preprint,superscriptaddress,nofootinbib,showpacs]{revtex4-1}

\usepackage[utf8]{inputenc}
\usepackage{amsmath}
\usepackage{amsfonts}
\usepackage{amssymb}
\usepackage{amsthm}
\usepackage{graphicx}
\usepackage{bm}
\usepackage{bbm}
\usepackage{hyperref}
\allowdisplaybreaks

\renewcommand{\k}{\bm{k}}
\newcommand{\q}{\bm{q}}
\newcommand{\p}{\bm{p}}

\newcommand{\Tr}{\mathrm{Tr}}
\newcommand{\A}{\bm{A}}
\newcommand{\E}{\bm{E}}

\newcommand{\D}{\bm{D}}
\renewcommand{\d}{\bm{\nabla}}

\begin{document}

\preprint{TUM-EFT 67/15}

\title{The Polyakov loop at next-to-next-to leading order}

\author{Matthias~Berwein}
\affiliation{Physik-Department, Technische Universit\"{a}t M\"{u}nchen, James-Franck-Stra\ss{}e 1, 85748 Garching, Germany}

\author{Nora~Brambilla}
\affiliation{Physik-Department, Technische Universit\"{a}t M\"{u}nchen, James-Franck-Stra\ss{}e 1, 85748 Garching, Germany}
\affiliation{Institute for Advanced Study, Technische Universit\"{a}t M\"{u}nchen, Lichtenbergstra\ss{}e 2a, 85748 Garching, Germany}

\author{P\'{e}ter~Petreczky}
\affiliation{Physics Department, Brookhaven National Laboratory, Upton, New York 11973, USA}

\author{Antonio~Vairo}
\affiliation{Physik-Department, Technische Universit\"{a}t M\"{u}nchen, James-Franck-Stra\ss{}e 1, 85748 Garching, Germany}

\date{\today}

\begin{abstract}
We calculate the next-to-next-to-leading correction to the expectation value of the Polyakov loop or equivalently to the free energy of a static charge. This correction is of order $g^5$. We show that up to this order the free energy of the static charge is proportional to the quadratic Casimir operator of the corresponding representation. We also compare our perturbative result with the most recent lattice results in SU(3) gauge theory.
\end{abstract}


\maketitle

\section{Introduction}

The Polyakov loop is an order parameter for deconfinement in pure SU(N) gauge theories at nonzero temperature $T$. It is defined as
\begin{equation}
L=\frac{1}{d_R}\,\Tr\biggl\langle\mathcal{P}\exp\biggl[ig\int_0^{1/T}d\tau A_0(\tau,x)\biggr]\biggr\rangle\,,
\end{equation}
where $\mathcal{P}$ denotes path ordering of the exponential of the zero component of the gauge field $A_0$ integrated along the compactified imaginary time direction, and $g$ is the coupling constant. Here we have defined the Polyakov loop in a general representation $R$ of SU(N), so the gauge fields are understood as matrices in this representation $R$, and $d_R$ is the dimension of this representation. The thermal expectation value of a single Polyakov loop is invariant under translations, so we can choose it to be at the origin in the following.

The nonzero expectation value of the Polyakov loop above some temperature indicates the onset of color screening and thus deconfinement~\cite{Polyakov:1978vu}. Early lattice studies of the Polyakov loop and its correlators were instrumental in establishing the existence of a deconfinement transition in non-Abelian gauge theories from first principle calculations~\cite{Kuti:1980gh,McLerran:1981pb}. The physical interpretation of the logarithm of the Polyakov loop expectation value is the free energy of a static quark $F_Q/T=-\ln L$ (see, e.g., discussions in Ref.~\cite{McLerran:1981pb}). The free energy of a static quark in a gluonic plasma is finite due to color screening, but becomes infinite below the phase transition temperature~$T_c$.

While in the presence of $n_f>0$ flavors of light quarks the Polyakov loop is no longer an order parameter for deconfinement~\cite{Fukushima:2002bk}, its value at sufficiently high temperatures is still a measure of the screening properties of the deconfined medium. It is easy to see that at leading nontrivial order the Polyakov loop expectation value is $L=1+C_R\alpha_\mathrm{s}m_D/2T$, or equivalently $F_Q=-C_R\alpha_\mathrm{s}m_D/2$, where $C_R$ is the quadratic Casimir of the representation $R$. The Debye mass $m_D$ is given by
\begin{equation}
 m_D^2=\frac{C_A+T_Fn_f}{3}g^2T^2\,,
 \label{mD}
\end{equation}
where $C_A=2T_FN$ is the quadratic Casimir of the adjoint representation and $T_F$ is the normalization constant of the fundamental representation, for which usually the value $1/2$ is taken. The next-to-leading-order (NLO) contribution to the Polyakov loop is of $\mathcal{O}\left(g^4\right)$. The first calculation of the NLO contribution was performed long ago~\cite{Gava:1981qd}. However, several years later, it was shown that this calculation was not correct and the correct NLO contribution was calculated independently by two groups~\cite{Burnier:2009bk,Brambilla:2010xn}.

The Polyakov loop has been studied in lattice QCD both in SU(N) gauge theories~\cite{Kaczmarek:2002mc,Kaczmarek:2004gv,Digal:2003jc,Mykkanen:2012ri} and in the physically relevant case of $2+1$ flavor QCD~\cite{Aoki:2006br,Cheng:2007jq,Aoki:2009sc,Bazavov:2009zn,Borsanyi:2010bp,Bazavov:2011nk,Bazavov:2013yv,Borsanyi:2015yka}. For the understanding of the screening properties of the deconfined medium it is important to connect lattice calculations with perturbative calculations at high temperatures and to see to what extent these calculations agree. In this perspective it is important to compute next-to-next-to-leading order (NNLO) corrections, which will considerably reduce the uncertainties of the NLO result by fixing the scale dependence of the coupling constant at leading order. The computation of the Polyakov loop at NNLO accuracy is the purpose of the present work.

One feature of the lattice results on the Polyakov loop is Casimir scaling~\cite{Mykkanen:2012ri}. One outcome of our analysis is that Casimir scaling holds up to $\mathcal{O}\left(g^7\right)$. This is important for understanding the lattice results for the Polyakov loops in higher representations~\cite{Gupta:2007ax,Mykkanen:2012ri,Petreczky:2015yta}.

The rest of the paper is organized as follows. In the next section we outline our strategy for the perturbative calculation to $\mathcal{O}\left(g^5\right)$ and discuss the power counting. The calculation of the necessary loop integrals is presented in section~\ref{calculation}, which also contains the main result of the paper. In section~\ref{hoc}, we comment on the higher order perturbative terms discussing Casimir scaling and outlining the $\mathcal{O}\left(g^6\right)$ calculation. In section~\ref{lattice}, we compare the perturbative $\mathcal{O}\left(g^5\right)$ result with available lattice results. Finally, section~\ref{conclusion} contains our conclusions. Several technical details of the calculations are presented in appendices.

\section{Outline of the perturbative calculation}
\label{outline}

In this section, we will outline the perturbative calculation of the Polyakov loop. We will perform calculations directly in QCD as well as using the effective field theory approach. The direct calculation of the NNLO correction to the Polyakov loop is rather involved and its details will be discussed in the next section. On the other hand, as we will see, the calculation that relies on the effective field theory approach is rather simple, because we can draw on previous results.

\subsection{The structure of the perturbative series}

The following way of defining the path ordered exponential is particularly suited for perturbative expansions:
\begin{align}
 L&=\frac{1}{d_R}\Tr\biggl\langle\mathcal{P}\exp\biggl[ig\int_0^{1/T}d\tau A_0(\tau,0)\biggr]\biggr\rangle\notag\\
 &=\sum_{n=0}^\infty\,(ig)^n\int_0^{1/T}d\tau_1\int_0^{\tau_1}d\tau_2\cdots\int_0^{\tau_{n-1}}d\tau_n\,\frac{1}{d_R}\Tr\bigl\langle A_0(\tau_1,0)A_0(\tau_2,0)\cdots A_0(\tau_n,0)\bigr\rangle\,.
\end{align}
The Feynman diagrams for the Polyakov loop can then be drawn as a straight line from $0$ to $1/T$ in the imaginary time direction to which $n$ gluons are attached. The line represents the contour integrations over the gauge fields.  In the gauges we are going to use for this calculation, where the gluon propagator is diagonal in color space, it is possible to split each diagram into a color coefficient and a loop integral. The color coefficient contains the trace over the color matrices from the gauge fields and any structure constants coming from interaction vertices as well as symmetry factors, while the loop integral contains the integrations over Euclidean time, spatial momenta, etc., as well as the propagators and the Lorentz structures.\footnote{Since three- and four-gluon vertices contain a sum over several terms, it may be necessary for some diagrams to split each term separately into color coefficient and loop integral. This is not required for any diagram appearing in this paper, only in the case of tadpoles there appear two terms from the vertex, but they give the same contribution, so we just include a factor $2$ in the color color coefficient.}

It has been shown in~\cite{Gatheral:1983cz,Frenkel:1984pz} that the perturbative series for any closed Wilson line can be rearranged such that it takes the form of an exponential of a series over the same diagrams but with altered color coefficients, several of which are zero. This result has been generalized in~\cite{Gardi:2010rn,*Gardi:2013ita} for the exponentiation of any Wilson line operator (for an application in the context of heavy quarks in thermal QCD, see~\cite{Berwein:2012mw,*Berwein:2013xza}). In the case of the Polyakov loop we have
\begin{align}
 L&=1+C_R\begin{minipage}[b]{0.1\linewidth}\includegraphics[width=\linewidth]{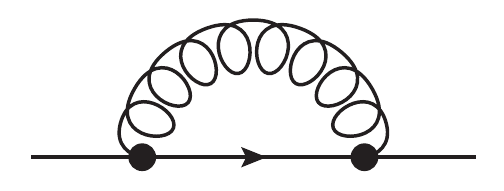}\end{minipage}+C_R^2\begin{minipage}[b]{0.1\linewidth}\includegraphics[width=\linewidth]{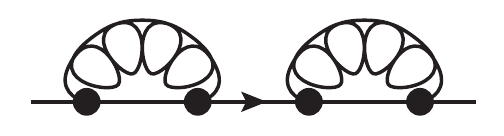}\end{minipage}+C_R\left(C_R-\frac{1}{2}C_A\right)\begin{minipage}{0.1\linewidth}\includegraphics[width=\linewidth]{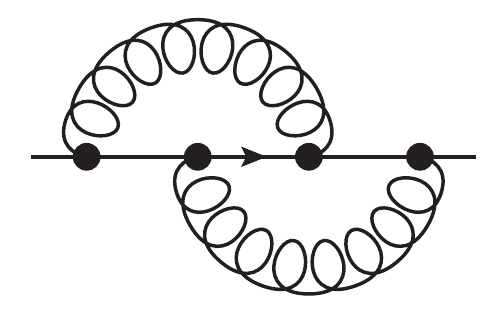}\end{minipage}+C_R^2\begin{minipage}[b]{0.1\linewidth}\includegraphics[width=\linewidth]{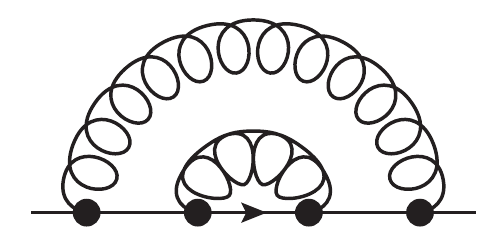}\end{minipage}+\dots\notag\\
 &=\exp\left[C_R\begin{minipage}[b]{0.1\linewidth}\includegraphics[width=\linewidth]{Poly11.pdf}\end{minipage}-\frac{1}{2}C_RC_A\begin{minipage}{0.1\linewidth}\includegraphics[width=\linewidth]{Poly13.pdf}\end{minipage}+\dots\right]=\exp\left(D_1+D_2+\dots\right)\,,
 \label{exponentiation}
\end{align}
where we have written the color coefficients explicitly. $C_R$ and $C_A$ are the quadratic Casimirs of the representation of the Polyakov loop and the adjoint representation respectively. The gluon propagators are understood as resummed. The dots represent diagrams with three or more gluons, which are at least $\mathcal{O}\left(g^6\right)$ and therefore beyond our accuracy. There is also a diagram with three propagators connected by a three-gluon vertex, which would be $\mathcal{O}\left(g^4\right)$ at leading order, but since a three-gluon vertex with only temporal indices vanishes, this diagram gives no contribution in all gauges where the propagator is block diagonal in the temporal and spatial components, so it has been neglected in the expression above. We note that the free energy of the static charge corresponding to the above expression is proportional to $C_R$. This property is known as Casimir scaling.

First, we perform the integral over the Euclidean time in the expression for $D_1$ and $D_2$ to get
\begin{align}
 D_1&=C_R(ig)^2\int_0^{1/T}d\tau_1\int_0^{\tau_1}d\tau_2\,\sum_K\hspace{-16pt}\int\,e^{ik_0(\tau_1-\tau_2)}D_{00}(K)=-\frac{C_Rg^2}{2T}\int_kD_{00}(0,\k)\,,\\
 D_2&=-\frac{1}{2}C_RC_A(ig)^4\int_0^{1/T}d\tau_1\int_0^{\tau_1}d\tau_2\int_0^{\tau_2}d\tau_3\int_0^{\tau_3}d\tau_4\,\sum_{K,\,Q\,}\hspace{-15pt}\int\,e^{ik_0(\tau_1-\tau_3)}e^{iq_0(\tau_2-\tau_4)}D_{00}(K)D_{00}(Q)\notag\\
 &=-\frac{C_RC_Ag^4}{4T}\int_{k,\,q}\left[\frac{1}{12T}D_{00}(0,\k)D_{00}(0,\q)-{\sum_{k_0}}'\frac{1}{k_0^2}D_{00}(K)\bigl(2D_{00}(0,\q)-D_{00}(k_0,\q)\bigr)\right],
\label{eq:D2}
\end{align}
where $K$ and $Q$ include both the spatial momenta $\k$ and $\q$ and the Matsubara frequencies $k_0$ and $q_0$, which are given by $2\pi T n$ with $n\in\mathbb{Z}$. We use boldface letters to denote vectors in $d$~dimensions and regular font letters for the absolute value, so $k^2=\bm{k}^2$. The sum-integral symbols are a shorthand for the Matsubara sums and the $d$-dimensional integrals, which are defined in the following way:
\begin{equation}
 \sum_K\hspace{-16pt}\int f(k_0,\bm{k})=\sum_{k_0}\int_k f(k_0,\bm{k})=T\sum_{n\in\mathbb{Z}}\int\frac{d^dk}{(2\pi)^d}\,f(2\pi Tn,\bm{k})\,.
\end{equation}
The sum with a prime denotes a Matsubara sum without the zero mode, i.e., $n\neq0$. Up to this point the discussion is independent of the choice of gauge for the perturbative calculations. In this paper we will use Feynman gauge, static gauge, and Coulomb gauge. In Appendix~\ref{propagators}, we discuss the gluon propagators and self-energies in these gauges.

The integration momenta $\k$ and $\q$ can either be of the order of the temperature scale $T$ or of the scale of the Debye mass $m_D$. In principle, they may also scale with the nonperturbative magnetic mass $m_M$. The magnetic mass enters the temporal propagators not directly but only through self-energies. Hence, as we will show at the end of this section, momentum regions scaling with $m_M$ contribute only to $\mathcal{O}\left(g^7\right)$. We use dimensional regularization to treat both infrared and ultraviolet divergences. In this regularization scheme the different momentum scales can be separated by expanding the integrand according to the hierarchy $T\gg m_D\gg m_M$.

We start considering $D_1$. Separating out the contributions from the scales $T$, $m_D$, and $m_M$ we write
\begin{align}
 D_1&=-\frac{C_Rg^2}{2T}\int_k\,\frac{1}{k^2+\Pi(0,k)}\notag\\
 &=-\frac{C_Rg^2}{2T}\left\{\int_{k\sim T}\frac{-\Pi^{(1)}_T(0,k)}{k^4}+\int_{k\sim m_D}\left[\frac{1}{k^2+m_D^2}-\frac{\Pi^{(1)}_{m_D}(0,k)}{\left(k^2+m_D^2\right)^2}+\frac{\Pi^{(1)}_{m_D}(0,k)^2}{\left(k^2+m_D^2\right)^3}\right.\right.\notag\\
 &\hspace{13pt}-\left.\left.\frac{1}{\left(k^2+m_D^2\right)^2}\left(\frac{d\Pi^{(1)}_T}{dk^2}(0,0)\,k^2+\Pi^{(2)}_T(0,0)+\Pi^{(2)}_{m_D}(0,k)+\Pi^{(1)}_{m_M}(0,k)\right)\,\right]\right\}+\mathcal{O}\left(g^6\right).
 \label{eq:D1}
\end{align}
Here $\Pi_T^{(i)}$, $\Pi_{m_D}^{(i)}$, and $\Pi_{m_M}^{(i)}$ denote the contributions to the self-energy of the $A_0$ field at $i$-loop order coming from loop momenta of order $T$, $m_D$, and $m_M$. There can also be self-energies where the loop momenta are not all of the same scale, but these do not contribute until $\mathcal{O}\left(g^6\right)$. The self-energies entering the above equation depend on the choice of gauge. The terms proportional to $\Pi_T^{(1)}$ and $\Pi_{m_D}^{(1)}$, together with tree-level $D_2$, give rise to the known $\mathcal{O}\left(g^4\right)$ term in the expression of the Polyakov loop~\cite{Burnier:2009bk,Brambilla:2010xn}. The 2-loop scale $T$ contribution to the self-energy $\Pi_T^{(2)}$ as well as the $\dfrac{d\Pi^{(1)}_T}{dk^2}(0,0)\,k^2$ term give rise to terms of $\mathcal{O}\left(g^5\right)$, some of which have been identified in Ref.~\cite{Brambilla:2010xn} and are related to the running of the coupling constant. The terms proportional to $\Pi^{(2)}_{m_D}$  and $\bigl(\Pi^{(1)}_{m_D}\bigr)^2$ are new and also contribute at $\mathcal{O}\left(g^5\right)$ to the Polyakov loop. In section~\ref{calculation}, we will discuss the calculation of these terms in detail. Finally, the term proportional to $\Pi_{m_M}^{(1)}$ does not contribute to the Polyakov loop at $\mathcal{O}\left(g^5\right)$ and $\mathcal{O}\left(g^6\right)$. This will also be shown in section~\ref{calculation}.

Concerning $D_2$, it is easy to see that the leading order contribution of the first term in Eq.~\eqref{eq:D2} in any gauge comes only from the scale $m_D$ and is of $\mathcal{O}\left(g^6\right)$, which is beyond the accuracy of this calculation. It was already identified in Ref.~\cite{Brambilla:2010xn}. The second and third terms in Eq.~\eqref{eq:D2} do not contribute in static gauge because $D_{00}(K)$ vanishes for nonzero $k_0$. In Coulomb gauge the second term starts to contribute at $\mathcal{O}\left(g^7\right)$ and the third at $\mathcal{O}\left(g^8\right)$, since the leading order propagators with nonzero Matsubara frequencies are scaleless and need at least one loop insertion to not vanish. In Feynman gauge the second term in Eq.~\eqref{eq:D2} contributes already at $\mathcal{O}\left(g^5\right)$ for $k\sim T$ and $q\sim m_D$, while the third starts to contribute at $\mathcal{O}\left(g^4\right)$ when both momenta are of the scale $T$. There is no $\mathcal{O}\left(g^5\right)$ contribution from the third term, since the scale $m_D$ can only enter Feynman gauge propagators with nonzero frequencies through loops, which would at least be of $\mathcal{O}\left(g^7\right)$.

In summary we see that the $\mathcal{O}\left(g^5\right)$ contribution to the Polyakov loop receives two different contributions. The first comes from terms with mixed scales like the two-loop self-energy at a scale $m_D$ with loop momenta of order $T$ or two-gluon exchange. The second comes from the two-loop self energy with loop momenta of order $m_D$. In the effective field theory approach that will be discussed in the next subsection these two contributions correspond to two different steps of the calculation: the determination of the matching coefficients and the calculation of the correlators in the effective theory, respectively.

\subsection{The Polyakov loop in an effective field theory approach}

The separation between the scales $T$ and $m_D$ that was used in the calculation of the previous section can be incorporated in an effective theory, the so-called electrostatic QCD (EQCD)~\cite{Braaten:1995jr,Kajantie:1997tt} (for earlier and related works on this subject see Refs.~\cite{Ginsparg:1980ef,Appelquist:1981vg,D'Hoker:1981us,Nadkarni:1982kb,Landsman:1989be,Reisz:1991er,LaCock:1991hh,Karkkainen:1992wu,Karkkainen:1993bu,Patkos:1997cw,Karsch:1998tx,Kajantie:1997pd,Laine:1999hh,Hart:1999dj,Hart:2000ha,Cucchieri:2000cy,Cucchieri:2001tw}). In EQCD the scale $T$ is integrated out, which includes all fields with nonzero Matsubara frequencies. This means that quark fields are completely absent and the gluons do not depend on the imaginary time coordinate. So EQCD is a three-dimensional field theory where $A_0$ no longer plays the role of a gauge field, but becomes an adjoint scalar field. As a consequence, a mass term for $A_0$ does not break gauge invariance in EQCD.

The contribution from the nonperturbative magnetic scale $m_M$ can also be calculated in this effective field theory approach. Namely if $m_D\gg m_M$ the scale $m_D$ can also be integrated out leading to an effective field theory called magnetostatic QCD (MQCD), which is the three-dimensional Yang-Mills theory~\cite{Braaten:1995jr}. Using this sequence of effective theories the weak coupling expansion of the QCD pressure has been calculated~\cite{Braaten:1995jr} finding a solution to the well known infrared problem~\cite{Linde:1980ts}.

The imaginary time integration in the EQCD action just gives a factor $1/T$, since the fields no longer depend on the imaginary time coordinate. This factor can be absorbed in a rescaling of the fields and gauge coupling by a factor $\sqrt{T}$ (denoted by a tilde over the fields). The Lagrangian is then given by
\begin{align}
 \mathcal{L}_{EQCD}=&\,\frac{1}{4}\left(\widetilde{F}_{ij}^a\right)^2+\frac{1}{2}\left(\widetilde{D}_i^{ab}\widetilde{A}_0^b\right)^2+\frac{m_E^2}{2}\left(\widetilde{A}_0^a\right)^2\notag\\*
 &+\lambda_E\left(\Tr\left[\widetilde{A}_0^2\right]\right)^2+\bar{\lambda}_E\left(\Tr\left[\widetilde{A}_0^4\right]-\frac{1}{2}\left(\Tr\left[\widetilde{A}_0^2\right]\right)^2\right)+\dots\,,
\end{align}
where $\widetilde{F}_{ij}^a=\partial_i\widetilde{A}_j^a-\partial_j\widetilde{A}_i^b+g_Ef^{abc}\widetilde{A}_i^b\widetilde{A}_j^c$, $\widetilde{D}_i^{ab}\widetilde{A}_0^b =\partial_i\widetilde{A}_0^a+g_Ef^{abc}\widetilde{A}_i^b\widetilde{A}_0^c$, and the dots stand for higher dimensional operators. The fields $\widetilde{A}_0$ and $\widetilde{A}_i$ in the above Lagrangian have canonical dimension $1/2$. The gauge coupling $g_E$ of EQCD is dimensionful. At leading order we have $g_E=g\sqrt{T}$, $m_E=m_D$, $\lambda_E=(6+N-n_f)g^4T/24\pi^2$, and $\bar{\lambda}_E=(N-n_f)g^4T/12\pi^2$ (see, e.g., \cite{Kajantie:1997tt}). The second quartic interaction is a vanishing operator for $N=2$ or $N=3$, so any result can only depend on $\bar{\lambda}_E$ for $N>3$. The couplings $g_E$, $\lambda_E$, and $\bar{\lambda}_E$ have been calculated to next-to-leading order (NLO)~\cite{Kajantie:1997tt}. The three-dimensional gauge coupling $g_E$ is known to next-to-next-to-leading order~\cite{Laine:2005ai}. The NLO correction to $m_E^2$ has been calculated in Ref.~\cite{Braaten:1995jr}
\begin{align}
 m_E^2=&\,m_D^2\left[1+\frac{\alpha_\mathrm{s}}{4\pi}\left(\frac{5}{3}C_A+\frac{4}{3}T_Fn_f\left(1-4\ln2\right)+2\beta_0\left(\gamma_E+\ln\frac{\mu}{4\pi T}\right)\right)\right]-2C_FT_Fn_f\alpha_\mathrm{s}^2T^2\,,
\label{mE}
\end{align}
where $C_F=T_F(N^2-1)/N$ is the quadratic Casimir of the fundamental representation. Here and in the rest of this section we express all matching parameters in terms of the renormalized coupling. Thus the pole that appears with the first coefficient of the beta function $\beta_0=11C_A/3-4T_Fn_f/3$ has already been canceled. The relation between the bare coupling $g$ and the renormalized coupling $g_R$ at one-loop in the $\overline{\mathrm{MS}}$-scheme is
\begin{equation}
 g^2=g_R^2\left[1-\frac{\alpha_\mathrm{s}\beta_0}{4\pi}\left(\frac{1}{\epsilon}-\gamma_E+\ln4\pi\right)+\dots\right]\,,
\end{equation}
where $\epsilon$ is related to the number of spatial dimensions $d$ through $d=3-2\epsilon$.

In EQCD we can write the Polyakov loop in the following way~\cite{Burnier:2009bk}
\begin{equation}
 L=\mathcal{Z}_0-\mathcal{Z}_2\frac{g^2}{2\,d_R\,T}\mathrm{Tr}\bigl\langle\widetilde{A}_0^2\bigr\rangle+\mathcal{Z}_4\frac{g^4}{24\,d_R\,T^2}\mathrm{Tr}\bigl\langle\widetilde{A}_0^4\bigr\rangle+\dots\,.
\end{equation}
The matching coefficients $\mathcal{Z}_n$ are equal to $1$ at leading order, at higher orders they can be written as an expansion in $\alpha_\mathrm{s}$, i.e., only in even powers of $g$. In the power counting of EQCD, every power of $\widetilde{A}_0$ counts as $\sqrt{gT}$, so the term proportional to $\mathcal{Z}_4$ starts to contribute at $\mathcal{O}\left(g^6\right)$. In EQCD only the scales $m_D$ and $m_M$ are still dynamical, which means that no loop momenta of order $T$ appear in the evaluation of the Feynman diagrams. The contributions of such loops are contained in higher order corrections to the matching coefficients $\mathcal{Z}_n$ and the EQCD parameters.

For the determination of $\mathcal{Z}_0$ and $\mathcal{Z}_2$ from QCD it is convenient to use the static gauge. In this gauge we can write
\begin{equation}
 L=1-\frac{g^2}{2\,d_R\,T^2}\,\Tr\left\langle A_0^2\right\rangle+\frac{g^4}{24\,d_R\,T^4}\,\Tr\left\langle A_0^4\right\rangle+\dots\,.
\end{equation}
Now we can separate each contribution into a static and a nonstatic piece, i.e., we can write
\begin{equation}
 \left\langle A_0^2\right\rangle=\left\langle A_0^2\right\rangle_\mathrm{s}+\left\langle A_0^2\right\rangle_\mathrm{ns}. 
\end{equation}
The notation $\langle\dots\rangle_\mathrm{ns}$ means that there appear only loop momenta of order $T$ in the evaluation of the corresponding Feynman diagrams, which corresponds to a strict perturbative expansion in $g$ without any resummation of self-energies. The notation $\langle\dots\rangle_\mathrm{s}$ then means that some or all loop momenta are of order $m_D$ or $m_M$. The corresponding Matsubara frequencies have to be zero, hence the name ``static''. We can write down a similar decomposition for $\Tr\bigl\langle A_0^4\bigr\rangle$.

The sum over all nonstatic pieces exactly gives $\mathcal{Z}_0$. Since in static gauge the scale $T$ can enter the temporal propagators only through loops, the nonstatic part of the $A_0^4$ contribution will contribute first at $\mathcal{O}\left(\alpha_\mathrm{s}^4\right)$. The leading order result for the $A_0^2$ piece can be found in Ref.~\cite{Brambilla:2010xn} and gives
\begin{equation}
 \mathcal{Z}_0=1+\frac{C_R\alpha_s^2}{2}\left[C_A\left(\frac{1}{2\epsilon}+1-\gamma_E+\ln\frac{\pi\mu^2}{T^2}\right)-2T_Fn_f\ln2\right]+\mathcal{O}\left(\alpha_s^3\right)\,.
\end{equation}
The pole in $\epsilon$ is not related to charge renormalization, but corresponds to an infrared divergence in the nonstatic part that cancels against an ultraviolet divergence in the static piece, or equivalently in the EQCD calculation.

The sum over the static pieces then contains all contributions from the scales $m_D$ and $m_M$ and thus corresponds to the EQCD representation of the Polyakov loop without the unit operator. Up to $\mathcal{O}\left(g^5\right)$ it is sufficient to consider only the quadratic terms, i.e., $\bigl\langle A_0^2\bigr\rangle_\mathrm{s}=\mathcal{Z}_2\bigl\langle\widetilde{A}_0^2\bigr\rangle$. The two gauge fields in $\bigl\langle A_0^2\bigr\rangle_\mathrm{s}$ themselves can carry either momenta $k\ll T$ or $k\sim T$, however, in the latter case they only start to contribute to the static piece at three-loop order. The first loop from the two gauge fields in the correlator is scaleless, so another loop is needed to introduce the scale $T$ and a third one to include the scale $m_D$ in order to be counted towards the static piece. This corresponds to diagrams like $L_{10}$ and $L_{12}$ in Fig.~\ref{diagrams} when only the tadpole or the subloop momentum is of order $m_D$ and the two other momenta are of order $T$. The two scale $T$ integrations give a contribution of $\mathcal{O}\left(\alpha_\mathrm{s}^2\right)$ to $\mathcal{Z}_2$ and only one propagator with a momentum of order $m_D$ remains, which corresponds to the leading order of $\bigl\langle\widetilde{A}_0^2\bigr\rangle$.

In the former case we can relate the QCD field $A_0$ to $\sqrt{Z_2}\widetilde{A}_0$ in EQCD, where the wave function normalization constant $Z_2$ can be obtained from the small momentum expansion of the propagator:
\begin{equation}
 D_{00}^\mathrm{QCD}(k\ll T)=Z_2D_{00}^\mathrm{EQCD}(k)+\dots\,.
\end{equation}
The dots stand for higher powers in the small $k^2$ expansion, which correspond to higher order two-point interactions in EQCD. From this expression it follows that
\begin{equation}
 Z_2=\left(1+\frac{d\Pi_T}{dk^2}\left(k^2=0\right)\right)^{-1}\,,
\end{equation}
and with the result from~\cite{Brambilla:2010xn} we have up to corrections of $\mathcal{O}\left(\alpha_\mathrm{s}^2\right)$
\begin{equation}
 \mathcal{Z}_2=Z_2=1+\frac{\alpha_\mathrm{s}}{4\pi}\left[\frac{11}{3}C_A+\frac{4}{3}T_Fn_f(1-4\ln2)+2\beta_0\left(\gamma_E+\ln\frac{\mu}{4\pi T}\right)\right]\,.
 \label{Z2}
\end{equation}

Now that we have determined the matching coefficients to the desired order, we can write the weak coupling expansion for $\bigl\langle\widetilde{A}_0^2\bigr\rangle$ as
\begin{equation}
 \frac{1}{d_R}\Tr\bigl\langle\widetilde{A}_0^2\bigr\rangle=-\frac{C_Rm_E}{4\pi}\left(1+a_1\frac{g_E^2}{m_E}+a_2\frac{g_E^4}{m_E^2}+a_3\frac{g_E^6}{m_E^3}+a_4\frac{\lambda_E}{m_E}+\dots \right)\,,
\label{pert3d}
\end{equation}
using simple dimensional analysis. In the above expression we explicitly wrote down all the terms contributing up to $\mathcal{O}\left(g^6\right)$ and ignored the magnetic mass scale $m_M$. We will return to the contribution from the scale $m_M$ later.

In Eq.~\eqref{pert3d} the terms proportional to $a_i$ come from the $i$-loop self-energy of the $\widetilde{A}_0$ field. The coefficient $a_1$ is known~\cite{Burnier:2009bk,Brambilla:2010xn}. We are primarily interested in the NNLO, i.e., $\mathcal{O}\left(g^5\right)$ contribution to $L$. It is evident from Eqs.~\eqref{mE}, \eqref{Z2}, and \eqref{pert3d} that the mixed scale contributions from the previous section come from the $\mathcal{O}\left(\alpha_\mathrm{s}\right)$ corrections to $\mathcal{Z}_2$ and $m_E^2$, while the pure scale $m_D$ term comes from the two-loop self-energy contribution contained in the coefficient $a_2$. One can perform a similar analysis for $\bigl\langle\widetilde{A}_0^4\bigr\rangle$ and see that it contributes at orders $\alpha_\mathrm{s}^2m_E^2$, $\alpha_\mathrm{s}^3m_E$, etc. It is also easy to generalize the analysis for $\bigl\langle\widetilde{A}_0^{2n}\bigr\rangle$, $n\ge3$, and see that these terms do not contribute at $\mathcal{O}\left(g^5\right)$.

The only remaining task is now to calculate the coefficient $a_2$. This can be done using the EQCD calculation of the pressure~\cite{Braaten:1995jr}
\begin{equation}
 p=-T\left(f_E+f_M-\frac{1}{V}\ln Z_\mathrm{MQCD}\right).
\end{equation}
Here we use the same notation as in Ref.~\cite{Braaten:1995jr}, i.e., $f_E$ denotes the contribution from the scale $T$,
$f_M$ denotes the contribution from the scale $m_D$, and $Z_\mathrm{MQCD}$ is the partition function of MQCD, which
is completely nonperturbative. Ignoring the contribution from MQCD it is easy to see that since
\begin{equation}
 f_M=-\frac{1}{V}\ln\int\mathcal{D}\widetilde{A}_0^a\mathcal{D}\widetilde{A}_i^a\,\exp\left[-\int d^3x\,\mathcal{L}_\mathrm{EQCD}\right]\,,
\end{equation}
it follows that
\begin{equation}
 \frac{1}{d_R}\Tr\bigl\langle\widetilde{A}_0^2\bigr\rangle=\frac{C_R}{N^2-1}\bigl\langle\widetilde{A}_0^a\widetilde{A}_0^a\bigr\rangle=\frac{2C_R}{N^2-1}\,\frac{\partial f_M}{\partial m_E^2}\,.
 \label{dfmdmE}
\end{equation}
Using the expression for $f_M$ from~\cite{Braaten:1995jr} we get
\begin{align}
 \frac{1}{d_R}\Tr\bigl\langle\widetilde{A}_0^2\bigr\rangle=&-\frac{C_R m_E}{4 \pi}+\frac{C_RC_Ag_E^2}{(4\pi)^2}\left[\frac{1}{2\epsilon}+\frac{1}{2}-\gamma_E+\ln\frac{\pi\mu^2}{m_E^2}\right]\notag\\
 &+\frac{2C_RC_A^2}{(4\pi)^3}\frac{g_E^4}{m_E}\left(\frac{89}{48}-\frac{11}{12}\ln2+\frac{\pi^2}{12}\right)+\mathcal{O}\left(g^4\right)\,.
\label{static}
\end{align}

The first term corresponds to the well-known leading order result. The second term is identical to the $\mathcal{O}\left(g^4\right)$ static contribution to $\bigl\langle A_0^2\bigr\rangle$ (c.f.~Eq.~$(44)$ of Ref.~\cite{Brambilla:2010xn}). The $1/\epsilon$ pole in this term is exactly the ultraviolet divergence that cancels against the infrared pole in the nonstatic contribution to $\bigl\langle A_0^2\bigr\rangle$~\cite{Brambilla:2010xn}. The scale dependence cancels in the same way. The last term gives the coefficient $a_2$ we are interested in. 

We still need to calculate the $\mathcal{O}\left(g^5\right)$ contribution arising from the $\mathcal{O}\left(\alpha_\mathrm{s}\right)$ corrections to $m_E$ and $\mathcal{Z}_2$ times the leading order result for $\Tr\bigl\langle\widetilde{A}_0^2\bigr\rangle=-C_R m_D/4\pi$. Using Eqs.~\eqref{mE} and~\eqref{Z2} we find that this $\mathcal{O}\left(g^5\right)$ contribution is
\begin{equation}
 \frac{3C_R\alpha_\mathrm{s}^2m_D}{16\pi T}\left[3C_A+\frac{4}{3}T_Fn_f(1-4\ln 2)+2\beta_0\left(\gamma_E+\ln\frac{\mu}{4\pi T}\right)\right]-\frac{C_RC_Fn_fT_F\alpha_\mathrm{s}^2T}{2m_D}\,.
\label{z2mE}
\end{equation}
With this result and Eq.~\eqref{static} we find the $\mathcal{O}\left(g^5\right)$ contribution to the Polyakov loop
\begin{align}
 L\,\Bigr|_{g^5}=&\,\frac{3C_R\alpha_\mathrm{s}^2m_D}{16\pi T}\left[3C_A+\frac{4}{3}T_Fn_f(1-4\ln 2)+2\beta_0\left(\gamma_E+\ln\frac{\mu}{4\pi T}\right)\right]\notag\\
 &-\frac{C_RC_A^2\alpha_\mathrm{s}^3T}{m_D}\left(\frac{89}{48}-\frac{11}{12}\ln2+\frac{\pi^2}{12}\right)-\frac{C_RC_Fn_fT_F\alpha_\mathrm{s}^3T}{2m_D}\,.
\label{Lg5}
\end{align}
The above equation is the main result of this paper. In the next section we will obtain this result via direct calculations 
in QCD.

The contribution from the scale $m_M$ to $\bigl\langle\widetilde{A}_0^2\bigr\rangle$, which we have neglected so far, can be calculated using MQCD, the effective theory obtained from EQCD by integrating out the electric scale $m_E\sim m_D$. The only scale in MQCD is the dimensionful coupling constant $g_M\sim \sqrt{m_M}$, which is given at leading order as $g_M=g_E$. Since in this theory we have only the three-dimensional gauge fields, we write
\begin{equation}
 L=\mathcal{Z}^M_0+\frac{\mathcal{Z}^M_1}{2m_D^3}\bigl\langle\widetilde{F}_{ij}^a\widetilde{F}_{ij}^a\bigr\rangle_\mathrm{MQCD}+\dots\,.
\end{equation}
The matching constant $\mathcal{Z}^M_0$ contains the contributions to $L$ from the scales $T$ and $m_D$, so $\mathcal{Z}^M_0=L$ up to $\mathcal{O}\left(g^5\right)$. The matching constant $\mathcal{Z}^M_1$ has been calculated in
Ref.~\cite{Braaten:1994qx} for the fundamental representation. We have repeated that calculation, but allowed for general representations; the result is
\begin{equation}
 \mathcal{Z}^M_1=\frac{C_RC_A\alpha_\mathrm{s}^2\pi}{12\left(N^2-1\right)}+\mathcal{O}\left(g^5\right)\,,
\end{equation}
and for $C_R=C_F$ one obtains the expression from~\cite{Braaten:1994qx}. Since $\bigl\langle\widetilde{F}_{ij}^a\widetilde{F}_{ij}^a\bigr\rangle\sim m_M^3\sim g_M^6$, we see that the contribution from the magnetic scale first appears at $\mathcal{O}\left(g^7\right)$. Through the explicit calculations presented in section~\ref{calculation} and appendix~\ref{scalemM} we will see that the magnetic contributions at $\mathcal{O}\left(g^5\right)$ and $\mathcal{O}\left(g^6\right)$ indeed vanish.

The $\mathcal{O}\left(g^7\right)$ contribution to the Polyakov loop can be obtained using lattice calculations in MQCD. However, for interesting temperature ranges the separation of the scales $m_D$ and $m_M$ is not obvious. Therefore, it is more practical to calculate $\bigl\langle \widetilde{A}_0^2\bigr\rangle$ using lattice calculations in EQCD. Such lattice calculations have been performed with the aim to estimate the QCD pressure using the EQCD approach in Ref.~\cite{Hietanen:2008tv}. We will use this lattice EQCD result when comparing the weak coupling expansion of the Polyakov loop with lattice results in QCD in section~\ref{lattice}.

\section{\texorpdfstring{Calculation of the $\bm{\mathcal{O}\left(g^5\right)}$ correction to the Polyakov loop}{Calculation of the O(g\^{}5) correction to the Polyakov loop}}
\label{calculation}

In this section we will present the calculations of the $\mathcal{O}\left(g^5\right)$ contribution to the Polyakov loop directly in QCD. From the discussion in the previous section it is clear that the diagrams that contribute at $\mathcal{O}\left(g^5\right)$ always have at least one momentum integral of order $m_D$, while the self-energy contributions may arise from the scales $T$, $m_D$, or $m_M$. In what follows we will refer to them as contributions from the scale $T$, $m_D$, or $m_M$, even though all the loop diagrams also involve the scale $m_D$. We will perform the calculations in Coulomb gauge and in Feynman gauge. The contribution from the diagram $D_2$ is only relevant in Feynman gauge. It involves one integral over the scale $m_D$ and another sum-integral over the scale $T$, so we will refer to it as a part of the contribution from the scale $T$.

\subsection{\texorpdfstring{Contribution from the scale $\bm{T}$}{Contribution from the scale T}}

All self-energies relevant for the contribution from the scale $T$ in Feynman gauge are known and can be found in Ref.~\cite{Braaten:1995jr} (they use a slightly different convention for the $\overline{\mathrm{MS}}$-scheme, which can be converted into our convention by replacing the renormalization scale $\Lambda^2$ in their expressions by $4\pi e^{-\gamma_E}\mu^2$):
\begin{align}
 \Pi^{(1)}_T(0,0)&\equiv m_D^2(\epsilon)=\frac{g^2T^2}{3}\left[\left(C_A+T_Fn_F\right)+C_A\left(-\gamma_E+2\frac{\zeta'(-1)}{\zeta(-1)}+\ln\frac{\mu^2}{4\pi T^2}\right)\epsilon\right.\notag\\
 &\hspace{96pt}+\left.T_Fn_f\left(1-\gamma_E+2\frac{\zeta'(-1)}{\zeta(-1)}+\ln\frac{\mu^2}{16\pi T^2}\right)\epsilon\,\right]\,,\\
 \frac{d\Pi^{(1)}_T}{dk^2}(0,0)&=-\frac{g^2}{(4\pi)^2}\hspace{-2pt}\left[\frac{5}{3}C_A\left(\frac{1}{\epsilon}-\frac{1}{5}+\gamma_E+\ln\frac{\mu^2}{4\pi T^2}\right)-\frac{4}{3}T_Fn_f\left(\frac{1}{\epsilon}-1+\gamma_E+\ln\frac{4\mu^2}{\pi T^2}\right)\right]\hspace{-2pt},\\
 \Pi^{(2)}_T(0,0)&=\frac{g^4T^2}{(4\pi)^2}\left[\frac{2}{3}C_A^2\left(\frac{1}{\epsilon}+1+2\frac{\zeta'(-1)}{\zeta(-1)}+2\ln\frac{\mu^2}{4\pi T^2}\right)\right.\notag\\
 &\hspace{50pt}+\left.\frac{2}{3}C_AT_Fn_f\left(\frac{1}{\epsilon}+2+2\frac{\zeta'(-1)}{\zeta(-1)}+2\ln\frac{\mu^2}{8\pi T^2}\right)-2C_FT_Fn_F\right]\notag\\
 &=\frac{g^2}{(4\pi)^2}\left[2C_Am_D^2(\epsilon)\left(\frac{1}{\epsilon}+1+\gamma_E+\ln\frac{\mu^2}{4\pi T^2}\right)-2g^2T^2C_FT_Fn_f\right]\,.
\end{align}
In the last line we have re-expressed some terms through $m_D^2(\epsilon)$, i.e., the leading order Debye mass
with ${\cal O}(\epsilon)$ corrections.
This will be crucial for the cancellation of the $1/\epsilon$-poles. 
The $\mathcal{O}(\epsilon)$ terms of $m_D^2(\epsilon)$ are necessary at this point.

With these we can calculate the first scale $T$ contribution from diagram $D_1$ at $\mathcal{O}\left(g^5\right)$ in Feynman gauge (FG):
\begin{align}
 D_1\Bigr|^{FG}_{g^5,\,T}&=\frac{C_Rg^2}{2T}\int_{k\sim m_D}\frac{1}{\left(k^2+m_D^2\right)^2}\left(\frac{d\Pi^{(1)}_T}{dk^2}(0,0)\,k^2+\Pi^{(2)}_T(0,0)\right)\notag\\
 &=\frac{3C_R\alpha_\mathrm{s}^2m_D(\epsilon)}{16\pi T}\left[\frac{7}{3}C_A\left(\frac{1}{\epsilon}+\frac{23}{21}+2\ln\frac{\mu^2}{2Tm_D}\right)-\frac{4}{3}T_Fn_f\left(\frac{1}{\epsilon}+\frac{1}{3}+2\ln\frac{2\mu^2}{Tm_D}\right)\right]\notag\\*
 &\hspace{13pt}-\frac{C_RC_FT_Fn_f\alpha_\mathrm{s}^3T}{2m_D}\,.
\end{align}
The scale $T$ contribution from $D_2$ is given by
\begin{align}
 D_2\Bigr|^{FG}_{g^5,\,T}&=\frac{C_RC_Ag^4}{2T}{\sum_{q_0}}'\int_{q\sim T}\int_{k\sim m_D}\frac{1}{q_0^2\left(q_0^2+q^2\right)\left(k^2+m_D^2\right)}\notag\\
 &=\frac{C_RC_A\alpha_\mathrm{s}^2m_D(\epsilon)}{4\pi T}\left(\frac{1}{\epsilon}+4+2\ln\frac{\mu^2}{2Tm_D}\right)\,,
\end{align}
and together they give
\begin{align}
 &\left(D_1+D_2\right)\Bigr|^{FG}_{g^5,\,T}=-\frac{C_RC_FT_Fn_f\alpha_\mathrm{s}^3T}{2m_D}\notag\\
 &\hspace{15pt}+\frac{3C_R\alpha_\mathrm{s}^2m_D(\epsilon)}{16\pi T}\left[\frac{11}{3}C_A\left(\frac{1}{\epsilon}+\frac{71}{33}+2\ln\frac{\mu^2}{2Tm_D}\right)-\frac{4}{3}T_Fn_f\left(\frac{1}{\epsilon}+\frac{1}{3}+2\ln\frac{2\mu^2}{Tm_D}\right)\right]\,.
\end{align}

We see now that the coefficient of the $1/\epsilon$-terms is proportional to the first coefficient of the beta function $\beta_0=11C_A/3-4T_Fn_f/3$. This suggests that they are removed through charge renormalization, which is indeed the case. The first counter term comes from charge renormalization of the $\mathcal{O}\left(g^3\right)$ term. We need to be careful with the $\epsilon\to0$ limit, so we will keep the dimension $d$ general until the last step:
\begin{align}
 D_1\Bigr|_{g^3}&=-\frac{C_Rg^2}{2T}\int_{k\sim m_D}\frac{1}{k^2+m_D^2}=-\frac{C_Rg^2\Gamma\left(1-\frac{d}{2}\right)m_D^{d-2}\mu^{2\epsilon}}{2(4\pi)^{\frac{d}{2}}T}\notag\\
 &\stackrel{g_0\to g_R}{\longrightarrow}-\frac{C_Rg^2\Gamma\left(1-\frac{d}{2}\right)m_D^{d-2}(\epsilon)\mu^{2\epsilon}}{2(4\pi)^{\frac{d}{2}}T}\left[1-\frac{d}{2}\frac{\alpha_\mathrm{s}\beta_0}{4\pi}\left(\frac{1}{\epsilon}-\gamma_E+\ln4\pi\right)+\mathcal{O}\left(\alpha_\mathrm{s}^2\right)\right]\,.
\end{align}
The factor $d/2$ comes from the power of $\alpha_\mathrm{s}$: $g^2m_D^{d-2}\propto\alpha_\mathrm{s}^{\frac{d}{2}}$. 
Including the counter term for the charge renormalization we get the full contribution from the scale $T$:
\begin{align}
 \left(D_1+D_2\right)\Bigr|_{g^5,\,T}=&\,\frac{3C_R\alpha_\mathrm{s}^2m_D}{16\pi T}\left[3C_A+\frac{4}{3}T_Fn_f\left(1-4\ln2\right)+2\beta_0\left(\gamma_E+\ln\frac{\mu}{4\pi T}\right)\right]\notag\\
 &-\frac{C_RC_FT_Fn_f\alpha_\mathrm{s}^3T}{2m_D}\,.
\end{align}
We no longer indicate Feynman gauge in this final result for the scale $T$ contribution, because it is gauge invariant.

The corresponding calculation goes the same way for both Coulomb (CG) and static gauge (SG). $D_2$ is scaleless at $\mathcal{O}\left(g^5\right)$, so only $D_1$ contributes. It has been shown in~\cite{Braaten:1995jr} that the electric mass parameter $m_E$ of EQCD is given up to $\mathcal{O}\left(\alpha_\mathrm{s}^2\right)$ by
\begin{equation}
 m_E^2=\Pi_T^{(1)}(0,0)+\Pi_T^{(2)}(0,0)-\Pi_T^{(1)}(0,0)\frac{d\Pi_T^{(1)}}{dk^2}(0,0)\,.
\end{equation}
Since $m_E$ is a gauge invariant parameter, we can eliminate $\Pi_T^{(2)}(0,0)$ in Coulomb or static gauge from this equation and express it through the Feynman gauge results and $\dfrac{d\Pi_T^{(1)}}{dk^2}(0,0)$, which is the same for Coulomb and static gauge and can be found, e.g., in~\cite{Brambilla:2010xn}:
\begin{equation}
 \frac{d\Pi_T^{(1)}}{dk^2}(0,0)=-\frac{g^2}{(4\pi)^2}\left[\frac{11}{3}C_A+\frac{4}{3}T_Fn_f\left(1-4\ln2\right)+\beta_0\left(\frac{1}{\epsilon}+\gamma_E+\ln\frac{\mu^2}{4\pi T^2}\right)\right]\,.
\end{equation}

With this we have
\begin{align}
 \Pi^{(2)}_T\Bigr|^{CG/SG}&=\biggl(\Pi^{(2)}_T-m_D^2(\epsilon)\frac{d\Pi^{(1)}_T}{dk^2}\biggr)\biggr|^{FG}+m_D^2(\epsilon)\frac{d\Pi^{(1)}_T}{dk^2}\biggr|^{CG/SG}\notag\\
 &=-\frac{2g^2}{(4\pi)^2}\left(C_Am_D^2+g^2T^2C_FT_Fn_f\right)\,.
\end{align}
The contributions from the scale $T$ are now
\begin{align}
 D_1\Bigr|^{CG/SG}_{g^5,\,T}&=\frac{C_Rg^2}{2T}\int_{k\sim m_D}\frac{1}{k^2+m_D^2}\left(\frac{d\Pi^{(1)}}{dk^2}(0,0)+\Pi^{(2)}(0,0)\right)\notag\\
 &=\frac{3C_R\alpha_\mathrm{s}^2m_D(\epsilon)}{16\pi T}\left[\frac{71}{9}C_A-\frac{4}{9}T_Fn_f\left(1+12\ln2\right)+\beta_0\left(\frac{1}{\epsilon}+2\ln\frac{\mu^2}{2Tm_D}\right)\right]\notag\\
 &\hspace{13pt}-\frac{C_RC_FT_Fn_f\alpha_\mathrm{s}^3T}{2m_D}\,.
\end{align}
This is the same result that we got in Feynman gauge from $D_1+D_2$, so including the counter term we obtain the same scale $T$ contribution in Coulomb and static gauge:
\begin{align}
 \left(D_1+D_2\right)\Bigr|_{g^5,\,T}=&\,\frac{3C_R\alpha_\mathrm{s}^2m_D}{16\pi T}\left[3C_A+\frac{4}{3}T_Fn_f\left(1-4\ln2\right)+2\beta_0\left(\gamma_E+\ln\frac{\mu}{4\pi T}\right)\right]\notag\\
 &-\frac{C_RC_FT_Fn_f\alpha_\mathrm{s}^3T}{2m_D}\,.
\end{align}

\subsection{\texorpdfstring{Contribution from the scale $\bm{m_D}$}{Contribution from the scale m\_D}}

\begin{figure}[t]
 \includegraphics[width=\linewidth]{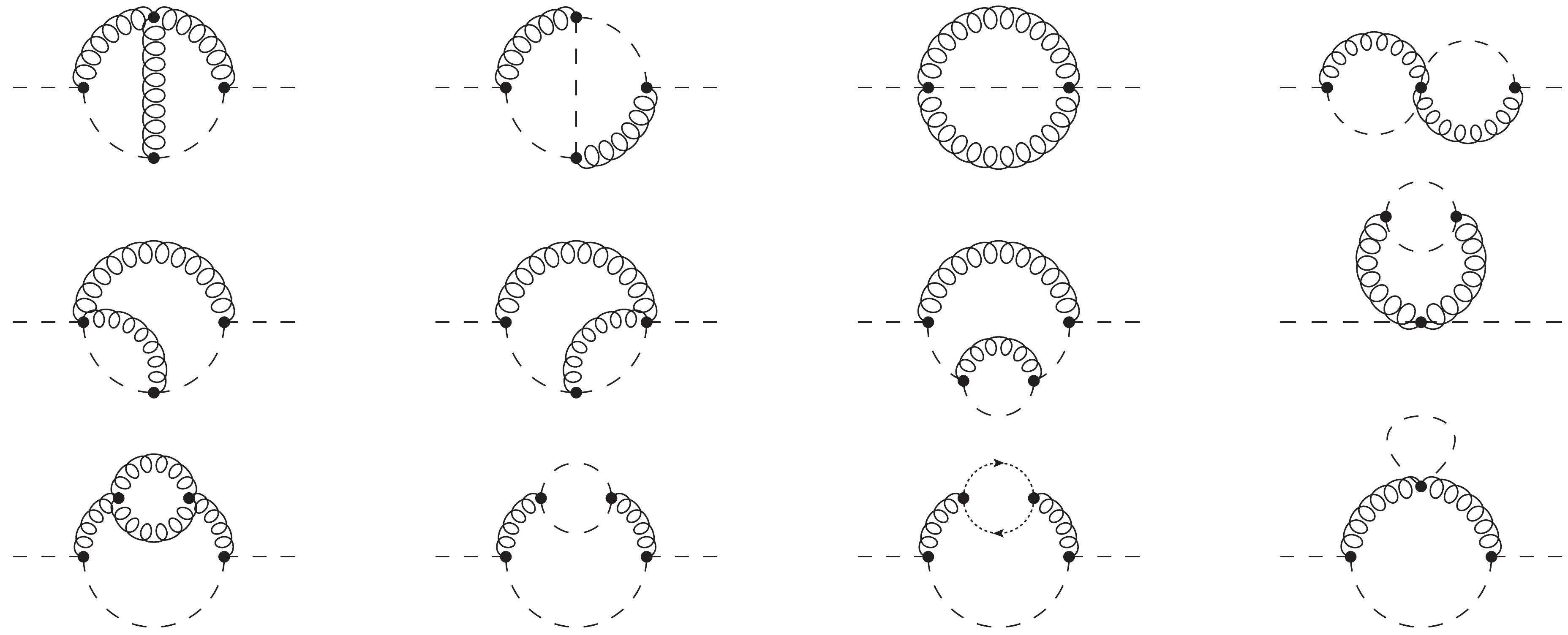}
 \caption{All Feynman diagrams contributing to $\Pi^{(2)}_{m_D}(0,k\sim m_D)$. Dashed lines represent temporal gluons, curly lines spatial gluons, and dotted lines with arrows are ghost propagators. The diagrams are labeled $L_1,\dots,L_{12}$ from top-left to bottom-right. $L_7,\dots,L_{12}$ are self-energy insertions into one-loop diagrams, while $L_1,\dots,L_6$ are new two-loop configurations.}
 \label{diagrams}
\end{figure}

The contribution from the scale $m_D$ consists of the two-loop self-energy and the square of the one-loop self-energy. It corresponds to the full $g_E^4$ contribution of $\bigl\langle \widetilde{A}_0^2\bigr\rangle$ in EQCD. The relevant diagrams for the two-loop self-energy are given in Fig.~\ref{diagrams}. The contribution from the square of the one-loop self-energy is not displayed, it corresponds to two one-loop insertions into the temporal gluon propagator. Together they also give a gauge invariant result:
\begin{align}
 D_1\Bigr|_{g^5,\,m_D}&=-\frac{C_Rg^2}{2T}\int_{k\sim m_D}\left[-\frac{\Pi_{m_D}^{(2)}(0,k)}{\left(k^2+m_D^2\right)^2}+\frac{\Pi_{m_D}^{(1)}(0,k)^2}{\left(k^2+m_D^2\right)^3}\right]\notag\\
 &=-\frac{C_RC_A^2\alpha_\mathrm{s}^3T}{m_D}\left[\frac{89}{48}+\frac{\pi^2}{12}-\frac{11}{12}\ln2\right]\,.
\end{align}

The calculation itself is quite involved, so we will not go into further details here. We use the method of integration by parts to reduce the three-loop integrals corresponding to each diagram down to a handful of known master integrals. A list of all integrals and their results in different gauges is given in appendix~\ref{scalemD}.

\subsection{\texorpdfstring{Contribution from the scale $\bm{m_M}$}{Contribution from the scale m\_M}}
\label{mMcancel}

Finally, we have to consider the contribution from the scale $m_M$. The temporal gluon momentum $k$ cannot be of order $m_M$, because then the propagator would have to be expanded in $1/m_D^2$ and the $k$ integration would be scaleless. But the loop momenta in the self-energy diagrams may be of order $m_M$ and such diagrams start to contribute at $\mathcal{O}\left(g^5\right)$. However, by the arguments from EQCD and MQCD in the previous chapter we expect the scale $m_M$ to enter the Polyakov loop only at $\mathcal{O}\left(g^7\right)$, so the $\mathcal{O}\left(g^5\right)$ contributions have to vanish, which is indeed the case.

\begin{figure}[t]
 \includegraphics[width=0.5\linewidth]{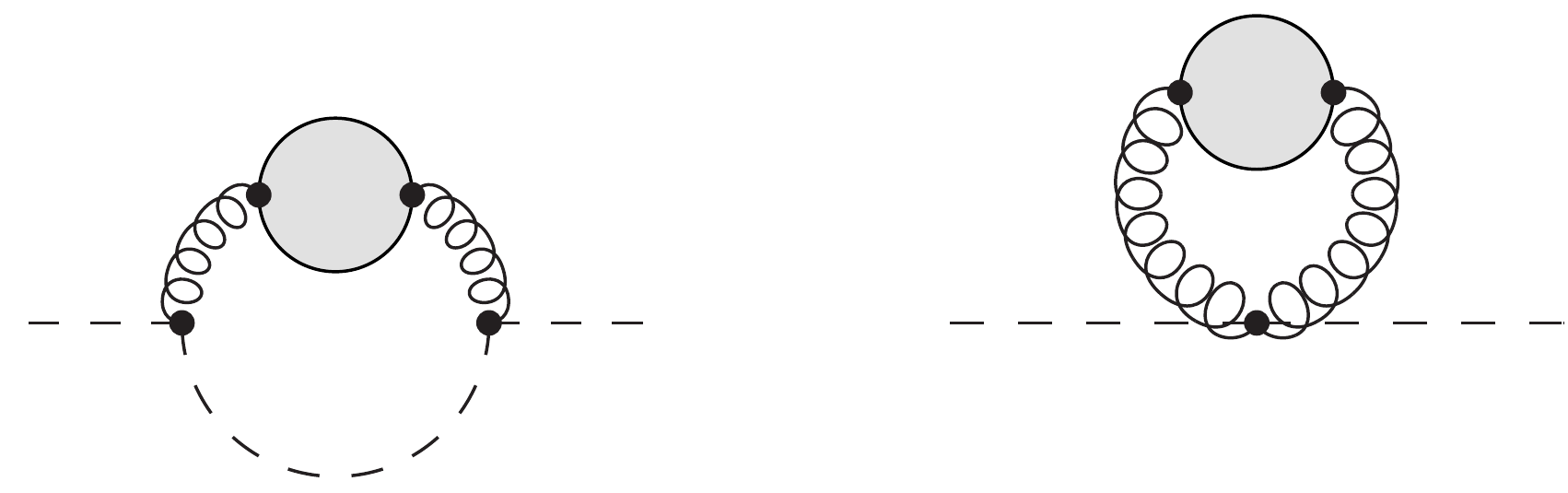}
 \caption{All $\mathcal{O}\left(g^5\right)$ diagrams that can give a contribution from the scale $m_M$. The bubble stands for the resummed propagator.}
 \label{scalemMdiag}
\end{figure}

There are two diagrams at this order (cf.\ Fig.~\ref{scalemMdiag}), both have one spatial gluon that carries the momentum of order $m_M$. The first is the diagram, where the spatial gluon connects at two three-gluon vertices, and the second is the tadpole diagram, where the spatial gluon connects at a four-gluon vertex. From the three-gluon vertices there comes a factor $(2k+q)_i(2k+q)_j$, where $k$ is the momentum of order $m_D$ and $q$ is of order $m_M$, but only $4k_ik_j$ needs to be kept, because the rest is of higher order. According to the power counting, each power of $q$ in the numerator adds a power of $g$ to the result, while terms with odd powers of the momenta in the numerator vanish. So the first higher order contributions (i.e., the terms quadratic in $q$) are of $\mathcal{O}\left(g^7\right)$. For the same reason, we only have to expand the propagator of the temporal gluon with momentum $\bm{k}+\bm{q}$ in the left diagram of Fig.~\ref{scalemMdiag} to leading order in $q$. Then we have
\begin{align}
 D_1\Bigr|_{g^5,\,m_M}&=\frac{C_Rg^2}{2T}\int_{k\sim m_D}\frac{\Pi^{(1)}_{m_M}(0,k)}{\left(k^2+m_D^2\right)^2}\notag\\
 &=\frac{C_RC_Ag^4}{2}\int_{q\sim m_M}D_{ij}(0,\q)\,\int_{k\sim m_D}\left[\frac{\delta_{ij}}{\left(k^2+m_D^2\right)^2}-\frac{4k_ik_j}{\left(k^2+m_D^2\right)^3}\right]\notag\\
 &=\frac{C_RC_Ag^4}{2}\int_{q\sim m_M}D_{ii}(0,\q)\,\frac{\Gamma\left(2-\frac{d}{2}\right)}{(4\pi)^{\frac{d}{2}}m_D^{4-d}}\left[1-\frac{4\,\Gamma\left(1+\frac{d}{2}\right)}{d\,\Gamma(3)\Gamma\left(\frac{d}{2}\right)}\right]=0\,.
\end{align}

Also the $\mathcal{O}\left(g^6\right)$ contributions from the scale $m_M$ need to vanish. These are the two-loop self-energy diagrams with one loop momentum of order $m_M$. We have also checked their cancellation explicitly, the details of this are given in appendix~\ref{scalemM}.

\subsection{Result}

Now we have all contributions to the Polyakov loop at $\mathcal{O}\left(g^5\right)$:
\begin{align}
 \ln L=&\,\frac{C_R\alpha_\mathrm{s}m_D}{2T}+\frac{C_R\alpha_\mathrm{s}^2}{2}\left[C_A\left(\frac{1}{2}+\ln\frac{m_D^2}{T^2}\right)-2T_Fn_f\ln2\right]\notag\\
 &+\frac{3C_R\alpha_\mathrm{s}^2m_D}{16\pi T}\left[3C_A+\frac{4}{3}T_Fn_f\left(1-4\ln2\right)+2\beta_0\left(\gamma_E+\ln\frac{\mu}{4\pi T}\right)\right]-\frac{C_RC_FT_Fn_f\alpha_\mathrm{s}^3T}{2m_D}\notag\\
 &-\frac{C_RC_A^2\alpha_\mathrm{s}^3T}{m_D}\left[\frac{89}{48}+\frac{\pi^2}{12}-\frac{11}{12}\ln2\right]+\mathcal{O}\left(g^6\right)\,.
\label{result}
\end{align}
The second line contains the contribution from the scale $T$ and the third line the contribution from the scale $m_D$.

\section{Higher order contributions}
\label{hoc}

\subsection{Casimir scaling}

It is known from lattice calculations that the logarithm of the Polyakov loop obeys Casimir scaling, at least approximately~\cite{Gupta:2007ax,Mykkanen:2012ri,Petreczky:2015yta}. Casimir scaling is observed by any quantity, in our case the free energy $F_Q$ of a static charge in representation $R$, if it is proportional to the quadratic Casimir operator $C_R$ of that representation. In other words, $F_Q/C_R$ should be independent of the representation $R$.

A necessary condition for the breaking of Casimir scaling is the appearance of a term not proportional to $C_R$. A term like that was identified for $L-1$ in Ref.~\cite{Brambilla:2010xn} at $\mathcal{O}\left(g^6\right)$. The term is
\begin{equation}
 \delta\langle L\rangle=\frac{1}{2}\left(\frac{C_R\alpha_\mathrm{s}m_D}{2T}\right)^2\,.
\end{equation}
This term, however, does not break the Casimir scaling of the free energy $F_Q$, since it is nothing else than the second order term in the expansion of $\exp (-F_Q/T)$, when $F_Q$ is taken at leading order. In fact, this term does not appear in $F_Q$. Note that the exponentiation formula given in~\eqref{exponentiation} provides a way of calculating $F_Q$ directly. It is then clear that at the level of two-gluon diagrams there is no breaking of Casimir scaling. Hence, we may ask, to which order of the perturbative series can Casimir scaling be observed?

There are several equivalent prescriptions on how the color coefficients in the logarithm of a closed Wilson line can be determined. It will not be necessary here to go into details on how they are calculated exactly (see appendix~\ref{Coefficients}), it is sufficient to know that for so-called connected diagrams, where every gluon is connected to every other gluon through gluon, ghost, or light quark propagators, the standard color factor and the one in the logarithm are the same.

At the three-gluon level we have several unconnected diagrams (cf.\ Fig.~\ref{Poly7}) and a few connected diagrams. By three-gluon diagrams we mean diagrams that correspond to three sum-integrals. We exclude sum-integrals from self-energy or vertex-function insertions from this definition, because if the corresponding tree-level diagram obeys Casimir scaling then also any self-energy or vertex-function insertion does.

The unconnected three-gluon diagrams are all scaleless in Coulomb or static gauge unless each gluon carries a momentum of order $m_D$, which means that they start to contribute at $\mathcal{O}\left(g^9\right)$. We will see below that Casimir scaling is already broken at a lower order, so we can ignore the unconnected three-gluon diagrams in Coulomb gauge on the basis of this argument. In other gauges these diagrams contribute at $\mathcal{O}\left(g^6\right)$,
but, as we will show in appendix~\ref{Coefficients}, their color coefficients obey Casimir scaling.

\begin{figure}
 \includegraphics[width=0.9\linewidth]{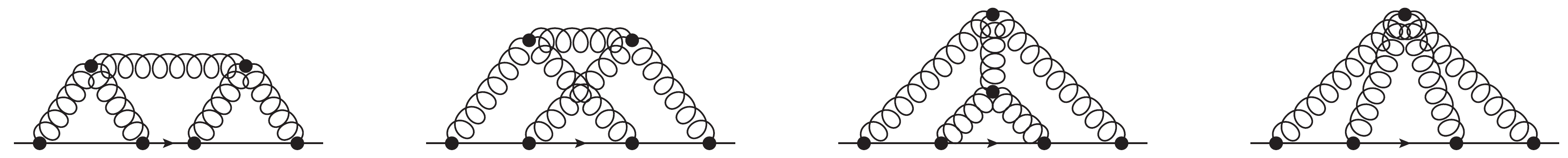}
 \caption{Connected three-gluon diagrams.}
 \label{Poly6}
\end{figure}

The connected three-gluon diagrams are shown in Fig.~\ref{Poly6}. Their color factors are all given by $-C_RC_A^2/4$, except for the second from left where it is $0$. All of these depend linearly on $C_R$, so at the three-gluon level Casimir scaling is still observed.

In general, the color factor of any diagram without light quarks is given as the trace over a product of color matrices in the respective representation divided by the dimension of the representation, where every color index is contracted with that of another color matrix or a structure constant from the interaction vertices. By repeated use of the commutation relation, the Jacobi identity or the quadratic Casimir
\begin{align}
 &\left[T_R^a,T_R^b\right]=if^{abc}T_R^c\,,& &f^{abe}f^{ecd}+f^{bce}f^{ead}+f^{bde}f^{eca}=0\,,\\
 &T_R^aT_R^a=C_R\mathbbm{1}\,,& &f^{acd}f^{bcd}=C_A\delta^{ab}\,,
\end{align}
one can express every such color factor as a combination of the following terms
\begin{equation}
 C_R^{(n)}=f^{i_1a_1i_2}f^{i_2a_2i_3}\cdots f^{i_na_ni_1}\frac{1}{d_R}\mathrm{Tr}\left[T_R^{a_1}T_R^{a_2}\cdots T_R^{a_n}\right]
 \label{Casin}
\end{equation}
and $C_R$ or $C_A$.

$C_R^{(1)}$ is trivially zero and $C_R^{(2)}$ and $C_R^{(3)}$ can be calculated independently of the representation:
\begin{equation}
 C_R^{(2)}=-C_RC_A\,,\hspace{50pt}C_R^{(3)}=-\frac{i}{4}C_RC_A^2\,.
 \label{Cn}
\end{equation}
But starting from $C_R^{(4)}$ there is no longer a simple unique formula like Eq.~\eqref{Cn} valid for all representations. For the fundamental and the adjoint representation one can replace every structure constant by color matrices and then use the Fierz identity to calculate the $C_R^{(n)}$ explicitly:
\begin{align}
 f^{abc}&=\frac{1}{iT_F}\mathrm{Tr}\left[T^a_F\left[T_F^b,T_F^c\right]\right]\,,\\
 \left(T_F^a\right)_{ij}\left(T_F^a\right)_{kl}&=T_F\left(\delta_{il}\delta_{kj}-\frac{1}{N}\delta_{ij}\delta_{kl}\right)\,.
\end{align}
Note that $T_F^a$ with a color index denotes the generators of the fundamental representation, while $T_F$ without a color index denotes the normalization constant of the fundamental representation. The two are related by
\begin{equation}
 \mathrm{Tr}\left[T_F^aT_F^b\right]=T_F\delta^{ab}\,,\hspace{20pt}\mathrm{or}\hspace{20pt}T_F^a=\sqrt{\frac{T_F}{2}}\lambda^a\,,
\end{equation}
where $\lambda^a$ are the Gell-Mann matrices.

\begin{figure}[t]
 \includegraphics[width=\linewidth]{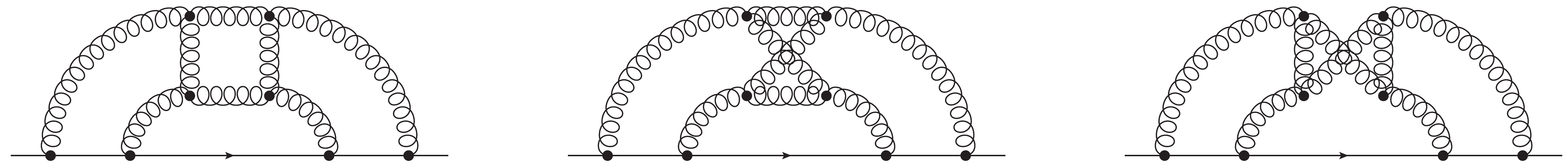}
 \caption{Diagrams at $\mathcal{O}\left(g^8\right)$ with a color coefficient $C_R^{(4)}$.}
 \label{Poly5}
\end{figure}

In this way we obtain
\begin{equation}
 C_F^{(4)}=\frac{1}{8}C_FC_A^2\left(3C_A-4C_F\right)\,,\hspace{50pt}C_A^{(4)}=\frac{1}{8}C_A^3\left(13C_A-24C_F\right)\,,
\end{equation}
or alternatively
\begin{equation}
 \frac{C_F^{(4)}}{C_A^{(4)}}=\frac{C_F}{C_A}\frac{N^2+2}{N^2+12}\,.
\end{equation}
If we can find a diagram whose color coefficient is given by $C_R^{(4)}$, then we have found a Casimir scaling breaking term. Such a diagram can appear only at $\mathcal{O}\left(g^8\right)$ or higher, because in the Feynman rules of QCD every color matrix and structure constant comes with a factor of $g$. Fig.~\ref{Poly5} shows some similar diagrams where the dependence on $C_R^{(4)}$ is immediately apparent. The diagram on the left has the color coefficient $C_R^{(4)}$ exactly and the other two have $C_R^{(4)}+C_RC_A^3/8$, because in both cases two color matrices have to be commuted to get the form of $C_R^{(4)}$ and the commutator gives $iC_R^{(3)}C_A/2$.

If we add up the contributions proportional to $C_R^{(4)}$ from all three diagrams, then the contour integrations simplify a lot and we get Kronecker deltas for the Matsubara frequencies of each of the four gluon propagators attached to the Polyakov loop contour times a coefficient of $1/8T^4$. One of these four Kronecker deltas is redundant and the Matsubara frequency in the internal loop (the square in the left diagram, or the twisted square in the other two diagrams) remains different from zero. So the momentum integrals are not scaleless, because the scale $T$ remains in the calculation, and we have a possible genuine nonvanishing contribution at $\mathcal{O}\left(g^8\right)$ that breaks Casimir scaling.

There are other diagrams similar to these three, which can be obtained from Fig.~\ref{Poly5} by contracting one or two propagators in the internal loop to a four-gluon vertex. Their color coefficients also involve $C_R^{(4)}$, so they will give other terms of $\mathcal{O}\left(g^8\right)$ that break Casimir scaling. In principle, light quark loops can also give rise to color factors that break Casimir scaling. If such a light quark loop has to two or three external gluon legs, then it can be included as a contribution to the self-energy or the vertex function and it will not affect Casimir scaling. With four or more external legs the color factor is no longer proportional to the quadratic Casimir, which can be checked in a similar calculation to the one above replacing the internal gluons in Fig.~\ref{Poly5} with light quark propagators, but such diagrams also start to contribute at $\mathcal{O}\left(g^8\right)$.

One could in principle imagine that all those terms cancel and only Casimir scaled terms remain, but that would imply some underlying mechanism that enforces Casimir scaling to all orders of perturbation theory. Such a mechanism, if it exists, has not been discovered so far. The approximate Casimir scaling observed in lattice calculations may be explained by the strong suppression of the $\mathcal{O}\left(g^8\right)$ contributions that possibly violate Casimir scaling.

\subsection{\texorpdfstring{Outline of the $\bm{\mathcal{O}\left(g^6\right)}$ calculation}{Outline of the O(g\^{}6) calculation}}

We will outline here the necessary steps for the calculation of the $\mathcal{O}\left(g^6\right)$ contributions, the last order accessible by perturbation theory. The amount of work one has to do is greatly reduced by choosing the appropriate gauge. As explained above, all the unconnected three-gluon diagrams (see Fig.~\ref{Poly7}) are scaleless at leading order and start to contribute only at $\mathcal{O}\left(g^9\right)$ in Coulomb or static gauge. In Feynman gauge, however, the six diagrams of Fig.~\ref{Poly7} whose modified color coefficients in the logarithm of the Polyakov loop do not vanish all contribute at $\mathcal{O}\left(g^6\right)$, so this is not the most efficient gauge to perform this calculation.

There are also unconnected three-gluon diagrams consisting of only two unconnected pieces, a single gluon and a piece of three propagators connected by a three-gluon vertex. These are not displayed in Fig.~\ref{Poly7}, because in gauges that are diagonal in temporal and spatial indices they vanish on account of the three-gluon vertex with three temporal indices, just like the corresponding two gluon diagram, but in nondiagonal gauges they also have to be considered.

The connected diagrams of Fig.~\ref{Poly6} can only contribute at $\mathcal{O}\left(g^6\right)$ when all momenta are of the scale $T$, the scale $m_D$ contributions are of higher order. However, in Coulomb or static gauge all of them vanish. The first three diagrams are essentially the same gluonic configuration, but with different path ordering prescriptions along the Polyakov loop contour, so we will only discuss the leftmost diagram, the others are analogous (apart from the second having a vanishing color coefficient). In static gauge all Matsubara frequencies have to be zero because of the temporal propagators, so the integrals are scaleless and vanish. In Coulomb gauge the Matsubara frequencies are not necessarily all zero, but the integrand vanishes by itself: Call $\k$ the momentum flowing from the first to the last point on the Polyakov loop contour, $\p$ the momentum flowing from the first to the second point, and $\q$ the momentum flowing from the third to the fourth point. The results of the $\p$ and $\q$ integrations can only be proportional to $\k$ because of rotational symmetry, where each vector comes from the three-gluon vertices. But these vectors $\k$ are then contracted with the transverse projector from the spatial propagator, which gives zero.

In the second diagram from the right in Fig.~\ref{Poly6} the Matsubara frequencies of the propagators connecting to the Polyakov loop contour have to be zero. In static gauge this is again a consequence of the temporal propagator, while in Coulomb gauge it follows after computing the contour integrations (this involves some convenient momentum shifts). There is no constraint on the frequency from the loop momentum that flows around the gluon triangle, but through the three-gluon vertices it appears to either linear or cubic power in the numerator, so the remaining Matsubara sum is odd and cancels. The same is true if the gluon triangle is replaced by a fermion loop. The rightmost diagram in Fig.~\ref{Poly6} vanishes because of the four-gluon vertex in all gauges where the gluon propagator is diagonal in temporal and spatial indices.

The two-gluon diagram $D_2$ has already been discussed above, only the first term in Eq.~\eqref{eq:D2} contributes in Coulomb or static gauge at $\mathcal{O}\left(g^6\right)$. It gives a contribution of $-C_RC_A\alpha_\mathrm{s}^2m_D^2/48T^2$ when both momenta are of the scale $m_D$. When one or both momenta are of the scale $T$, then the first nonvanishing contribution is of $\mathcal{O}\left(g^7\right)$ or $\mathcal{O}\left(g^8\right)$, respectively.

The $\mathcal{O}\left(g^6\right)$ contribution from diagram $D_1$ contains several different elements: the three-loop self-energy with all momenta of order $m_D$, products of one-loop and two-loop self-energies [essentially the last line of Eq.~\eqref{eq:D1} times $\Pi_{m_D}^{(1)}(k)/\left(k^2+m_D^2\right)$] or the one-loop self-energy cubed from the expansion of the resummed propagator, the two-loop self-energy at the scale $m_D$ with one loop momentum of order $T$ and the other of order $m_D$, and the two-loop or square of one-loop self-energy with all momenta of order $T$.

Fortunately, most of these contributions can be inferred in the EFT approach from an already existing EQCD calculation. As explained previously [see Eq.~\eqref{dfmdmE}], the correlator of two $\widetilde{A}_0$ fields in EQCD can be obtained from the pressure or vacuum energy density, which has been calculated at the four-loop level in~\cite{Kajantie:2003ax}. From this we get
\begin{align}
 \frac{1}{d_R}\Tr\bigl\langle\widetilde{A}_0^2\bigr\rangle_s=&-\frac{C_R m_E}{4 \pi}+\frac{C_RC_Ag_E^2}{(4\pi)^2}\left[\frac{1}{2\epsilon}+\frac{1}{2}-\gamma_E+\ln\frac{\pi\mu^2}{m_E^2}\right]\notag\\
 &+\frac{2C_RC_A^2}{(4\pi)^3}\frac{g_E^4}{m_E}\left(\frac{89}{48}-\frac{11}{12}\ln2+\frac{\pi^2}{12}\right)\notag\\
 &+\frac{2C_RT_F^2\left(N^2+1\right)\lambda_E}{(4\pi)^2}+\frac{C_RT_F^2\bar{\lambda}_E}{(4\pi)^2}\left(\frac{4N^2-6}{N}-N^2-1\right)\notag\\
 &+\frac{C_RC_A^3g_E^6}{(4\pi)^4m_E^2}\left(\frac{43}{4}-\frac{491\pi^2}{768}\right)+\mathcal{O}\left(g^5\right)\,.
\label{4loop}
\end{align}
If we now insert the explicit expression for $\lambda_E$ and $\bar{\lambda}_E$ in terms of $g$ and the one-loop corrections to $g_E$, $m_E$, and $\mathcal{Z}_2$, then we have almost the full $\mathcal{O}\left(g^6\right)$ contribution to the logarithm of the Polyakov loop. The only thing that is missing is the contribution from $D_2$ given above and the two-loop and square of one-loop self-energy contributions with all momenta of order $T$ in Coulomb or static gauge.

The one-loop correction to the EQCD coupling constant is given by~\cite{Kajantie:1997tt}:
\begin{equation}
 g_E^2=g^2T\left\{1+\frac{\alpha_\mathrm{s}}{4\pi}\left[\frac{1}{3}C_A-\frac{16}{3}T_Fn_f\ln2+\beta_0\left(\frac{1}{\epsilon}+\gamma_E+\ln\frac{\mu^2}{4\pi T^2}\right)\right]\right\}\,.
\end{equation}
Because of the $1/\epsilon$ pole in the $g_E^2$ term in Eq.~\eqref{4loop} we also need the $\mathcal{O}(\epsilon)$ terms of both $g_E^2$~\cite{Laine:2005ai} and $\mathcal{Z}_2$:
\begin{align}
 g_E^2\Bigr|_{\mathcal{O}(\epsilon)}=\frac{g^4T\epsilon}{(4\pi)^2}&\Biggl[\beta_0\left(\frac{1}{2}\left(\gamma_E+\ln\frac{\mu^2}{4\pi T^2}\right)^2+\frac{\pi^2}{4}-2\gamma_E^2-4\gamma_1\right)\notag\\
 &+\left(\frac{1}{3}C_A-\frac{16}{3}T_Fn_f\ln2\right)\left(\gamma_E+\ln\frac{\mu^2}{4\pi T^2}\right)-\frac{16}{3}T_Fn_f(\ln2)^2\Biggr]\,,
\end{align}
\begin{align}
 \mathcal{Z}_2\Bigr|_{\mathcal{O}(\epsilon)}=\frac{\alpha_\mathrm{s}\epsilon}{4\pi}&\Biggl[\beta_0\left(\frac{1}{2}\left(\gamma_E+\ln\frac{\mu^2}{4\pi T^2}\right)^2+\frac{\pi^2}{4}-2\gamma_E^2-4\gamma_1\right)\notag\\
 &+\left(\frac{11}{3}C_A+\frac{4}{3}T_Fn_f(1-4\ln2)\right)\left(\gamma_E+\ln\frac{\mu^2}{4\pi T^2}\right)+\frac{23}{3}C_A-\frac{16}{3}T_Fn_f(\ln2)^2\Biggr]\,.
\end{align}
Here $\gamma_1$ is the coefficient of the linear term in the expansion of $\zeta(1-x)$ for small $x$. Combining all these terms and inserting the renormalized coupling, we then have
\begin{align}
 \ln L\Bigr|_{\mathcal{O}\left(g^6\right)}=&-\frac{C_R\alpha_\mathrm{s}^3}{8\pi}\Biggl[\beta_0C_A\left(\frac{1}{2}\left(\gamma_E+\ln\frac{\mu^2}{4\pi T^2}\right)^2+\frac{\pi^2}{4}-2\gamma_E^2-4\gamma_1\right)\notag\\
 &+C_A\left(2C_A+\frac{2}{3}T_Fn_f(1-8\ln2)\right)\left(\gamma_E+\ln\frac{\mu^2}{4\pi T^2}\right)\notag\\
 &+8C_A^2+\frac{2}{3}C_AT_Fn_f\left(1+4\ln2-8(\ln2)^2\right)+8C_FT_Fn_f\notag\\
 &+C_A\left(4C_A+\frac{4}{3}T_Fn_f(1-8\ln2)+4\beta_0\left(\gamma_E+\ln\frac{\mu}{4\pi T}\right)\right)\left(\frac{1}{2\epsilon}-\gamma_E+\ln\frac{\pi\mu^2}{m_D^2}\right)\Biggr]\notag\\
 &-\frac{C_RC_A\alpha_\mathrm{s}^4T^2}{m_D^2}\left(C_A^2\left(\frac{43}{8}-\frac{491\pi^2}{1536}\right)+C_FT_Fn_f\right)-\frac{C_RC_A\alpha_\mathrm{s}^2m_D^2}{48T^2}\notag\\
 &-\frac{C_Rg^2}{2T}\int_{k\sim T}\frac{1}{k^6}\left[\left(\Pi_T^{(1)}(0,k)+\frac{\beta_0\alpha_\mathrm{s}}{4\pi}\left(\frac{1}{\epsilon}-\gamma_E+\ln4\pi\right)k^2\right)^2-k^2\,\Pi_T^{(2)}(0,k)\right]\,.
\label{g6}
\end{align}
The first $1/\epsilon$ pole, a UV divergence from the scale $m_D$, has to cancel against a corresponding IR divergence in the scale $T$ integrals. The $1/\epsilon$ pole in the last line comes from the charge renormalization in the $\overline{\mathrm{MS}}$-scheme of the $\mathcal{O}\left(g^4\right)$ contribution from the scale $T$, i.e., the first term in Eq.~\eqref{eq:D1}, and it cancels the UV divergence in the one-loop vacuum part of the self energy.

There are also $\mathcal{O}\left(g^6\right)$ contributions from two-loop diagrams with two momenta of order $m_D$ and one momentum of order $m_M$. From the MQCD analysis of the Polyakov loop we know that the scale $m_M$ can only appear first at $\mathcal{O}\left(g^7\right)$, so these contributions ultimately have to cancel. We have checked this cancellation explicitly in appendix~\ref{scalemM}.

\section{Convergence of the perturbative series and comparison with the lattice results}
\label{lattice}

In this section, we discuss the convergence of the perturbative series for the Polyakov loop, or equivalently the free energy of a static quark, and compare the weak coupling results with lattice QCD results. For a reliable comparison of the lattice and the weak coupling results we need to consider a temperature range that extends to sufficiently high temperatures. So far, it is only in pure SU(3) gauge theory, i.e., in QCD with zero light quark flavors ($n_f=0$), that we have lattice results at sufficiently high temperatures to perform such a comparison. Namely, the renormalized Polyakov loop has been calculated up to temperatures of $24T_c$~\cite{Gupta:2007ax}, with $T_c$ being the deconfinement transition temperature.

\begin{figure}
\includegraphics[width=12cm]{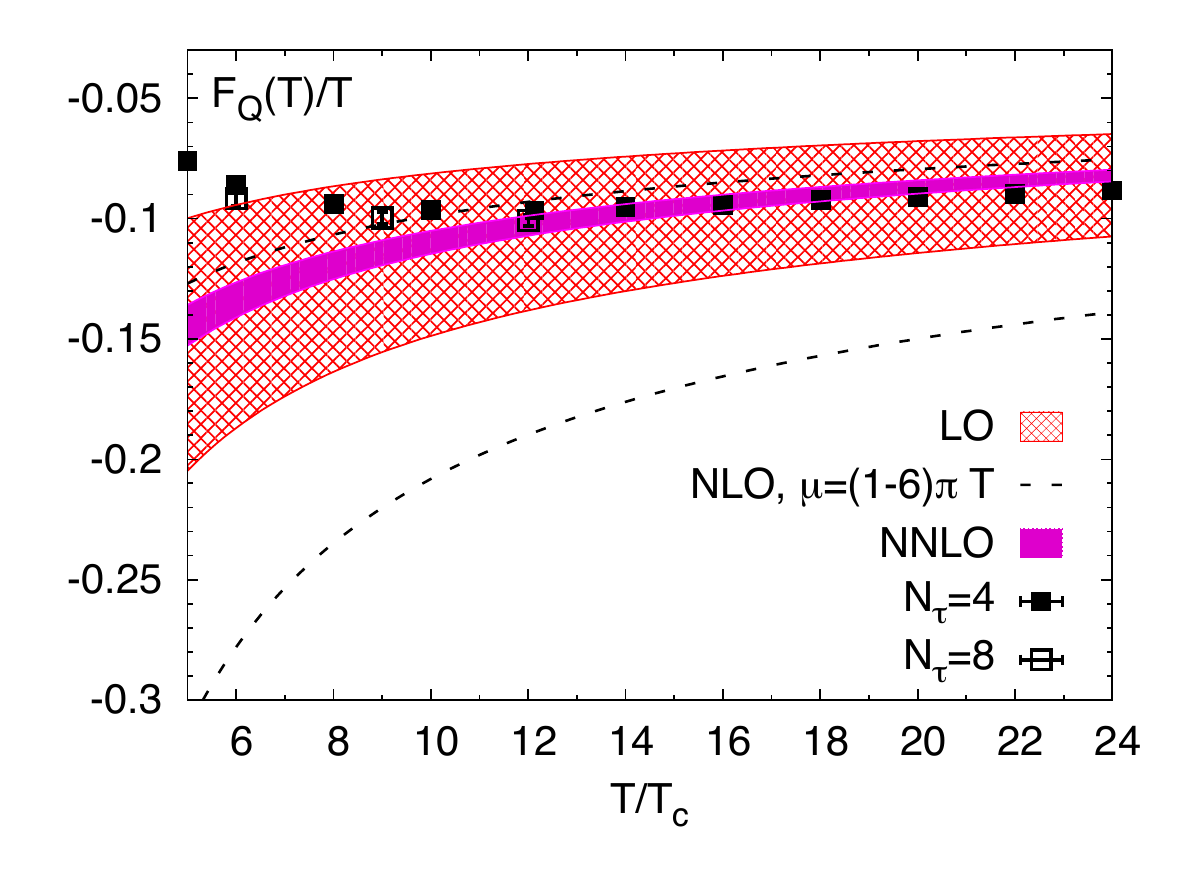}
\caption{The free energy of a static quark $F_Q$ for the SU(3) gauge theory in weak coupling expansion at LO, NLO and NNLO. The bands are obtained by varying the renormalization scale $\mu$ between $\pi T$ and $6\pi T$. Also shown are the lattice data for $F_Q$ obtained on lattices with temporal extent $N_\tau=4$ and $8$~\cite{Gupta:2007ax}.}
\label{fig:fqcomp}
\end{figure}

In Fig.~\ref{fig:fqcomp} we show the perturbative results for the free energy of a static quark at various orders in perturbation theory for pure SU(3) gauge theory ($n_f=0$). We use one-loop running for $\alpha_\mathrm{s}$. To determine the renormalization scale for different values of $T/T_c$ we used the relation $r_0 T_c=0.7498(50)$~\cite{Necco:2003vh}, where $r_0$ is the Sommer scale~\cite{Sommer:1993ce}. The value of $\Lambda_{\overline{\mathrm{MS}}}$ was determined in Ref.~\cite{Brambilla:2010pp}: $r_0 \Lambda_{\overline{\mathrm{MS}}}=0.637^{+0.032}_{-0.030}$. With this we get $T_c/\Lambda_{\overline{\mathrm{MS}}}=1.177$. One can see that the scale dependence of the leading order (LO) results is quite large and becomes even larger at NLO. The scale dependence of $F_Q$ is first reduced at NNLO and is, in fact, quite small, making a meaningful comparison with the lattice results possible. In Fig.~\ref{fig:fqcomp} we also show the lattice results for the static quark free energy for $n_f=0$ from Ref.~\cite{Gupta:2007ax}. The lattice results appear to agree with the LO and NLO results, given their large scale uncertainty, but are slightly larger than the NNLO results at small $T$.

We should keep in mind, however, that the comparison of the lattice and the perturbative results for $F_Q$ is not as straightforward as Fig.~\ref{fig:fqcomp} may suggest. This fact seems to be generally overlooked in the literature. The perturbative calculations are performed in the $\overline{\mathrm{MS}}$-scheme, while on the lattice the calculation is performed in a scheme in which the static quark-antiquark energy at zero temperature is normalized such that it is equal to the string potential $V(r)=-\pi/(12 r)+\sigma r$ at large distances, with $\sigma$ being the string tension. To match the two schemes one has to normalize the static energy at zero temperature in the perturbative calculation at each order to the lattice potential at short distances. This then corresponds to a constant shift $C_\mathrm{shift}$ in physical units of the perturbative static energy, which is different at different orders of perturbation theory.

This matching has been carried out for both $n_f=0$~\cite{Brambilla:2010pp} and $n_f=3$~\cite{Bazavov:2012ka}. The shift of the static energy implies that one has to add $C_\mathrm{shift}/2$ to the perturbative result for $F_Q$ before the comparison with the lattice results can be made. However, $C_\mathrm{shift}$ is sensitive to the perturbative order, to the resummation of the logarithms associated with the running coupling constant, as well as to the ultra-soft scale (see, e.g., discussions in Ref.~\cite{Bazavov:2014soa}). Thus, the uncertainty in the determination of $C_\mathrm{shift}$ will be the dominant systematic uncertainty in the comparison of the weak coupling and lattice calculations for $F_Q$. For this reason we did not add $C_\mathrm{shift}$ in the comparison of the lattice and the perturbative result for $F_Q$ in Fig.~\ref{fig:fqcomp}.

We can avoid this problem by calculating the entropy of the static quark defined as
\begin{equation}
S_Q=-\frac{\partial F_Q(T)}{\partial T}.
\end{equation}
In this quantity the normalization constant $C_{\rm shift}$ drops out. In perturbation theory it is straightforward to calculate the entropy of a static quark by taking the temperature derivative of Eq.~\eqref{result} times $T$. In order to calculate the entropy of a static quark on the lattice, we use the lattice data on the renormalized Polyakov loop obtained on $N_\tau=4$ lattices in Ref.~\cite{Gupta:2007ax}. We interpolate these data using different smoothing splines and calculate the derivatives of the splines using the R package~\cite{Rpackage}. The statistical errors of the interpolation and the derivative were calculated using the bootstrap method. Furthermore, we considered different spline interpolations, varying the number of knots and the value of the smoothing parameter. We enlarged the statistical error to take into account the difference between the different splines, if those were outside the statistical error. In this way we obtained the total error for the entropy in lattice QCD.

\begin{figure}
\includegraphics[width=12cm]{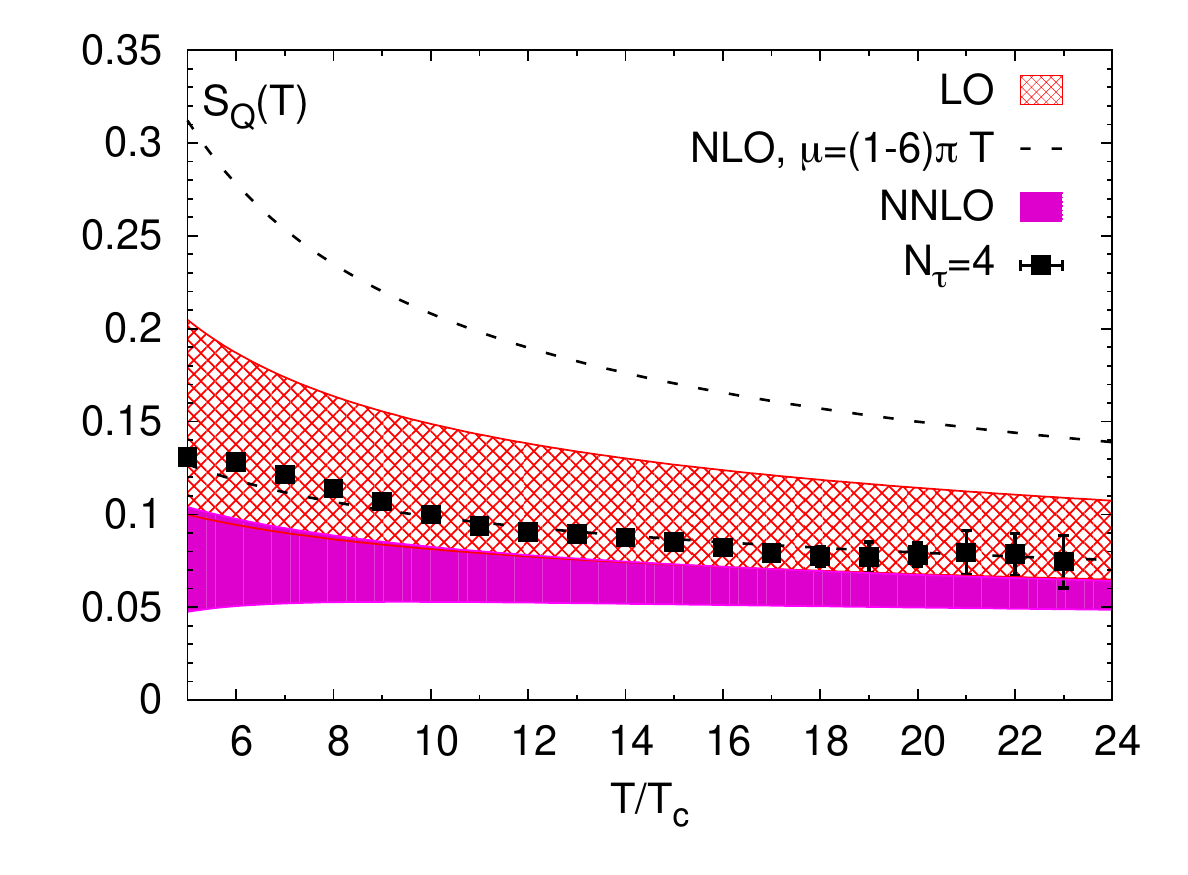}
\caption{The entropy of a static quark $S_Q$ for the SU(3) gauge theory in weak coupling expansion at LO, NLO and NNLO. The bands are obtained by varying the renormalization scale $\mu$ between $\pi T$ and $6\pi T$. Also shown are the lattice results for $S_Q$, cf.\ the description in the text.}
\label{fig:sqcomp}
\end{figure}

In Fig.~\ref{fig:sqcomp} we compare the entropy of a static quark estimated in lattice QCD and in the weak coupling calculations. As in the case of the static quark free energy, the scale dependence of the LO and NLO results is quite large. Within this large scale uncertainty the perturbative calculations and the lattice data agree. 
The scale dependence of the NNLO result is much smaller. The NNLO result, however, lies below the lattice data. 
This implies that higher order corrections in the weak coupling expansion may still be important. In view of this, below we discuss some higher order terms in the weak coupling expansion of the static quark free energy and have a closer look on the convergence of the perturbative series.

As discussed above, in the weak coupling expansion we have three types of contributions, purely nonstatic, i.e., arising from the scale $T$, purely static contributions corresponding to the scales $m_D$ and $m_M$, which can be calculated within EQCD, and mixed contributions, where some loop momenta are of order $m_D$ or $m_M$ and others are of order $T$. Here we will discuss the latter two types of contributions, referring to them as EQCD type and mixed type contributions, respectively. Together they have been called the static contribution in the previous sections, but here we want to distinguish between them.

The EQCD type contributions arise from the weak coupling expansion of $\mathrm{Tr}\bigl\langle\widetilde{A}_0^2\bigr\rangle$ with the expansion parameter $C_Ag_E^2/(4\pi m_E)$ [c.f.\ Eq.~\eqref{4loop}], using only the leading order results for the matching parameters $\mathcal{Z}_2$, $g_E$, and $m_E$ and neglecting quartic or higher order interactions. Beyond four-loop order the condensate $\mathrm{Tr}\bigl\langle\widetilde{A}_0^2\bigr\rangle$ contains a nonperturbative contribution of order $g_E^8/m_E^3$, which was calculated using lattice simulations of EQCD~\cite{Hietanen:2008tv}. Furthermore, in Ref.~\cite{Hietanen:2008tv} a simple parametrization of those higher order contributions to the condensate beyond four-loop order was given.

\begin{figure}
\includegraphics[width=12cm]{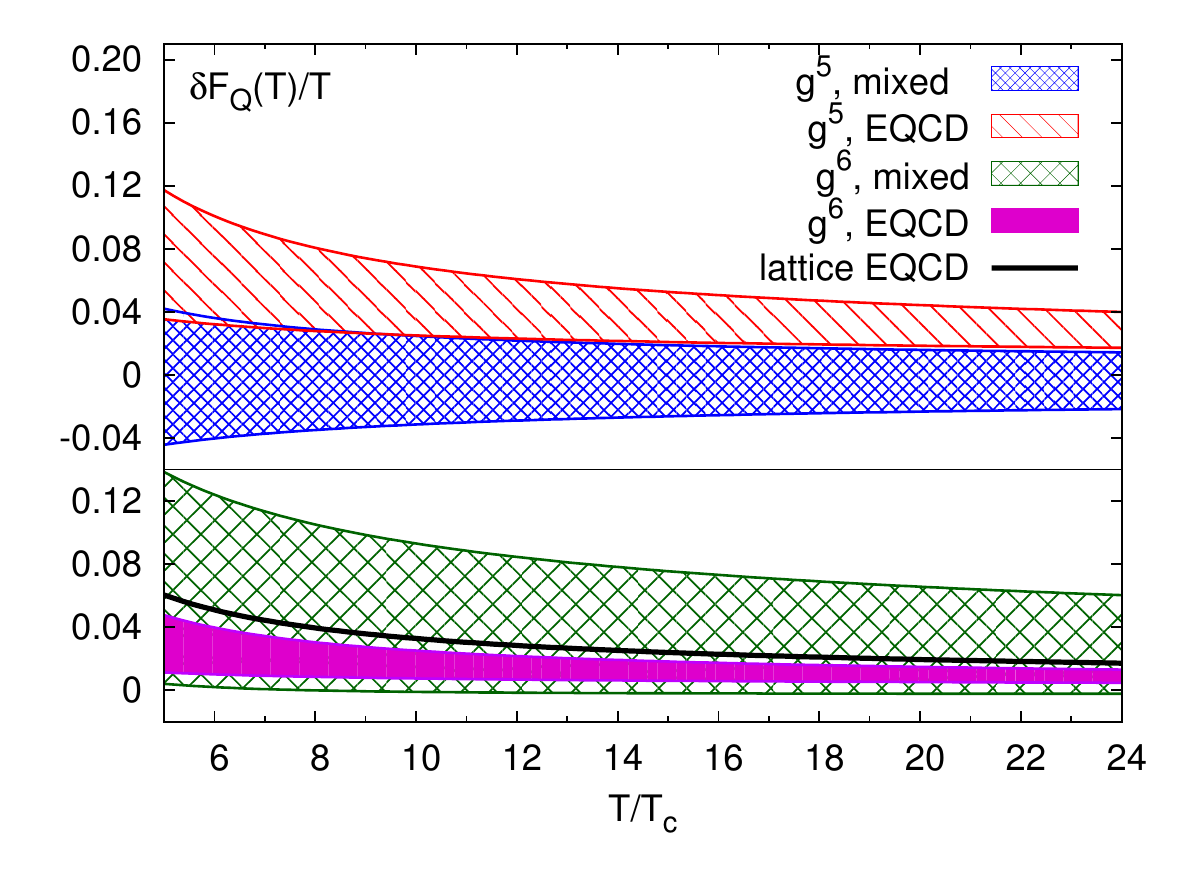}
\vspace{-15pt}
\caption{EQCD type and mixed contributions to $F_Q$ at $\mathcal{O}\left(g^5\right)$ (upper panel) and $\mathcal{O}\left(g^6\right)$ (lower panel). The bands correspond to the variation of the renormalization scale from $\pi T$ to $6\pi T$. The thick black line corresponds to the higher order EQCD type contributions from lattice EQCD estimated in Ref.~\cite{Hietanen:2008tv} for the renormalization scale $\mu=4\pi T$, cf.\ the description in the text.}
\label{fig:g6}
\end{figure}

In Fig.~\ref{fig:g6} we show the EQCD type contributions at $\mathcal{O}\left(g^5\right)$ and $\mathcal{O}\left(g^6\right)$ as well as the sum of all higher order contributions calculated in lattice EQCD, which we plot using Eq.~(4.1) of Ref.~\cite{Hietanen:2008tv}. The bands shown in the figure correspond to the variation of the renormalization scale $\mu$ from $\pi T$ to $6\pi T$. The magnitude of the different contributions is decreasing with increasing order, the $\mathcal{O}\left(g^6\right)$ contribution is smaller than the $\mathcal{O}\left(g^5\right)$ contribution, and the sum of all the higher order contributions to $g^2/(2Td_R)\times\mathrm{Tr}\bigl\langle\widetilde{A}_0^2\bigr\rangle$ [starting from $\mathcal{O}\left(g^7\right)$], which includes the nonperturbative contributions, is about the same size as the $\mathcal{O}\left(g^6\right)$ contribution. Thus, we conclude that the weak coupling expansion for the purely static contribution is converging reasonably well and there are no large nonperturbative corrections to the Polyakov loop from the static chromomagnetic sector. Furthermore, as shown in Fig.~\ref{fig:g6}, the sum of the higher order corrections to the static quark free energy is positive and thus would shift the perturbative result away from the lattice data. 

Now let us discuss the mixed contributions, which come from higher order corrections to the matching parameters and higher interaction terms in EQCD. In Fig.~\ref{fig:g6} we show the $\mathcal{O}\left(g^5\right)$ and $\mathcal{O}\left(g^6\right)$ mixed contributions. The latter is evaluated by using Eq.~\eqref{g6} and omitting the last two lines as well as the $1/\epsilon$ pole. In contrast to the EQCD type contributions, the mixed contributions can be positive or negative depending on the choice of the renormalization scale. At $\mathcal{O}\left(g^5\right)$ the mixed contribution is smaller than the EQCD type contribution, while at $\mathcal{O}\left(g^6\right)$ the mixed contribution is of the same size or larger (depending on the renormalization scale). Furthermore, the two mixed contributions are about the same size, which means that the full $\mathcal{O}\left(g^6\right)$ contribution might be large. Clearly, for rigorous statements about the convergence of the weak coupling expansion and comparison with lattice QCD results a complete calculation of the $\mathcal{O}\left(g^6\right)$ contribution will be necessary.

\section{Conclusions}
\label{conclusion}

In this paper, we have calculated the next-to-next-to-leading-order contribution to the Polyakov loop or equivalently to the static quark free energy. This contribution is of $\mathcal{O}\left(g^5\right)$. The calculations have been performed directly in QCD as well as through an effective theory approach using known results from EQCD. The effective theory approach based on EQCD also allowed us to calculate some of the higher order contributions of $\mathcal{O}\left(g^6\right)$ as well as to have an estimate of some nonperturbative contributions starting to appear at $\mathcal{O}\left(g^7\right)$. The weak coupling expansion in EQCD seems to converge reasonably well, but there could be potentially large contributions from nonstatic modes at $\mathcal{O}\left(g^6\right)$. 

While the scale dependence of the $\mathcal{O}\left(g^5\right)$ result is reasonably small, we do not find a very good agreement between the lattice data and the weak coupling expansion. It is possible that the observed discrepancy between the lattice results and the weak coupling expansion is due to the missing $\mathcal{O}\left(g^6\right)$ term. Therefore, the calculation of the complete $\mathcal{O}\left(g^6\right)$ contribution is important. 

Finally, we discussed the Casimir scaling of the static quark free energy. We have shown that Casimir scaling holds up to $\mathcal{O}\left(g^7\right)$, but at $\mathcal{O}\left(g^8\right)$ there may appear terms that break Casimir scaling. The fact that the breaking of Casimir scaling happens only at $\mathcal{O}\left(g^8\right)$ in the weak coupling expansion may explain the lattice results on the Polyakov loop in higher representations, which show approximate Casimir scaling in the high temperature region~\cite{Gupta:2007ax,Petreczky:2015yta}.

\acknowledgments
This work has been supported by the DFG grant BR 4058/1-2 ``Effective field theories for heavy probes of hot plasma'', and by the DFG cluster of excellence ``Origin and structure of the universe'' (www.universe-cluster.de). P.P.\ was supported by U.S.\ Department of Energy under Contract No.\ DE-SC0012704.

\appendix
\section{Gluon propagators}
\label{propagators}

\subsection{Feynman gauge}

Feynman gauge is obtained by adding the gauge fixing term $\left(\partial_\mu A_\mu^a\right)^2/2$ to the Lagrangian, as well as the ghost Lagrangian $\left(\partial_\mu\bar{c}^a\right)D_\mu^{ab} c^b$. Then the free propagators for gluons $D_0$ and ghosts $G_0$ are given by
\begin{equation}
 D_0=\frac{\delta_{\mu\nu}}{k_0^2+k^2}\hspace{10pt}\mathrm{and}\hspace{10pt}G_0=\frac{1}{k_0^2+k^2}\,.
\end{equation}
We will not explicitly display color indices, because they only appear in Kronecker deltas.

For the resummed gluon propagators we need to sum over all one-particle reducible diagrams, i.e., over all bubble insertions in a propagator, where the bubbles define the self-energy tensor $-\Pi_{\mu\nu}$. We can parametrize the self-energy tensor in the following way
\begin{equation}
 \Pi=\begin{pmatrix} \Pi_{00} & \Pi_Ak_0k_j \\ \Pi_Ak_ik_0 & \Pi_B\delta_{ij}+\Pi_Ck_ik_j \end{pmatrix}\,,
\end{equation}
which comprises all tensor structures allowed by rotational symmetry. Even though Feynman gauge is designed to be fully covariant under Lorentz transformations, the existence of the medium explicitly breaks the full Lorentz symmetry down to the rotational symmetry in the rest frame of the medium, so that the temporal and mixed components of the self-energy tensor $\Pi_{00}$ and $\Pi_{i0}=\Pi_{0i}$ may have different coefficients than the corresponding tensor structures in the spatial components $\Pi_{ij}$. In other words, $\Pi_{00}\neq\Pi_B+\Pi_Ck_0^2$ and $\Pi_A\neq\Pi_C$.

The sum over one-particle reducible diagrams constitutes a geometric series. So the resummed propagators are given by
\begin{equation}
 D=D_0\sum_{n=0}^\infty(-\Pi D_0)^n=D_0(1+\Pi D_0)^{-1}=\left(D_0^{-1}+\Pi\right)^{-1}\,,
\end{equation}
and similarly for the ghosts. By inverting this matrix we get
\begin{align}
 D_{00}&=\frac{k_0^2+k^2+\Pi_B+\Pi_Ck^2}{\left(k_0^2+k^2+\Pi_{00}\right)\left(k_0^2+k^2+\Pi_B+\Pi_Ck^2\right)-\Pi_A^2k_0^2k^2}\,,\\
 D_{i0}&=\frac{-\Pi_Ak_ik_0}{\left(k_0^2+k^2+\Pi_{00}\right)\left(k_0^2+k^2+\Pi_B+\Pi_Ck^2\right)-\Pi_A^2k_0^2k^2}\,,\\
 D_{ij}&=\frac{1}{k_0^2+k^2+\Pi_B}\left(\delta_{ij}-\frac{k_ik_j}{k^2}\right)\notag\\
 &\hspace{13pt}+\frac{k_0^2+k^2+\Pi_{00}}{\left(k_0^2+k^2+\Pi_{00}\right)\left(k_0^2+k^2+\Pi_B+\Pi_Ck^2\right)-\Pi_A^2k_0^2k^2}\frac{k_ik_j}{k^2}\,.
\end{align}

We can rewrite these expressions in terms of the self-energy tensor as
\begin{equation}
 D_{00}=\frac{1}{k_0^2+k^2+\Pi}\hspace{10pt}\mathrm{and}\hspace{10pt}D_{ij}=\frac{1}{k_0^2+k^2+\Sigma_1}\left(\delta_{ij}-\frac{k_ik_j}{k^2}\right)+\frac{1}{k_0^2+k^2+\Sigma_2}\frac{k_ik_j}{k^2}\,,
\end{equation}
where
\begin{align}
 \Pi&=\Pi_{00}-\frac{\Pi_A^2k_0^2k^2}{k_0^2+k^2+\Pi_B+\Pi_Ck^2}=\Pi_{00}-\frac{\Pi_{0i}\Pi_{i0}}{k_0^2+k^2+\Pi_{ij}k_ik_j/k^2}\,,\\
 \Sigma_1&=\Pi_B=\frac{1}{d-1}\left(\Pi_{ii}-\frac{\Pi_{ij}k_ik_j}{k^2}\right)\,,\\
 \Sigma_2&=\Pi_B+\Pi_Ck^2-\frac{\Pi_A^2k_0^2k^2}{k_0^2+k^2+\Pi_{00}}=\frac{\Pi_{ij}k_ik_j}{k^2}-\frac{\Pi_{0i}\Pi_{i0}}{k_0^2+k^2+\Pi_{00}}\,,
\end{align}
and
\begin{equation}
 D_{0l}=\frac{-\Pi_{0l}}{\left(k_0^2+k^2+\Pi_{00}\right)\left(k_0^2+k^2+\Pi_{ij}k_ik_j/k^2\right)-\Pi_{0i}\Pi_{i0}}\,.
\end{equation}
We see that, although the free Feynman propagator is diagonal, the resummed propagator is not.

The free ghost propagator $G_0$ as well as the ghost self-energy $\Gamma$ are scalar functions, so the resummation of the geometric series for the full ghost propagator $G$ is trivial:
\begin{equation}
 G=\left(G_0^{-1}+\Gamma\right)^{-1}=\frac{1}{k_0^2+k^2+\Gamma}\,.
\end{equation}

\subsection{Static gauge}

Static gauge~\cite{D'Hoker:1981us} satisfies the gauge condition $\partial_0 A_0=0$, but this condition alone does not give an invertible propagator, so we need to modify it in order to fix the gauge also for the spatial gluons. This can be done by adding the gauge fixing term $\bigl(\partial_0A_0+\sqrt{\alpha/\xi}\d\cdot\A^a\bigr)^2/2\alpha$ and taking the limit $\alpha\to0$, which gives back the original gauge condition. This limit would diverge in the Lagrangian, but leads to a finite propagator. The freedom in how to fix the gauge for the spatial gluons is reflected in the residual gauge fixing parameter $\xi$. The gauge condition on the spatial gluons is lifted for $\xi\to\infty$ and accordingly the propagator diverges in this limit.

The inverse of the free propagator can be read from the Lagrangian:
\begin{equation}
 D_{0}^{-1}=\begin{pmatrix} \dfrac{k_0^2}{\alpha}+k^2 & & -\left(1-\dfrac{1}{\sqrt{\alpha\xi}}\right)k_0k_j \\ \vspace{-10pt} \\ -\left(1-\dfrac{1}{\sqrt{\alpha\xi}}\right)k_ik_0 & & \left(k_0^2+k^2\right)\delta_{ij}-\left(1-\dfrac{1}{\xi}\right)k_ik_j \end{pmatrix}\,,
\end{equation}
which can be inverted to
\begin{align}
 D_{0}&=\begin{pmatrix} \dfrac{\alpha\left(\xi k_0^2+k^2\right)}{\left(\sqrt{\xi}k_0^2+\sqrt{\alpha}k^2\right)^2} & & \dfrac{\left(\alpha\xi-\sqrt{\alpha\xi}\right)k_0k_j}{\left(\sqrt{\xi}k_0^2+\sqrt{\alpha}k^2\right)^2} \\ \vspace{-10pt} \\ \dfrac{\left(\alpha\xi-\sqrt{\alpha\xi}\right)k_ik_0}{\left(\sqrt{\xi}k_0^2+\sqrt{\alpha}k^2\right)^2} & & \dfrac{1}{k_0^2+k^2}\left(\delta_{ij}-\dfrac{k_ik_j}{k^2}\right)+\dfrac{\xi\left(k_0^2+\alpha k^2\right)}{\left(\sqrt{\xi}k_0^2+\sqrt{\alpha}k^2\right)^2}\dfrac{k_ik_j}{k^2} \end{pmatrix}\notag\\\notag\\
 &\stackrel{\alpha\to0}{=}\begin{pmatrix} \dfrac{\delta_{k_0}}{k^2} & & 0 \\ \vspace{-10pt} \\ 0 & & \dfrac{1-\delta_{k_0}}{k_0^2+k^2}\left(\delta_{ij}-\dfrac{k_ik_j}{k_0^2}\right)+\dfrac{\delta_{k_0}}{k^2}\left(\delta_{ij}-(1-\xi)\dfrac{k_ik_j}{k^2}\right) \end{pmatrix}\,,
\end{align}
where by $\delta_{k_0}$ we mean for $k_0=2\pi Tn$ with $n\in\mathbb{Z}$ that $\delta_{k_0}=\delta_{0n}$, i.e., selecting only the zero mode in the Matsubara sum. We see that the free propagator explicitly distinguishes between zero and nonzero modes. In particular, the $00$ component of the propagator contains only the zero mode, which means that in position space it does not depend on the imaginary time coordinate, as required by the gauge condition.

The ghost Lagrangian is given by
\begin{equation}
 \mathcal{L}_{gh}=\frac{1}{\sqrt{\alpha}}\left(\partial_0\bar{c}^a\right)D_0^{ab}c^b+\frac{1}{\sqrt{\xi}}\left(\d\bar{c}^a\right)\cdot\D^{ab}c^b\,,
 \label{staticghost}
\end{equation}
from which it follows that the free ghost propagator is
\begin{equation}
 G_0=\frac{\sqrt{\alpha\xi}}{\sqrt{\xi}k_0^2+\sqrt{\alpha}k^2}\stackrel{\alpha\to0}{=}\frac{\sqrt{\xi}\delta_{k_0}}{k^2}\,.
\end{equation}

There is a ghost vertex with a temporal gluon that is proportional to $1/\sqrt{\alpha}$, so the $\alpha\to0$ limit may potentially be problematic in this interaction. However, this vertex is also proportional to the Matsubara frequency $k_0$ of the outgoing ghost propagator, which means that only nonzero modes can participate in this interaction. The number of ghost propagators and ghost-gluon vertices is always the same in any loop diagram, so in the most singular diagrams, where all vertices are with a temporal gluon, the powers of $\sqrt{\alpha}$ cancel exactly between the vertices and the numerators of the propagators. Then the $\alpha\to0$ limit can be taken without problems and all propagators are given by $1/k_0^2$, which is not singular because the zero-modes do not contribute. This makes all loop integrations scaleless and therefore vanish. If there are some vertices with spatial gluons, then there are more powers of $\sqrt{\alpha}$ in the numerator than in the denominator and the diagram vanishes trivially in the $\alpha\to0$ limit.

So we see that the ghosts completely decouple from the temporal gluons. For the interactions with the spatial gluons the $\alpha\to0$ limit is unproblematic. There is a factor of $1/\sqrt{\xi}$ at each vertex, which exactly cancels the $\sqrt{\xi}$ factor in the ghost propagators. So we can in fact simplify the ghost sector considerably, because as we have just shown the nonzero Matsubara frequencies, the parameter $\xi$, or interactions with temporal gluons are irrelevant. Therefore the modified ghost Lagrangian and free propagator
\begin{equation}
 \mathcal{L}_{gh}=\left(\d\bar{c}^a\right)\cdot\D^{ab}c^b\hspace{20pt}\mathrm{and}\hspace{20pt}G_0=\frac{\delta_{k_0}}{k^2}
\end{equation}
with static (i.e., independent of the imaginary time coordinate) ghost fields give exactly the same contributions as the more complicated Lagrangian given above.

For the resummed propagator we can use the same parametrization of the self-energy tensor as in Feynman gauge. Then we get
\begin{align}
 D_{00}&=\frac{k_0^2+k^2/\xi+\Pi_B+\Pi_Ck^2}{\left(k_0^2/\alpha+k^2+\Pi_{00}\right)\bigl(k_0^2+k^2/\xi+\Pi_B+\Pi_Ck^2\bigr)-\bigl(1-1/\sqrt{\alpha\xi}-\Pi_A\bigr)^2k_0^2k^2}\notag\\
 &\stackrel{\alpha\to0}{=}\frac{\delta_{k_0}}{k^2+\Pi_{00}}\,,\\
 D_{i0}&=\frac{\left(1-1/\sqrt{\alpha\xi}-\Pi_A\right)k_ik_0}{\left(k_0^2/\alpha+k^2+\Pi_{00}\right)\bigl(k_0^2+k^2/\xi+\Pi_B+\Pi_Ck^2\bigr)-\bigl(1-1/\sqrt{\alpha\xi}-\Pi_A\bigr)^2k_0^2k^2}\stackrel{\alpha\to0}{=}0\,,\\
 D_{ij}&=\frac{1}{k_0^2+k^2+\Pi_B}\left(\delta_{ij}-\frac{k_ik_j}{k^2}\right)\notag\\
 &\hspace{13pt}+\frac{k_0^2/\alpha+k^2+\Pi_{00}}{\left(k_0^2/\alpha+k^2+\Pi_{00}\right)\bigl(k_0^2+k^2/\xi+\Pi_B+\Pi_Ck^2\bigr)-\bigl(1-1/\sqrt{\alpha\xi}-\Pi_A\bigr)^2k_0^2k^2}\frac{k_ik_j}{k^2}\notag\\
 &\stackrel{\alpha\to0}{=}\frac{1-\delta_{k_0}}{k_0^2+k^2+\Pi_B}\left(\delta_{ij}+\frac{\left(1-\Pi_C\right)k_ik_j}{k_0^2+\Pi_B+\Pi_Ck^2}\right)+\frac{\delta_{k_0}}{k^2+\Pi_B}\left(\delta_{ij}-\frac{\left(1-\xi+\xi\Pi_C\right)k_ik_j}{k^2+\xi\left(\Pi_B+\Pi_Ck^2\right)}\right)\,.
\end{align}
Or in analogy to the functions $\Pi$, $\Sigma_1$, and $\Sigma_2$ that we defined in Feynman gauge we can also write
\begin{align}
 D_{00}&=\frac{\delta_{k_0}}{k^2+\Pi}\,,\\
 D_{i0}&=D_{0j}=0\,,\\
 D_{ij}&=\frac{1-\delta_{k_0}}{k_0^2+k^2+\Sigma_1}\left(\delta_{ij}-\frac{k_ik_j}{k^2}\right)+\frac{1-\delta_{k_0}}{k_0^2+\Sigma_2}\frac{k_ik_j}{k^2}+\frac{\delta_{k_0}}{k^2+\Sigma_1}\left(\delta_{ij}-\frac{k_ik_j}{k^2}\right)+\frac{\xi\delta_{k_0}}{k^2+\xi\Sigma_2}\frac{k_ik_j}{k^2}\,,
\end{align}
where now
\begin{equation}
 \Pi=\Pi_{00}\,,\hspace{10pt}\Sigma_1=\Pi_B=\frac{1}{(d-1)k^2}\left(k^2\Pi_{ii}-\Pi_{ij}k_ik_j\right)\,,\hspace{10pt}\mathrm{and}\hspace{10pt}\Sigma_2=\Pi_B+\Pi_Ck^2=\frac{\Pi_{ij}k_ik_j}{k^2}\,.
\end{equation}

The resummed ghost propagator follows trivially from the modified ghost Lagrangian.
\begin{equation}
 G=\frac{\delta_{k_0}}{k^2+\Gamma}\,.
\end{equation}

For $\xi=1$ the static part of the gluon propagator (i.e., $k_0=0$) has the same form as in Feynman gauge, which is why this choice is also called Feynman static gauge. The self-energy functions still differ between the two gauges. For $\xi=0$ the static part of the propagator has the same form as in Coulomb gauge, so this choice could be called Coulomb static gauge.

\subsection{Coulomb gauge}

Coulomb gauge is defined by the gauge condition $\d\cdot\A^a=0$. It can be implemented by adding the gauge fixing term $\left(\bm{\nabla}\cdot\A^a\right)^2/2\xi$ to the Lagrangian as well as the ghost Lagrangian $\left(\bm{\nabla}\bar{c}^a\right)\cdot\D^{ab} c^b$ with the limit $\xi\to0$. If we compare this to the gauge fixing term in static gauge, we see that Coulomb gauge can also be obtained from there by first taking the limit $\alpha\to\infty$ and then $\xi\to0$, so we can reuse all results from the previous section.

The free propagator is then given by
\begin{equation}
 D_0\stackrel{\xi\to0}{=}\begin{pmatrix} \dfrac{1}{k^2} & 0 \\\vspace{-10pt}\\ 0 & \dfrac{1}{k_0^2+k^2}\left(\delta_{ij}-\dfrac{k_ik_j}{k^2}\right) \end{pmatrix}\,,
\end{equation}
and the resummed propagator by
\begin{equation}
 D\stackrel{\xi\to0}{=}\begin{pmatrix} \dfrac{1}{k^2+\Pi} & & 0 \\\vspace{-10pt}\\ 0 & & \dfrac{1}{k_0^2+k^2+\Sigma_1}\left(\delta_{ij}-\dfrac{k_ik_j}{k^2}\right)\end{pmatrix}\,,
\end{equation}
where the self-energy functions $\Pi$ and $\Sigma_1$ are defined as in static gauge.

The temporal component of the propagator is the same as in static gauge, except that in Coulomb gauge also the nonzero Matsubara frequencies are allowed (although they do not appear explicitly in the free propagator). The spatial part of the propagator is transversely polarized with respect to $\k$ and the mixed temporal and spatial components vanish, such that the gauge condition is explicitly satisfied as $k_iD_{i\nu}=0$. This relation holds for both the free and the resummed propagator, and only the coefficient $\Sigma_1$ of the transversely polarized part of the self-energy tensor remains in the propagator after the resummation.

After a redefinition of the ghost fields $(\bar{c},c)\to\xi^{1/4}(\bar{c},c)$, the limit $\xi\to0$ eliminates the first term in~\eqref{staticghost} and the free and resummed propagators are given by
\begin{equation}
 G_0=\frac{1}{k^2}\hspace{10pt}\mathrm{and}\hspace{10pt}G=\frac{1}{k^2+\Gamma}\,.
\end{equation}
The ghosts only couple to spatial gluons like in static gauge.

Quantization in Coulomb gauge generates the so-called Schwinger-Christ-Lee term~\cite{Schwinger:1962wd,Christ:1980ku}. This term is an $\alpha_\mathrm{s}^2$ suppressed term that involves a nonlocal interaction with transverse gluons. It is beyond the accuracy of the present work.

\subsection{Phase-space Coulomb gauge}

There exists an alternative formulation of Coulomb gauge QCD that is defined in the so-called phase-space formalism~\cite{Andrasi:2003zf}, which we will adapt here to the Euclidean space of the imaginary time formalism. An auxiliary field $\E$ is introduced in the action $S$:
\begin{align}
 e^{-S}&=\exp\left[-\int_0^{1/T}\hspace{-4pt}d\tau\int d^3x\left(\frac{1}{4}F_{ij}^aF_{ij}^a+\frac{1}{2}F_{0i}^aF_{0i}^a\right)\right]\notag\\
 &=\mathcal{N}^{-1}\int\mathcal{D}E_i\exp\left[-\int_0^{1/T}\hspace{-4pt}d\tau\int d^3x\left(\frac{1}{4}F_{ij}^aF_{ij}^a+iE_i^aF_{0i}^a+\frac{1}{2}E_i^aE_i^a\right)\right]\,.
\end{align}
This step can be interpreted such that now the chromoelectric field is treated as a dynamical variable. This interpretation originates from the equations of motion for the $\E$-field, which are $E_i^a=-iF_{0i}^a$ (the factor $i$ is an effect of the imaginary time formalism, in Minkowski space it is absent). So we will call $\E$ the electric field for the rest of this section. One can easily return to the original action, up to some irrelevant constant $\mathcal{N}$, by explicitly carrying out the path integral over the electric field, which is possible because it only appears in quadratic terms in the exponential.

With this new action we can calculate as if there was a seven-component gluon field $A_\alpha$, where $\alpha=0$ corresponds to $A_0$, $\alpha=1,2,3$ to $\A$, and $\alpha=4,5,6$ to $\E$. The free propagator $\left(D_0\right)_{\alpha\beta}$ will be the $7\times7$ matrix given through the quadratic terms of this gluon field as $A_\alpha \left(D_0^{-1}\right)_{\alpha\beta}A_\beta$. In order to distinguish between the spatial gluon and electric field components in this unified description, we will use the propagator indices $i$, $j$ and $m$, $n$ exclusively for $\alpha,\beta=1,2,3$ and $\alpha,\beta=4,5,6$ respectively.

In order to fix the gauge we again introduce the terms $\left(\bm{\nabla}\cdot\A^a\right)^2/2\xi$ and $\left(\bm{\nabla}\bar{c}^a\right)\cdot\D^{ab} c^b$ into the Lagrangian. The ghost sector remains unchanged compared to standard Coulomb gauge, so we will only discuss the gluonic sector. Going from position to momentum space in the free action,
\begin{align}
 S_0&=\int_0^{1/T}\hspace{-4pt}d\tau\int d^3x\,\left[\frac{1}{2}\left(\partial^{\,}_iA_j^a\right)\left(\partial^{\,}_iA_j^a\right)-\frac{1}{2}\left(\partial^{\,}_iA_j^a\right)\left(\partial^{\,}_jA_i^a\right)+\frac{1}{2\xi}\left(\partial^{\phantom{b}}_iA_i^a\right)\left(\partial^{\,}_jA_j^a\right)\right.\notag\\
 &\hspace{95pt}+\left.iE_i^a\partial_0A_i^a-iE_i^a\partial^{\,}_iA_0^a+\frac{1}{2}E_i^aE_i^a\right]\notag\\
 &=\sum_K\hspace{-16pt}\int\,\frac{1}{2}\left[A_i^a(-K)\left(k^2\delta_{ij}-\frac{\xi-1}{\xi}k_ik_j\right)A_j^a(K)-E_i^a(-K)k_0A_i^a(K)+A_i^a(-K)k_0E_i^a(K)\right.\notag\\
 &\hspace{50pt}+\left.E_i^a(-K)k_iA_0^a(K)-A_0^a(-K)k_iE_i^a(K)+\frac{1}{2}E_i(-K)^aE_i^a(K)\right]\notag\\
 &=\sum_K\hspace{-16pt}\int\,\left[\frac{1}{2}A_\alpha^a(-K)\left(D_0^{-1}\right)_{\alpha\beta}A_\beta^a(K)\right]\,,
\end{align}
we get the inverse of the propagator as
\begin{equation}
 D_0^{-1}=\begin{pmatrix} 0 & 0 & -k_n \\ 0 & k^2\,\delta_{ij}-\left(1-1/\xi\right)k_ik_j & k_0\,\delta_{in} \\ k_m & -k_0\,\delta_{mj} & \delta_{mn} \end{pmatrix}\,,
\end{equation}
where we have written the $7\times7$ matrix in terms of $(1,3,3)\times(1,3,3)$ blocks. Inverting this and taking the $\xi\to0$ limit, we get the free propagator:
\begin{equation}
 D_0=\begin{pmatrix} \dfrac{1}{k^2} & & 0 & & \dfrac{k_n}{k^2} \\\vspace{-10pt}\\ 0 & & \dfrac{1}{k_0^2+k^2}\left(\delta_{ij}-\dfrac{k_ik_j}{k^2}\right) & & -\dfrac{k_0}{k_0^2+k^2}\left(\delta_{in}-\dfrac{k_ik_n}{k^2}\right) \\\vspace{-10pt}\\ -\dfrac{k_m}{k^2} & & \dfrac{k_0}{k_0^2+k^2}\left(\delta_{mj}-\dfrac{k_mk_j}{k^2}\right) & & \dfrac{k^2}{k_0^2+k^2}\left(\delta_{mn}-\dfrac{k_mk_n}{k^2}\right) \end{pmatrix}\,.
\end{equation}
We see that the temporal and spatial components still have the same propagators as in the standard formalism, in particular they do not mix with each other for $\xi=0$, but both do mix with the electric field. Also note that $D_0^T(K)=D_0(-K)$. This is of relevance for the off-diagonal terms, which have odd powers of the momentum in the numerator (the reason is that $A_0$ and $\A$ are of mass dimension $1$, while $\E$ is of dimension $2$).

\begin{figure}[t]
 \includegraphics[width=\linewidth]{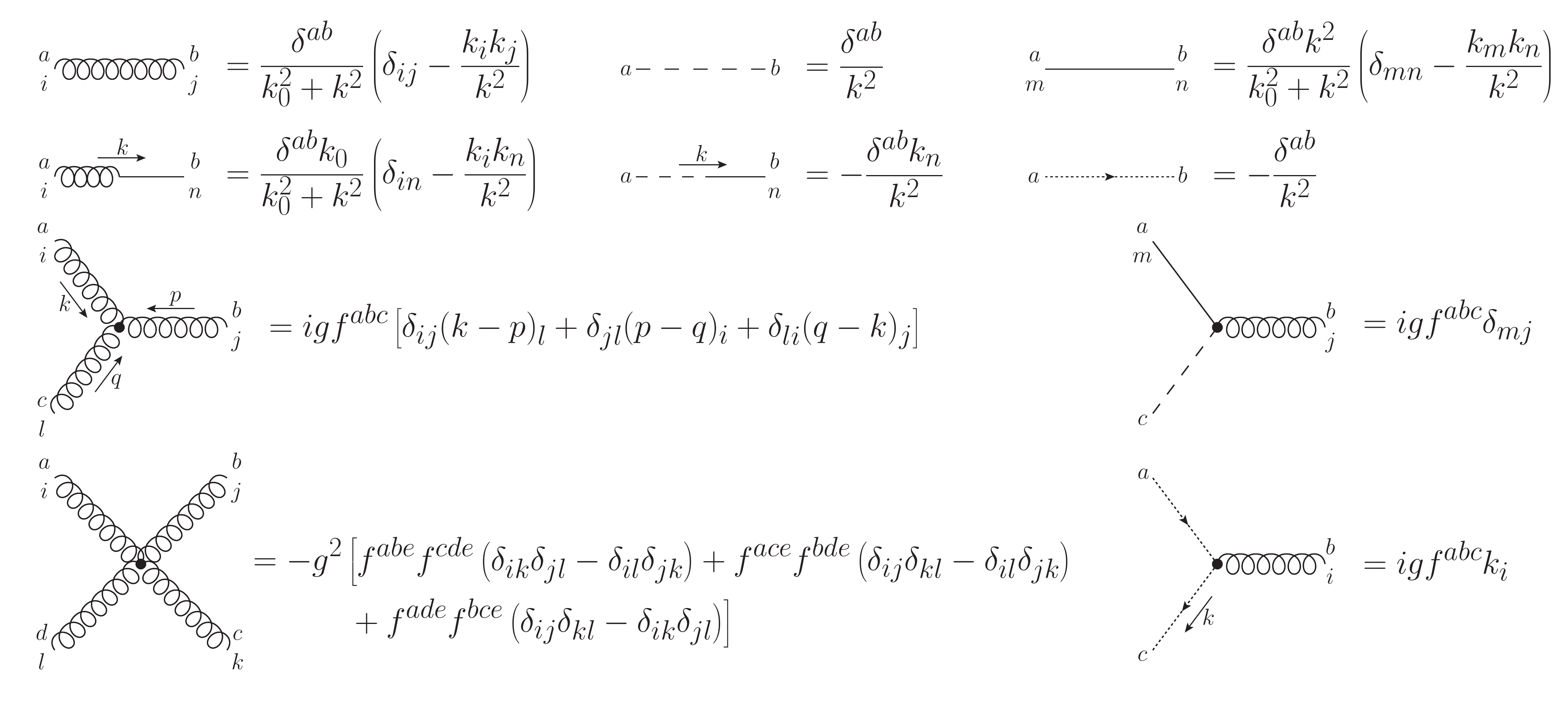}
 \caption{All free propagators and interaction vertices in phase-space Coulomb gauge. Whenever there is an arrow specifying the direction of a momentum over a mixed propagator, opposite momenta will give the negative propagator.}
 \label{PSFeynRules}
\end{figure}

The interaction part of the action is given by
\begin{equation}
 S_{int}=\int_0^{1/T}\hspace{-4pt}d\tau\int d^3x\,\left[gf^{abc}\left(\partial^{\,}_iA_j^a\right)A_i^bA_j^c+\frac{g^2}{4}f^{abe}f^{cde}A_i^aA_j^bA_i^cA_j^d-igf^{abc}A_0^aE_i^bA_i^c\right]\,.
\end{equation}
This gives the same three- and four-gluon vertices as in standard Coulomb gauge if only spatial gluons are involved, but the temporal gluons now interact with the spatial gluons only through a three-field vertex with an additional electric field and the simple coefficient $igf^{abc}\delta_{im}$. All Feynman rules of phase-space Coulomb gauge are shown in Fig.~\ref{PSFeynRules}.

For the resummed propagator we need to introduce a new parametrization of the self-energy tensor in the form a $7\times7$-matrix:
\begin{equation}
 \Pi=\begin{pmatrix} \Pi_{tt} &  k_0k_j\,\Pi_{ts} & -k_n\,\Pi_{te} \\\vspace{-10pt}\\  k_ik_0\,\Pi_{ts} & \Pi_{ss1}\,\delta_{ij}+\Pi_{ss2}\,\dfrac{k_ik_j}{k^2} &  k_0\left(\Pi_{se1}\,\delta_{in}+\Pi_{se2}\,\dfrac{k_ik_n}{k^2}\right) \\\vspace{-10pt}\\  k_m\,\Pi_{te} & -k_0\left(\Pi_{se1}\,\delta_{mj}+\Pi_{se2}\,\dfrac{k_mk_j}{k^2}\right) & \Pi_{ee1}\,\delta_{mn}+\Pi_{ee2}\,\dfrac{k_mk_n}{k^2} \end{pmatrix}\,,
\end{equation}
where the labels $t$, $s$, and $e$ stand for temporal, spatial, and electric respectively. Then the resummed propagators are
\begin{align}
 &D_{00}=\frac{1+\Pi_{ee1}+\Pi_{ee2}}{k^2\left(1+\Pi_{te}\right)^2+\left(1+\Pi_{ee1}+\Pi_{ee2}\right)\Pi_{tt}}\,,\label{PSCGin}\\
 &D_{0j}=D_{i0}=0\,,\\
 &D_{0n}=\frac{\left(1+\Pi_{te}\right)k_n}{k^2\left(1+\Pi_{te}\right)^2+\left(1+\Pi_{ee1}+\Pi_{ee2}\right)\Pi_{tt}}\,,\\
 &D_{m0}=\frac{-\left(1+\Pi_{te}\right)k_m}{k^2\left(1+\Pi_{te}\right)^2+\left(1+\Pi_{ee1}+\Pi_{ee2}\right)\Pi_{tt}}\,,\\
 &D_{ij}=\frac{1+\Pi_{ee1}}{k_0^2\left(1+\Pi_{se1}\right)^2+k^2\left(1+\Pi_{ee1}\right)+\left(1+\Pi_{ee1}\right)\Pi_{ss1}}\left(\delta_{ij}-\frac{k_ik_j}{k^2}\right)\,,\\
 &D_{in}=\frac{-\left(1+\Pi_{se1}\right)k_0}{k_0^2\left(1+\Pi_{se1}\right)^2+k^2\left(1+\Pi_{ee1}\right)+\left(1+\Pi_{ee1}\right)\Pi_{ss1}}\left(\delta_{in}-\frac{k_ik_n}{k^2}\right)\,,\\
 &D_{mj}=\frac{\left(1+\Pi_{se1}\right)k_0}{k_0^2\left(1+\Pi_{se1}\right)^2+k^2\left(1+\Pi_{ee1}\right)+\left(1+\Pi_{ee1}\right)\Pi_{ss1}}\left(\delta_{mj}-\frac{k_mk_j}{k^2}\right)\,,\\
 &D_{mn}=\frac{k^2+\Pi_{ss1}}{k_0^2\left(1+\Pi_{se1}\right)^2+k^2\left(1+\Pi_{ee1}\right)+\left(1+\Pi_{ee1}\right)\Pi_{ss1}}\left(\delta_{mn}-\frac{k_mk_n}{k^2}\right)\notag\\
 &\hspace{40pt}+\frac{\Pi_{tt}}{k^2\left(1+\Pi_{te}\right)^2+\left(1+\Pi_{ee1}+\Pi_{ee2}\right)\Pi_{tt}}\frac{k_mk_n}{k^2}\,.\label{PSCGfin}
\end{align}

We see that the self-energy components that are proportional to $k_i$ or $k_j$ (i.e., $\Pi_{ts}$, $\Pi_{ss2}$, and $\Pi_{se2}$) do not appear at all, while the ones that are proportional only to $k_m$ or $k_n$ (i.e., $\Pi_{te}$ and $\Pi_{ee2}$), appear only in $D_{00}$, $D_{m0}$, $D_{0n}$, and $D_{mn}$. The reason for this is that every free propagator with a spatial gluon index $i$ or $j$ is proportional to the transverse projector $\delta_{ij}-k_ik_j/k^2$, so the self-energy components $\Pi_{ts}$, $\Pi_{ss2}$, and $\Pi_{se2}$ drop out of the geometric series. Since only the $\delta_{ij}$ self-energy terms remain in the geometric series for $D_{ij}$, $D_{in}$, and $D_{mj}$, also the resummed propagators are proportional to the transverse projector. A mixing of temporal and spatial gluons is still not possible, because $\left(D_0\right)_{i0}$ and $\left(D_0\right)_{0j}$ are zero from the outset and intermediate electric field contributions like, e.g., $\left(D_0\right)_{in}\Pi_{nm}\left(D_0\right)_{m0}$ or $\left(D_0\right)_{in}\Pi_{n0}\left(D_0\right)_{00}$ always involve a contraction of the transverse projector with the momentum $k_m$, either from the self-energy or the $\left(D_0\right)_{m0}$ propagator. In the case of the propagators $D_{00}$, $D_{m0}$, $D_{0n}$, and $D_{mn}$, there appear terms in the geometric series without any transverse projectors, so those propagators also depend on the self-energy terms $\Pi_{te}$ and $\Pi_{ee2}$. Also note that, in contrast to the free propagator, the resummed $D_{mn}$ contains a part that is not proportional to the transverse projector, which comes, e.g., from terms like $\left(D_0\right)_{m0}\Pi_{00}\left(D_0\right)_{0n}$.

\subsection{Expansions of the propagators}
\label{Expprop}

In the small coupling case the two energy scales $T$ and $m_D\sim gT$ are well separated, so we expand the propagators accordingly. The Matsubara frequencies are always of order $T$ and the momentum $k$ can be either of order $T$ or $m_D$. The self-energy functions are at least of order $g^2T^2$, so if $k$ is of order $T$ then the propagators have to be expanded in the self-energy, which is equivalent to using free propagators instead of resummed propagators.

If $k$ is of order $m_D$ but $k_0$ is not zero, then the propagators also have to be expanded in $k^2/k_0^2$, which leads to scaleless integrals in most cases (and in all integrals appearing in this paper). An exception to this are the temporal propagators in static and Coulomb gauge, which do not have a $k_0^2$ term in the denominator.

If $k$ is of order $m_D$ and $k_0$ is zero, then the leading term of the self-energy may be of the same order as $k^2$ and the propagator has to be expanded in the next-to-leading terms. It is known that only the self-energy in the temporal propagator has a term of order $g^2T^2$, which is gauge invariant and given by the square of the Debye mass $m_D^2$, see Eq.~\eqref{mD}. In Coulomb gauge, the free propagator is independent of the Matsubara frequencies. The self-energy, however, is such that it is of order $g^2T^2$ for the zero mode, while it is of higher order for the other frequencies. The self-energies in the spatial propagator start at order $g^4T^2$, therefore the spatial propagator has to be expanded and we can use the free one.

It is a straightforward calculation to show that also in the phase-space Coulomb gauge (PSCG) only $\Pi_{tt}$ has a term of order $g^2T^2$ and this is again given by $m_D^2$. All other self-energies need to be expanded, see Eqs.~\eqref{PSCGin}-\eqref{PSCGfin}. Therefore, the spatial and mixed spatial-electric propagators remain massless, but the electric and mixed temporal-electric propagators also get massive denominators.

We summarize here the propagators in different gauges in the leading order expansion for $k_0=0$ and $k\sim m_D$.
\begin{align}
 &D^{FG}=\begin{pmatrix} \dfrac{1}{k^2+m_D^2} & & 0 \\\vspace{-10pt}\\ 0 & & \dfrac{\delta_{ij}}{k^2}\end{pmatrix}\,,\hspace{10pt}
 D^{SG}=\begin{pmatrix} \dfrac{1}{k^2+m_D^2} & & 0 \\\vspace{-10pt}\\ 0 & & \dfrac{1}{k^2}\left(\delta_{ij}-(1-\xi)\dfrac{k_ik_j}{k^2}\right)\end{pmatrix}\,,\\
 &D^{CG}=D^{SG}\bigr|_{\xi=0}\,,\hspace{10pt}D^{PSCG}=\begin{pmatrix} \dfrac{1}{k^2+m_D^2} & & 0 & & \dfrac{k_n}{k^2+m_D^2} \\\vspace{-10pt}\\ 0 & & \dfrac{1}{k^2}\left(\delta_{ij}-\dfrac{k_ik_j}{k^2}\right) & & 0 \\\vspace{-10pt}\\ -\dfrac{k_m}{k^2+m_D^2} & & 0 & & \delta_{mn}-\dfrac{k_mk_n}{k^2+m_D^2} \end{pmatrix}\,.\label{PSCGprop}
\end{align}

\section{Electric scale two-loop integrals}
\label{scalemD}

In this appendix, we will explicitly write down the integrals and their results for all the two-loop self-energy diagrams at the scale $m_D$. In order to calculate the integrals we make use of an algorithm that systematically reduces the integrals to a handful of master integrals by the method of integration by parts and then replaces these master integrals by their known values. More details on this algorithm can be found in appendix~\ref{M3LI}.

All relevant diagrams for $\Pi^{(2)}_{m_D}(0,k\sim m_D)$ are shown in Fig.~\ref{diagrams}. As explained in appendix~\ref{Expprop}, only temporal gluons carry the Debye mass in the propagator, so it makes sense to visually distinguish between temporal and spatial gluons in the diagrams. All Matsubara frequencies are assumed to be zero, which means that a vertex with one temporal gluon and two spatial gluons or ghosts (if they are required by the chosen gauge) cannot appear, because it would be proportional to the Matsubara frequencies. This is why there are no three-gluon vertices with just one temporal gluon in all the diagrams of Fig.~\ref{diagrams}. Tadpole diagrams with only spatial gluons or ghosts are scaleless and therefore have been omitted in Fig.~\ref{diagrams}. Fermion propagators do not have zero-modes, so also light quark loops cannot contribute to $\Pi^{(2)}_{m_D}(0,k\sim m_D)$.

We will do the calculation explicitly in Feynman, Coulomb, and phase-space Coulomb gauge. In the case of the static gauge we will not perform the calculation for a generic gauge fixing parameter $\xi$. For $\xi=1$ and $\xi=0$ the calculation is identical to the one in Feynman and Coulomb gauge, respectively.

The color factors of the two-loop self-energy can be calculated using the quadratic Casimir of the adjoint representation and the Jacobi identity:
\begin{equation}
 \mathrm{Tr}\bigl[T^a_AT^b_A\bigr]=(-if^{acd})(-if^{bdc})=f^{acd}f^{bcd}=C_A\,\delta^{ab}\,,
\end{equation}
\begin{equation}
 f^{abe}f^{ecd}+f^{bce}f^{ead}+f^{cae}f^{ebd}=0\,.
\end{equation}
With these we get
\begin{equation}
 f^{acd}f^{dce}f^{egh}f^{hgb}=(-C_A\,\delta^{ae})(-C_A\,\delta^{eb})=C_A^2\,\delta^{ab}\,,
\end{equation}
\begin{equation}
 f^{acd}f^{dgh}f^{hge}f^{ecb}=-C_Af^{acd}f^{dcb}=C_A^2\,\delta^{ab}\,,
\end{equation}
\begin{equation}
 f^{acd}f^{cge}f^{deh}f^{hgb}=-\frac{1}{2}f^{acd}\bigl(f^{cge}f^{edh}+f^{che}f^{egd}\bigr)f^{hgb}=\frac{1}{2}f^{acd}f^{dce}f^{egh}f^{hgb}=\frac{1}{2}C_A^2\,\delta^{ab}\,.
\end{equation}

All color factors are given by these expressions or combinations thereof. Symmetry factors appear only when gluons of the same type (temporal or spatial) can be exchanged, which is the case for $L_3$, $L_8$, $L_9$, $L_{10}$, and $L_{12}$, although in the case of $L_3$, $L_8$, and $L_{12}$ one symmetry factor $1/2$ is compensated by a factor $2$ from the four-gluon vertices. From the vertices we either get $(ig)^4$, $(ig)^2(-g^2)$ or $(-g^2)^2$, which is equal to $g^4$ in each case. So no additional signs arise from the vertices, but the ghost loop gets a minus due to its Grassmann nature. Then we have
\begin{equation}
 C(L_1)=C(L_2)=-C(L_8)=C(L_9)=C(L_{10})=\frac{1}{2}C_A^2\,,
\end{equation}
\begin{equation}
 C(L_7)=-C(L_{11})=-C(L_{12})=C_A^2\,,
\end{equation}
\begin{equation}
 C(L_3)=-C(L_4)=-C(L_5)=-C(L_6)=\frac{3}{2}C_A^2\,.
\end{equation}

\subsection{Feynman gauge}

We will call the momenta in the diagrams of Fig.~\ref{diagrams} in such a way that $k$ appears in each temporal gluon propagator (even in the temporal gluon loops in $L_8$, $L_{10}$ and $L_{12}$ through a shift of the loop momentum by $k$), while the additional loop momenta will be called $p$ and $q$. In the denominator only the combinations $\k+\p$, $\k+\q$, and either $\k+\p+\q$ or $\p-\q$ can appear. The reason for this choice is that with this momentum configuration the integrals are already in the form required by the algorithm described in appendix~\ref{M3LI}. We will use the abbreviation $P(\k)=\k^2+m_D^2$. $L_5$ and $L_6$ are the same up to a relabeling of the momenta, so we will calculate them together. Then we have
\begin{align}
 L_1&=-\frac{C_RC_A^2g^6}{4}\int\hspace{-18pt}\int_{k,\,p,\,q\sim m_D}\limits\hspace{-18pt}\int\frac{2T(2k_i+p_i)\hspace{-2pt}\left(p^2\delta_{ij}-p_ip_j+q^2\delta_{ij}-q_iq_j-(\p\cdot\q)\delta_{ij}+p_iq_j\right)\hspace{-2pt}(2k_j+q_j)}{\p^2\,(\p-\q)^2\,\q^2\,P(\k+\p)P(\k+\q)P(\k)^2}\notag\\
 &=\frac{C_RC_A^2\alpha_\mathrm{s}^3T}{m_D}\left[-\frac{1}{32\epsilon}-\frac{7}{32}+\frac{3}{32}\gamma_E-\frac{\pi^2}{24}-\frac{3}{32}\ln\frac{\pi\mu^2}{m_D^2}+\mathcal{O}(\epsilon)\right]\,,\\
 L_2&=-\frac{C_RC_A^2g^6}{4}\int\hspace{-18pt}\int_{k,\,p,\,q\sim m_D}\limits\hspace{-18pt}\int\,\frac{T\,\bigl[(2\k+\p)\cdot(2\k+\p+2\q)\bigr]\,\bigl[(2\k+\q)\cdot(2\k+2\p+\q)\bigr]}{\p^2\,\q^2\,P(\k+\p)\,P(\k+\q)\,P(\k+\p+\q)\,P(\k)^2}\notag\\
 &=\frac{C_RC_A^2\alpha_\mathrm{s}^3T}{m_D}\left[-\frac{1}{32\epsilon}-\frac{1}{4}+\frac{3}{32}\gamma_E-\frac{\pi^2}{12}-\frac{3}{32}\ln\frac{\pi\mu^2}{256m_D^2}+\mathcal{O}(\epsilon)\right]\,,\\
 L_3&=-\frac{3C_RC_A^2g^6}{4}\int\hspace{-18pt}\int_{k,\,p,\,q\sim m_D}\limits\hspace{-18pt}\int\,\frac{T\,d}{\p^2\,(\p-\q)^2\,P(\k+\q)\,P(\k)^2}\notag\\
 &=\frac{C_RC_A^2\alpha_\mathrm{s}^3T}{m_D}\left[-\frac{9}{32\epsilon}-\frac{3}{8}+\frac{27}{32}\gamma_E-\frac{27}{32}\ln\frac{\pi\mu^2}{m_D^2}+\mathcal{O}(\epsilon)\right]\,,\\
 L_4&=\frac{3C_RC_A^2g^6}{4}\int\hspace{-18pt}\int_{k,\,p,\,q\sim m_D}\limits\hspace{-18pt}\int\,\frac{T\,(2\k+\p)\cdot(2\k+\q)}{\p^2\,\q^2\,P(\k+\p)\,P(\k+\q)\,P(\k)^2}\notag\\
 &=\frac{C_RC_A^2\alpha_\mathrm{s}^3T}{m_D}\left[\frac{39}{32}+\mathcal{O}(\epsilon)\right]\,,\\
 L_5&+L_6=\frac{3C_RC_A^2g^6}{4}\int\hspace{-18pt}\int_{k,\,p,\,q\sim m_D}\limits\hspace{-18pt}\int\,\frac{2T\,(2\k+\p)\cdot(2\k+\p+\q)}{\p^2\,(\p-\q)^2\,P(\k+\p)\,P(\k+\q)\,P(\k)^2}\notag\\
 &=\frac{C_RC_A^2\alpha_\mathrm{s}^3T}{m_D}\left[\frac{9}{32\epsilon}-\frac{3}{16}-\frac{27}{32}\gamma_E+\frac{\pi^2}{8}+\frac{27}{32}\ln\frac{\pi\mu^2}{m_D^2}+\mathcal{O}(\epsilon)\right]\,,\\
 L_7&=-\frac{C_RC_A^2g^6}{2}\int\hspace{-18pt}\int_{k,\,p,\,q\sim m_D}\limits\hspace{-18pt}\int\,\frac{T\,(2\k+\p)^2\,(2\k+\p+\q)^2}{\p^2\,(\p-\q)^2\,P(\k+\p)^2\,P(\k+\q)\,P(\k)^2}\notag\\
 &=\frac{C_RC_A^2\alpha_\mathrm{s}^3T}{m_D}\left[-\frac{1}{8\epsilon}-\frac{9}{8}+\frac{3}{8}\gamma_E-\frac{3}{8}\ln\frac{\pi\mu^2}{m_D^2}+\mathcal{O}(\epsilon)\right]\,,\\
 L_8&=\frac{C_RC_A^2g^6}{4}\int\hspace{-18pt}\int_{k,\,p,\,q\sim m_D}\limits\hspace{-18pt}\int\,\frac{T\,(2\k+\p+\q)^2}{\left((\p-\q)^2\right)^2P(\k+\p)\,P(\k+\q)\,P(\k)^2}\notag\\
 &=\frac{C_RC_A^2\alpha_\mathrm{s}^3T}{m_D}\left[-\frac{1}{32\epsilon}+\frac{3}{32}\gamma_E-\frac{3}{32}\ln\frac{\pi\mu^2}{m_D^2}+\mathcal{O}(\epsilon)\right]\,,\\
 L_9&=-\frac{C_RC_A^2g^6}{4}\int\hspace{-18pt}\int_{k,\,p,\,q\sim m_D}\limits\hspace{-18pt}\int\,\frac{T\,(2k_i+p_i)\,(2k_j+p_j)}{P(\k+\p)\,P(\k)^2}\notag\\
 &\hspace{68pt}\times\frac{\left(5p^2\delta_{ij}-(6-d)p_ip_j+2q^2\delta_{ij}-(6-4d)q_iq_j-2(\p\cdot\q)\delta_{ij}+(6-4d)p_iq_j\right)}{\left(\p^2\right)^2(\p-\q)^2\,\q^2}\notag\\
 &=\frac{C_RC_A^2\alpha_\mathrm{s}^3T}{m_D}\left[\frac{13}{64\epsilon}-\frac{19}{32}-\frac{39}{64}\gamma_E+\frac{39}{64}\ln\frac{\pi\mu^2}{m_D^2}+\mathcal{O}(\epsilon)\right]\,,\\
 L_{10}&=-\frac{C_RC_A^2g^6}{4}\int\hspace{-18pt}\int_{k,\,p,\,q\sim m_D}\limits\hspace{-18pt}\int\,\frac{T\,\left[(2\k+\p)\cdot(2\k+\p+2\q)\right]^2}{\left(\p^2\right)^2P(\k+\p)\,P(\k+\q)\,P(\k+\p+\q)\,P(\k)^2}\notag\\
 &=\frac{C_RC_A^2\alpha_\mathrm{s}^3T}{m_D}\left[-\frac{5}{48}+\frac{1}{6}\ln2+\mathcal{O}(\epsilon)\right]\,,\\
 L_{11}&=\frac{C_RC_A^2g^6}{2}\int\hspace{-18pt}\int_{k,\,p,\,q\sim m_D}\limits\hspace{-18pt}\int\,\frac{T\,\left[(2\k+\p)\cdot(-\q)\right]\,\left[(2\k+\p)\cdot(\p-\q)\right]}{\left(\p^2\right)^2(\p-\q)^2\,\q^2\,P(\k+\p)\,P(\k)^2}\notag\\
 &=\frac{C_RC_A^2\alpha_\mathrm{s}^3T}{m_D}\left[\frac{1}{64\epsilon}-\frac{1}{32}-\frac{3}{64}\gamma_E+\frac{3}{64}\ln\frac{\pi\mu^2}{m_D^2}+\mathcal{O}(\epsilon)\right]\,,\\
 L_{12}&=\frac{C_RC_A^2g^6}{2}\int\hspace{-18pt}\int_{k,\,p,\,q\sim m_D}\limits\hspace{-18pt}\int\,\frac{T\,(2\k+\p)^2}{\left(\p^2\right)^2P(\k+\p)\,P(\k+\q)\,P(\k)^2}\notag\\
 &=\frac{C_RC_A^2\alpha_\mathrm{s}^3T}{m_D}\left[\frac{1}{8}+\mathcal{O}(\epsilon)\right]\,.
\end{align}

We also have to include the contribution from the square of the one-loop self-energy in order to get a gauge invariant result. This contribution can also be put into the form required by the algorithm:
\begin{align}
 -\frac{C_Rg^2}{2T}\int_{k\sim m_D}\frac{\bigl(\Pi^{(1)}_{m_D}(0,k)\bigr)^2}{\left(k^2+m_D^2\right)^3}&=-\frac{C_RC_A^2g^6}{2}\int\hspace{-18pt}\int_{k,\,p,\,q\sim m_D}\limits\hspace{-18pt}\int\,\frac{T\,(2\k+\p)^2\,(2\k+\q)^2}{\p^2\,\q^2\,P(\k+\p)\,P(\k+\q)\,P(\k)^3}\notag\\
 &=\frac{C_RC_A^2\alpha_\mathrm{s}^3T}{m_D}\left[-\frac{5}{16}-\frac{\pi^2}{12}+\mathcal{O}(\epsilon)\right]\,.
\end{align}
The sum of all these terms then gives the $\mathcal{O}\left(g^5\right)$ contribution from the scale $m_D$:
\begin{equation}
 D_1\Bigr|_{g^5,\,m_D}=-\frac{C_RC_A^2\alpha_\mathrm{s}^3T}{m_D}\left[\frac{89}{48}+\frac{\pi^2}{12}-\frac{11}{12}\ln2\right]\,.
\end{equation}

\subsection{Coulomb gauge}

In Coulomb gauge we have
\begin{align}
 L_1&=-\frac{C_RC_A^2g^6}{4}\int\hspace{-18pt}\int_{k,\,p,\,q\sim m_D}\limits\hspace{-18pt}\int\frac{-16T}{\p^2\,(\p-\q)^2\,\q^2\,P(\k+\p)\,P(\k+\q)\,P(\k)^2}\notag\\
 &\times\left[\left(\k\cdot\q-\frac{(\k\cdot\p)(\p\cdot\q)}{\p^2}\right)\left(\k\cdot\p-\frac{(\k\cdot\q)(\p\cdot\q)}{\q^2}\right)\left(1+\frac{\k\cdot(\p+\q)}{(\p-\q)^2}\right)\right.\notag\\
 &\hspace{12pt}+\left(\k^2-\frac{(\k\cdot\p)^2}{\p^2}\right)\left(\k\cdot\p-\frac{(\k\cdot\q)(\p\cdot\q)}{\q^2}\right)\left(1-\frac{\q^2}{(\p-\q)^2}\right)\notag\\
 &\hspace{12pt}+\left(\k\cdot\q-\frac{(\k\cdot\p)(\p\cdot\q)}{\p^2}\right)\left(\k^2-\frac{(\k\cdot\q)^2}{\q^2}\right)\left(1-\frac{\p^2}{(\p-\q)^2}\right)\notag\\
 &\hspace{12pt}-\frac{\p^2\q^2-(\p\cdot\q)^2}{(\p-\q)^2}\left.\left(\k^2-\frac{(\k\cdot\p)^2}{\p^2}-\frac{(\k\cdot\q)^2}{\q^2}+\frac{(\k\cdot\p)(\k\cdot\q)(\p\cdot\q)}{\p^2\,\q^2}\right)\right]\notag\\
 &=\frac{C_RC_A^2\alpha_\mathrm{s}^3T}{m_D}\left[\frac{1}{8}-\frac{\pi^2}{24}+\mathcal{O}(\epsilon)\right]\,,\\
 L_2&=-\frac{C_RC_A^2g^6}{4}\int\hspace{-18pt}\int_{k,\,p,\,q\sim m_D}\limits\hspace{-18pt}\int\frac{16T}{\p^2\,\q^2\,P(\k+\p)\,P(\k+\q)\,P(\k+\p+\q)\,P(\k)^2}\notag\\
 &\times\left[\k\cdot(\k+\q)-\frac{(\k\cdot\p)(\k\cdot\p+\p\cdot\q)}{\p^2}\right]\left[\k\cdot(\k+\p)-\frac{(\k\cdot\q)(\k\cdot\q+\p\cdot\q)}{\q^2}\right]\notag\\
 &=\frac{C_RC_A^2\alpha_\mathrm{s}^3T}{m_D}\left[\frac{1}{8}+\frac{3}{4}\ln2-\frac{\pi^2}{12}+\mathcal{O}(\epsilon)\right]\,,\\
 L_3&=-\frac{3C_RC_A^2g^6}{4}\int\hspace{-18pt}\int_{k,\,p,\,q\sim m_D}\limits\hspace{-18pt}\int\frac{T}{\p^2\,\q^2\,P(\k+\p+\q)\,P(\k)^2}\left[d-2+\frac{(\p\cdot\q)^2}{\p^2\,\q^2}\right]\notag\\
 &=\frac{C_RC_A^2\alpha_\mathrm{s}^3T}{m_D}\left[-\frac{9}{64\epsilon}-\frac{3}{64}+\frac{27}{64}\gamma_E-\frac{27}{64}\ln\frac{\pi\mu^2}{m_D^2}+\mathcal{O}(\epsilon)\right]\,,\\
 L_4&=\frac{3C_RC_A^2g^6}{4}\int\hspace{-18pt}\int_{k,\,p,\,q\sim m_D}\limits\hspace{-18pt}\int\frac{4T\left[\k^2\p^2\q^2-(\k\cdot\p)^2\q^2-(\k\cdot\q)^2\p^2+(\k\cdot\p)(\k\cdot\q)(\p\cdot\q)\right]}{\left(\p^2\right)^2\left(\q^2\right)^2P(\k+\p)\,P(\k+\q)\,P(\k)^2}\notag\\
 &=\frac{C_RC_A^2\alpha_\mathrm{s}^3T}{m_D}\left[\frac{15}{8}-\frac{\pi^2}{8}+\mathcal{O}(\epsilon)\right]\,,\\
 L_5&+L_6=\frac{3C_RC_A^2g^6}{2}\int\hspace{-18pt}\int_{k,\,p,\,q\sim m_D}\limits\hspace{-18pt}\int\frac{4T}{\p^2\,\q^2\,P(\k+\p)\,P(\k+\p+\q)\,P(\k)^2}\notag\\
 &\times\left[\k^2-\frac{(\k\cdot\p)^2}{\p^2}-\frac{(\k\cdot\q)(\k\cdot\q+\p\cdot\q)}{\q^2}+\frac{(\k\cdot\p)(\k\cdot\q+\p\cdot\q)(\p\cdot\q)}{\p^2\q^2}\right]\notag\\
 &=\frac{C_RC_A^2\alpha_\mathrm{s}^3T}{m_D}\left[-\frac{3}{2}+\frac{\pi^2}{4}+\mathcal{O}(\epsilon)\right]\,,\\
 L_7&=-\frac{C_RC_A^2g^6}{2}\int\hspace{-18pt}\int_{k,\,p,\,q\sim m_D}\limits\hspace{-18pt}\int\frac{16T\left[\k^2\p^2-(\k\cdot\p)^2\right]\left[(\k+\p)^2\q^2-(\k\cdot\q+\p\cdot\q)^2\right]}{\left(\p^2\right)^2\left(\q^2\right)^2P(\k+\p)^2\,P(\k+\p+\q)\,P(\k)^2}\notag\\
 &=\frac{C_RC_A^2\alpha_\mathrm{s}^3T}{m_D}\left[-\frac{9}{8}+\mathcal{O}(\epsilon)\right]\,,\\
 L_8&=\frac{C_RC_A^2g^6}{4}\int\hspace{-18pt}\int_{k,\,p,\,q\sim m_D}\limits\hspace{-18pt}\int\frac{4T\left[(\k+\q)^2\p^2-(\k\cdot\p+\p\cdot\q)^2\right]}{\left(\p^2\right)^3P(\k+\q)\,P(\k+\p+\q)\,P(\k)^2}\notag\\
 &=\frac{C_RC_A^2\alpha_\mathrm{s}^3T}{m_D}\left[-\frac{1}{32\epsilon}+\frac{3}{32}\gamma_E-\frac{3}{32}\ln\frac{\pi\mu^2}{m_D^2}+\mathcal{O}(\epsilon)\right]\,,\\
 L_9&=-\frac{C_RC_A^2g^6}{4}\int\hspace{-18pt}\int_{k,\,p,\,q\sim m_D}\limits\hspace{-18pt}\int\frac{16T}{\left(\p^2\right)^2(\p-\q)^2\,\q^2\,P(\k+\p)\,P(\k)^2}\notag\\
 &\hspace{13pt}\times\left[\left(\k^2-\frac{(\k\cdot\p)^2}{\p^2}\right)\left(\p^2\q^2-(\p\cdot\q)^2\right)\left(\frac{1}{(\p-\q)^2}+\frac{1}{\q^2}\right)\right.\notag\\
 &\hspace{13pt}\left.+\left(\k\cdot\q-\frac{(\k\cdot\p)(\p\cdot\q)}{\p^2}\right)^2\left(d-1-\frac{\p^2\q^2-(\p\cdot\q)^2}{(\p-\q)^2\,\q^2}\right)\right]\notag\\
 &=\frac{C_RC_A^2\alpha_\mathrm{s}^3T}{m_D}\left[\frac{5}{32\epsilon}-\frac{43}{64}-\frac{15}{32}\gamma_E+\frac{15}{32}\ln\frac{\pi\mu^2}{m_D^2}+\mathcal{O(\epsilon)}\right]\,,\\
 L_{10}&=-\frac{C_RC_A^2g^6}{4}\int\hspace{-18pt}\int_{k,\,p,\,q\sim m_D}\limits\hspace{-18pt}\int\frac{16T\left[\left(\k^2+\k\cdot\q\right)\p^2-(\k\cdot\p)(\k\cdot\p+\p\cdot\q)\right]^2}{\left(\p^2\right)^4P(\k+\p)\,P(\k+\q)\,P(\k+\p+\q)\,P(\k)^2}\notag\\
 &=\frac{C_RC_A^2\alpha_\mathrm{s}^3T}{m_D}\left[-\frac{5}{48}+\frac{1}{6}\ln2+\mathcal{O(\epsilon)}\right]\,,\\
 L_{11}&=\frac{C_RC_A^2g^6}{2}\int\hspace{-18pt}\int_{k,\,p,\,q\sim m_D}\limits\hspace{-18pt}\int\frac{4T\left[(\k\cdot\q)\p^2-(\k\cdot\p)(\p\cdot\q)\right]^2}{\left(\p^2\right)^4(\p-\q)^2\,\q^2\,P(\k+\p)\,P(\k)^2}\notag\\
 &=\frac{C_RC_A^2\alpha_\mathrm{s}^3T}{m_D}\left[\frac{1}{64\epsilon}-\frac{1}{32}-\frac{3}{64}\gamma_E+\frac{3}{64}\ln\frac{\pi\mu^2}{m_D^2}+\mathcal{O(\epsilon)}\right]\,,\\
 L_{12}&=\frac{C_RC_A^2g^6}{2}\int\hspace{-18pt}\int_{k,\,p,\,q\sim m_D}\limits\hspace{-18pt}\int\frac{4T\left[\k^2\p^2-(\k\cdot\p)^2\right]}{\left(\p^2\right)^3P(\k+\p)\,P(\k+\q)\,P(\k)^2}\notag\\
 &=\frac{C_RC_A^2\alpha_\mathrm{s}^3T}{m_D}\left[\frac{1}{8}+\mathcal{O(\epsilon)}\right]\,.
\end{align}

The square of the one-loop self-energy from the scale $m_D$ gives the contribution
\begin{align}
 -\frac{C_Rg^2}{2T}\int_{k\sim m_D}\frac{\bigl(\Pi^{(1)}_{m_D}(0,k)\bigr)^2}{\left(k^2+m_D^2\right)^3}=&\,-\frac{C_RC_A^2g^6}{2}\int\hspace{-18pt}\int_{k,\,p,\,q\sim m_D}\limits\hspace{-18pt}\int\frac{16T\left[\k^2\p^2-(\k\cdot\p)^2\right]\left[\k^2\q^2-(\k\cdot\q)^2\right]}{\left(\p^2\right)^2\left(\q^2\right)^2P(\k+\p)\,P(\k+\q)\,P(\k)^3}\notag\\
 =&\,\frac{C_A^2\alpha_\mathrm{s}^3T}{m_D}\left[-\frac{5}{8}-\frac{\pi^2}{12}+\mathcal{O}(\epsilon)\right]\,,
\end{align}
and after summing up all these terms, we again obtain the same result as in Feynman gauge:
\begin{equation}
 D_1\Bigr|_{g^5,\,m_D}=-\frac{C_RC_A^2\alpha_\mathrm{s}^3T}{m_D}\left[\frac{89}{48}+\frac{\pi^2}{12}-\frac{11}{12}\ln2\right]\,.
\end{equation}

There is a subtlety in Coulomb gauge regarding the nonzero modes. In Feynman gauge all Matsubara frequencies have to be zero, because otherwise the necessary expansions of the propagators only lead to scaleless or higher order contributions. But in Coulomb gauge the frequencies do not appear explicitly in the temporal gluon or ghost propagators, the only dependence on the frequencies is that the Debye mass appears in the temporal gluon propagator only for the zero mode. So, in principle, the propagators do not have to be expanded and there is nothing preventing also nonzero frequencies to appear in the Matsubara sums, as long as they do not appear in spatial gluon propagators.

In most diagrams there is only the zero mode because of the contour integration, but in diagrams $L_8$, $L_{10}$, $L_{11}$, and $L_{12}$ the momentum of the temporal gluon or ghost loop can have a nonzero frequency without it entering a spatial gluon propagator. This poses a problem, because those loops do not depend on the frequency, so the Matsubara sums contain an infinite sum over a constant, which is divergent and not regulated by dimensional regularization.

\begin{figure}[t]
 \includegraphics[width=\linewidth]{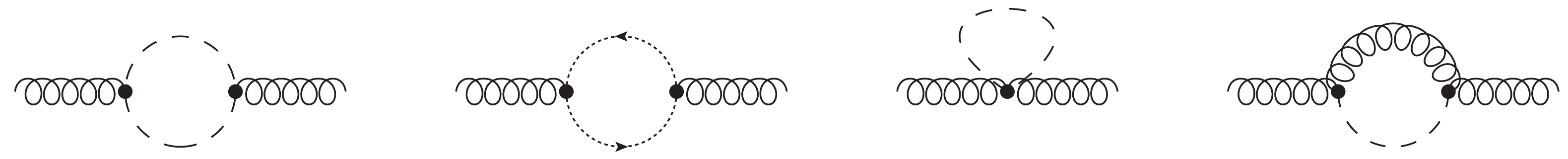}
 \caption{All diagrams relevant for the cancellation of the nonzero modes in the one-loop spatial gluon self-energy in Coulomb gauge.}
 \label{nonzero}
\end{figure}

However, these sums are canceled by a diagram that we could ignore so far, because it vanishes for the zero mode. This is the last diagram in Fig.~\ref{nonzero} and the Matsubara frequencies in the numerator from the vertices exactly cancel the denominator of the spatial gluon after it has been expanded. Then the sum over all diagrams of Fig.~\ref{nonzero} gives from left to right
\begin{equation}
 g^2C_A{\sum_Q}'\hspace{-18pt}\int\hspace{5pt}\left[\frac{1}{2}\frac{4q_iq_j}{(\p-\q)^2\q^2}-\frac{q_iq_j}{(\p-\q)^2\q^2}-\frac{\delta_{ij}}{(\p-\q)^2}+\frac{q_0^2}{(\p-\q)^2}\frac{1}{q_0^2}\left(\delta_{ij}-\frac{q_iq_j}{\q^2}\right)\right]=0\,.
\end{equation}
We have used the momentum $\p-\q$ instead of just $\q$ in the tadpole loop so that its cancellation becomes more apparent, and we do not have to consider the higher order expansion terms of the spatial gluon propagator in the last diagram, because they only contain scaleless integrals.

So even though each diagram contains a divergent series, the sum of all four of them is finite, because for each particular value of the frequency the sum cancels. In static gauge with $\xi=0$ this problem does not arise, because the temporal gluon and ghost propagators vanish for nonzero frequencies. Since the last diagram of Fig.~\ref{nonzero} gives no other contribution apart from canceling the nonzero-frequency contributions of the other diagrams in Coulomb gauge, the corresponding diagram has not been displayed in Fig.~\ref{diagrams}.

\subsection{Phase-space Coulomb gauge}

\begin{figure}[t]
 \includegraphics[width=\linewidth]{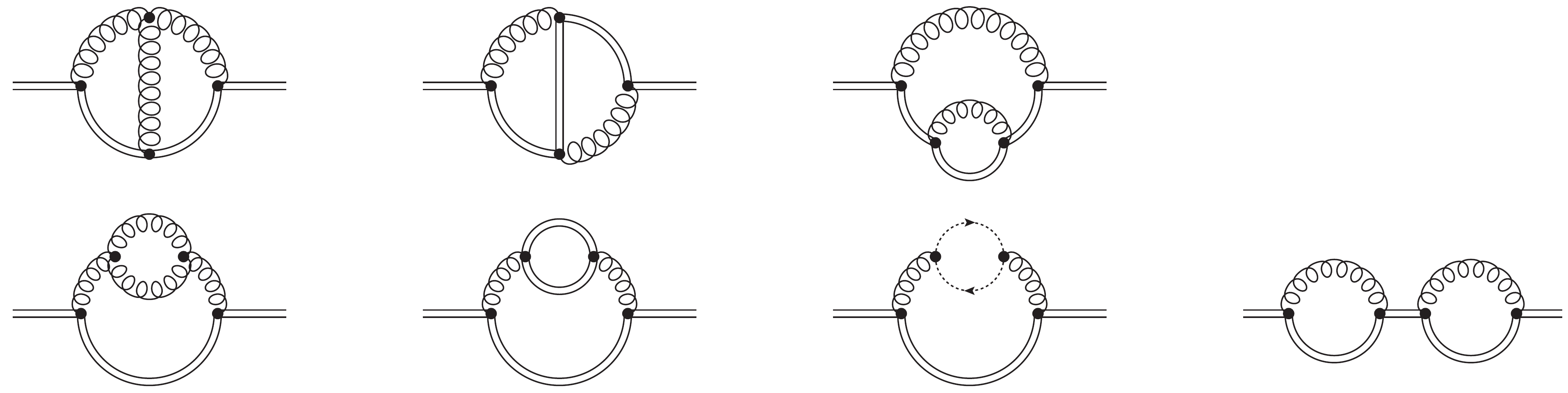}
 \caption{All two-loop diagram topologies in phase-space Coulomb gauge. The double-line propagators can represent either a temporal, an electric, or a mixed propagator. Also the diagram with two one-loop bubbles is displayed in the bottom-right corner. We will label the diagrams $\widetilde{L}_1,\dots,\widetilde{L}_7$ from top-left to bottom-right.}
 \label{PSdiagrams}
\end{figure}

In phase-space Coulomb gauge there are less diagram topologies, because temporal gluons only couple in a three-gluon vertex. These topologies are shown in Fig.~\ref{PSdiagrams}. But because the massive propagators now can be temporal, electric, or mixed, there are more diagrams in total. However, it is possible for each diagram topology to factorize the massive propagators from the spatial gluon propagators, so that we can sum over all possibilities for the massive propagators before multiplying them with the spatial gluons. This sum over all massive propagators is represented by the double-line propagators in Fig.~\ref{PSdiagrams}.

We have included the one-particle reducible diagram $\widetilde{L}_7$ in Fig.~\ref{PSdiagrams}, which corresponds to the second order expansion of the resummed propagator. In this case the re-expanded temporal propagator depends on several different self-energy functions, so it is easier to just calculate this diagram explicitly.

We will denote the sum over massive propagators by $D_{\alpha\beta}^{m_1m_2\dots}\left(\k_0,\k_1,\k_2,\dots\right)$. The indices $m_i$ correspond to the vector indices at each vertex $i$, which can then be contracted with the spatial gluon propagator. The initial momentum of the series of propagators is $\k_0$ and the $\k_i$ are the incoming momenta at each vertex $i$. The final and initial indices of the propagator series are $\alpha$ and $\beta$, respectively. We will need temporal indices for most diagrams, but also mixed indices for the double line loop in diagram $\widetilde{L}_5$.

\begin{figure}[t]
 \includegraphics[width=\linewidth]{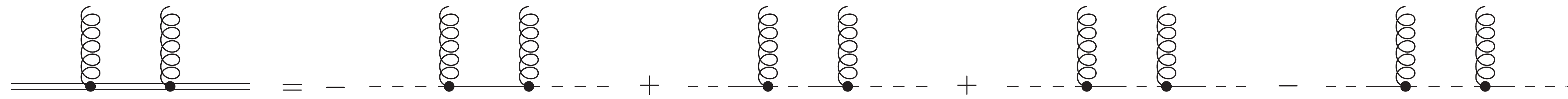}
 \caption{Explicit expression for a double line propagator with two vertices in terms of temporal, electric, and mixed propagators.}
 \label{doubleline}
\end{figure}

We will show the summation over massive propagators explicitly in one case for illustration and just give the result for the other relevant cases. Fig.~\ref{doubleline} shows the double line propagator with two vertices in terms of temporal, electric, and mixed propagators. By the phase-space Coulomb gauge Feynman rules this gives
\begin{align}
 D_{00}^{m_1m_2}(\k_0,\k_1,\k_2)=&-D_{00}(\k_0+\k_1+\k_2)D_{m_2m_1}(\k_0+\k_1)D_{00}(\k_0)\notag\\
 &+D_{00}(\k_0+\k_1+\k_2)D_{m_20}(\k_0+\k_1)D_{m_10}(\k_0)\notag\\
 &+D_{0m_2}(\k_0+\k_1+\k_2)D_{0m_1}(\k_0+\k_1)D_{00}(\k_0)\notag\\
 &-D_{0m_2}(\k_0+\k_1+\k_2)D_{00}(\k_0+\k_1)D_{m_10}(\k_0)\notag\\
 =&-\frac{1}{P(\k_0+\k_1+\k_2)}\left(\delta_{m_2m_1}-\frac{(k_0+k_1)_{m_2}(k_0+k_1)_{m_1}}{P(\k_0+\k_1)}\right)\frac{1}{P(\k_0)}\notag\\
 &+\frac{1}{P(\k_0+\k_1+\k_2)}\frac{-(k_0+k_1)_{m_2}}{P(\k_0+\k_1)}\frac{-(k_0)_{m_1}}{P(\k_0)}\notag\\
 &+\frac{(k_0+k_1+k_2)_{m_2}}{P(\k_0+\k_1+\k_2)}\frac{(k_0+k_1)_{m_1}}{P(\k_0+\k_1)}\frac{1}{P(\k_0)}\notag\\
 &-\frac{(k_0+k_1+k_2)_{m_2}}{P(\k_0+\k_1+\k_2)}\frac{1}{P(\k_0+\k_1)}\frac{-(k_0)_{m_1}}{P(\k_0)}\notag\\
 =&-\frac{1}{P(\k_0+\k_1+\k_2)}\left(\delta_{m_1m_2}-\frac{4(k_0+k_1)_{m_2}(k_0)_{m_1}}{P(\k_0+\k_1)}\right)\frac{1}{P(\k_0)}\,.
\end{align}
Here we have used the fact that all vector indices are contracted with gluon propagators that are proportional to the transverse propagator, which means that all terms $(k_i)_{m_i}$ cancel in the numerator and can be neglected. The different signs in front of the propagators come from the two color structure functions in the vertices, which are even or odd depending on whether the temporal, electric, and spatial fields are attached with the same ordering or the opposite one compared to the three-field vertex shown in Fig.~\ref{PSFeynRules}.

In the same way one can calculate double line propagators with more vertices or different initial and final indices:
\begin{align}
 &D_{00}^{m_1m_2m_3}(\k_0,\k_1,\k_2,\k_3)\notag\\
 &=\frac{2(k_0+k_1+k_2)_{m_3}\delta_{m_2m_1}}{P(\k_0+\k_1+\k_2+\k_3)P(\k_0+\k_1+\k_2)P(\k_0)}+\frac{2\delta_{m_3m_2}(k_0)_{m_1}}{P(\k_0+\k_1+\k_2+\k_3)P(\k_0+\k_1)P(\k_0)}\notag\\
 &-\frac{8(k_0+k_1+k_2)_{m_3}(k_0+k_1)_{m_2}(k_0)_{m_1}}{P(\k_0+\k_1+\k_2+\k_3)P(\k_0+\k_1+\k_2)P(\k_0+\k_1)P(\k_0)}\,,\\
 &D_{00}^{m_1m_2m_3m_4}(\k_0,\k_1,\k_2,\k_3,\k_4)\notag\\
 &=\frac{1}{P(\k_0+\k_1+\k_2+\k_3+\k_4)}\left(\delta_{m_4m_3}-\frac{4(k_0+k_1+k_2+k_3)_{m_4}(k_0+k_1+k_2)_{m_3}}{P(\k_0+\k_1+\k_2+\k_3)}\right)\notag\\
 &\hspace{13pt}\times\frac{1}{P(\k_0+\k_1+\k_2)}\left(\delta_{m_2m_1}-\frac{4(k_0+k_1)_{m_2}(k_0)_{m_1}}{P(\k_0+\k_1)}\right)\frac{1}{P(\k_0)}\notag\\
 &\hspace{13pt}-\frac{4(k_0+k_1+k_2+k_3)_{m_4}\delta_{m_3m_2}(k_0)_{m_1}}{P(\k_0+\k_1+\k_2+\k_3+\k_4)P(\k_0+\k_1+\k_2+\k_3)P(\k_0+\k_1)P(\k_0)}\,,\\
 &D_{0n}^{m_1}(\k_0,\k_1)=-\frac{1}{P(\k_0+\k_1)}\left(\delta_{m_1n}-\frac{2(k_0)_{m_1}(k_0)_n}{P(\k_0)}\right)\,.
\end{align}

With these we can write the phase-space Coulomb gauge diagrams in a rather compact form:
\begin{align}
 \widetilde{L}_1&=-\frac{C_RC_A^2g^6T}{4}\int\hspace{-18pt}\int_{k,\,p,\,q\sim m_D}\limits\hspace{-18pt}\int\,D_{00}^{ijl}(\k,\p,\q-\p,-\q)D_{ii'}(\p)D_{jj'}(\q-\p)D_{ll'}(-\q)\notag\\
 &\hspace{120pt}\times\left[\delta_{i'j'}(2p-q)_{l'}+\delta_{j'l'}(2q-p)_{i'}-\delta_{l'i'}(p+q)_{j'}\right]\notag\\
 &=\frac{C_RC_A^2\alpha_\mathrm{s}^3T}{m_D}\left[\frac{1}{8}-\frac{\pi^2}{24}+\mathcal{O}(\epsilon)\right]\,,\\
 \widetilde{L}_2&=-\frac{C_RC_A^2g^6T}{4}\int\hspace{-18pt}\int_{k,\,p,\,q\sim m_D}\limits\hspace{-18pt}\int\,D_{00}^{iji'j'}(\k,\p,\q,-\p,-\q)D_{ii'}(\p)D_{jj'}(\q)\notag\\
 &=\frac{C_RC_A^2\alpha_\mathrm{s}^3T}{m_D}\left[-\frac{3}{64\epsilon}+\frac{15}{64}+\frac{9}{64}\gamma_E-\frac{\pi^2}{24}+\frac{3}{4}\ln2-\frac{9}{64}\ln\frac{\pi\mu^2}{m_D^2}+\mathcal{O}(\epsilon)\right]\,,\\
 \widetilde{L}_3&=-\frac{C_RC_A^2g^6T}{2}\int\hspace{-18pt}\int_{k,\,p,\,q\sim m_D}\limits\hspace{-18pt}\int\,D_{00}^{ijj'i'}(\k,\p,\q,-\q,-\p)D_{ii'}(\p)D_{jj'}(\q)\notag\\
 &=\frac{C_RC_A^2\alpha_\mathrm{s}^3T}{m_D}\left[-\frac{3}{32\epsilon}-\frac{69}{32}+\frac{9}{32}\gamma_E+\frac{\pi^2}{6}-\frac{9}{32}\ln\frac{\pi\mu^2}{m_D^2}+\mathcal{O}(\epsilon)\right]\,,\\
 \widetilde{L}_4&=-\frac{C_RC_A^2g^6T}{2}\int\hspace{-18pt}\int_{k,\,p,\,q\sim m_D}\limits\hspace{-18pt}\int\,D_{00}^{ij}(\k,\p,-\p)D_{ii'}(\p)D_{jj'}(-\p)\,\frac{2}{(\p-\q)^2\q^2}\notag\\
 &\hspace{13pt}\times\left[\left(\p^2\q^2-(\p\cdot\q)^2\right)\left(\frac{1}{\q^2}+\frac{1}{(\p-\q)^2}\right)\delta_{i'j'}+\left(d-1-\frac{\p^2\q^2-(\p\cdot\q)^2}{(\p-\q)^2\q^2}\right)q_{i'}q_{j'}\right]\notag\\
 &=\frac{C_RC_A^2\alpha_\mathrm{s}^3T}{m_D}\left[\frac{5}{32\epsilon}-\frac{43}{64}-\frac{15}{32}\gamma_E+\frac{15}{32}\ln\frac{\pi\mu^2}{m_D^2}+\mathcal{O}(\epsilon)\right]\,,\\
 \widetilde{L}_5&=-\frac{C_RC_A^2g^6T}{2}\int\hspace{-18pt}\int_{k,\,p,\,q\sim m_D}\limits\hspace{-18pt}\int\,D_{00}^{ij}(\k,\p,-\p)D_{ii'}(\p)D_{jj'}(-\p)D_{0n}^{j'}(\k+\q,\p)\delta_{i'n}\notag\\
 &=\frac{C_RC_A^2\alpha_\mathrm{s}^3T}{m_D}\left[-\frac{1}{32\epsilon}+\frac{1}{48}+\frac{3}{32}\gamma_E+\frac{1}{6}\ln2-\frac{3}{32}\ln\frac{\pi\mu^2}{m_D^2}+\mathcal{O}(\epsilon)\right]\,,\\
 \widetilde{L}_6&=-\frac{C_RC_A^2g^6T}{2}\int\hspace{-18pt}\int_{k,\,p,\,q\sim m_D}\limits\hspace{-18pt}\int\,D_{00}^{ij}(\k,\p,-\p)D_{ii'}(\p)D_{jj'}(-\p)\frac{q_{i'}q_{j'}}{(\p-\q)^2\q^2}\notag\\
 &=\frac{C_RC_A^2\alpha_\mathrm{s}^3T}{m_D}\left[\frac{1}{64\epsilon}-\frac{1}{32}-\frac{3}{64}\gamma_E+\frac{3}{64}\ln\frac{\pi\mu^2}{m_D^2}+\mathcal{O}(\epsilon)\right]\,,\\
 \widetilde{L}_7&=-\frac{C_RC_A^2g^6T}{2}\int\hspace{-18pt}\int_{k,\,p,\,q\sim m_D}\limits\hspace{-18pt}\int\,D_{00}^{ii'jj'}(\k,\p,-\p,\q,-\q)D_{ii'}(\p)D_{jj'}(\q)\notag\\
 &=\frac{C_RC_A^2\alpha_\mathrm{s}^3T}{m_D}\left[\frac{5}{8}-\frac{\pi^2}{6}+\mathcal{O}(\epsilon)\right]\,.
\end{align}
We have changed the momenta in the double line loop of diagram $\widetilde{L}_5$ from $\p+\q$ and $\q$ to $\k+\p+\q$ and $\k+\q$ by a shift of the integration momentum $\q$, such that the integrals are all of the form required by the algorithm of appendix~\ref{M3LI}.

We see that the sum of these integrals gives the same result as in standard Coulomb gauge and in Feynman gauge. But we can make the correspondence between phase-space and standard Coulomb gauge even clearer. The propagator of the electric field [cf.\ Eq.~\eqref{PSCGprop}] contains a part that is just a Kronecker delta, which gives exactly the same contribution as if the electric propagator were contracted to a point and replaced by a four-gluon vertex in standard Coulomb gauge. The second part of the electric propagator contains components of the momentum in the numerator and a massive denominator. This term has the same form as a corresponding three-gluon vertex in Coulomb gauge, where the momentum components in the numerator come from the vertex and not the propagator. The same applies for mixed temporal-electric propagators.

The correspondence is not one-to-one, for example diagrams $\widetilde{L}_2$ and $\widetilde{L}_3$ both give diagram $L_3$ in standard Coulomb gauge when we replace the second and fourth double-line propagator by a Kronecker delta. But if we look at the color coefficients $C_RC_A^2/2$ of $\widetilde{L}_2$ and $C_RC_A^2$ of $\widetilde{L}_3$, we see that they add up exactly to the color coefficient $3C_RC_A^2/2$ of $L_3$. So ultimately it is only a matter of combinatorics to see that phase-space and standard Coulomb gauge generate exactly the same integrals.

A simpler check of this statement is to compare certain classes of diagrams between phase-space and standard Coulomb gauge, which have unique topologies in both gauge formulations. In our case, $L_1$ and $\widetilde{L}_1$ are the only diagrams with a vertex of three spatial gluons, and diagrams $L_8,\dots,L_{12}$, and $\widetilde{L}_4,\dots,\widetilde{L}_6$ are the only ones with a one-loop self-energy in a spatial propagator. So accordingly, we find the equalities
\begin{align}
 \widetilde{L}_1&=L_1\,,\\
 \widetilde{L}_2+\widetilde{L}_3+\widetilde{L}_7&=L_2+L_3+L_4+L_5+L_6+L_7+L_{13}\,,\\
 \widetilde{L}_4+\widetilde{L}_5+\widetilde{L}_6&=L_8+L_9+L_{10}+L_{11}+L_{12}\,,
\end{align}
where we used $L_{13}$ to denote the contribution from the square of the one-loop self-energy at the scale $m_D$.

The cancellation of the nonzero frequency contributions is a bit simpler in this formulation than in standard Coulomb gauge. The double-line loop in diagram $\widetilde{L}_5$ of Fig.~\ref{PSdiagrams} gives rise to two contributions, one where the loop contains a temporal and an electric propagator and one where both propagators are mixed temporal-electric. The first contribution is unproblematic, since the electric propagator for nonzero frequencies contains the denominator $q_0^2+q^2$, for which the Matsubara sum is finite. The second contribution exactly cancels the ghost loop diagram.

\section{\texorpdfstring{Magnetic scale cancellation at $\bm{\mathcal{O}\left(g^6\right)}$}{Magnetic scale cancellation at O(g\^{}6)}}
\label{scalemM}

We will list here all contributions at $\mathcal{O}\left(g^6\right)$ that involve the scale $m_M$. At $\mathcal{O}\left(g^5\right)$ those were one-loop diagrams where the spatial gluon carries a momentum of order $m_M$ and the temporal gluon carries a momentum of order $m_D$ (see section~\ref{mMcancel}). At $\mathcal{O}\left(g^6\right)$ it is the same principle: two-loop diagrams with all propagators carrying momenta of order $m_D$ except for one spatial gluon with a momentum of order $m_M$. In three-gluon vertices only the momenta of order $m_D$ are to be kept in the numerator.

\begin{figure}
 \includegraphics[width=0.6\linewidth]{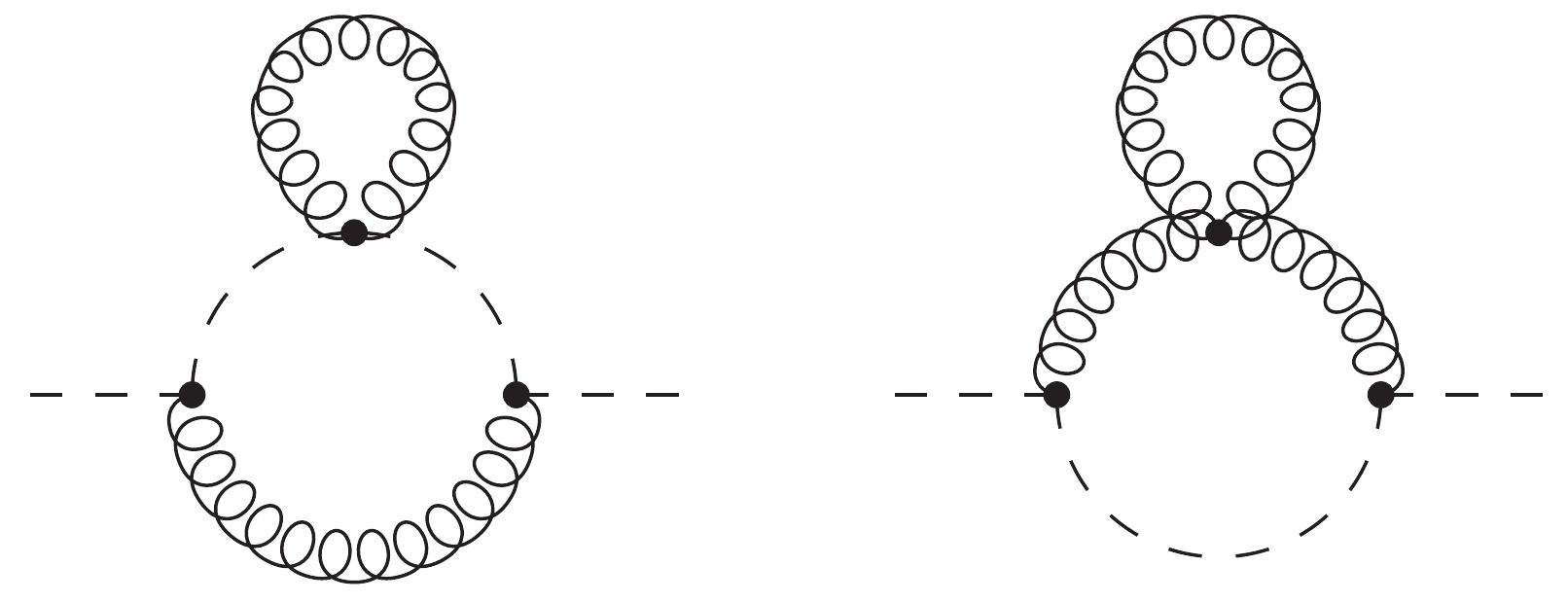}
 \caption{Additional diagrams carrying the scale $m_M$ at $\mathcal{O}\left(g^6\right)$.}
 \label{Poly9}
\end{figure}

We refer again to Fig.~\ref{diagrams}, which essentially gives all relevant diagrams for this calculation. In diagrams $L_1,\dots,L_7$ any of the spatial gluons can be the one that carries the scale $m_M$, in diagram $L_9$ it is only the gluons in the sub-loop. So $L_1$ contains three and $L_2,\dots,L_7,L_9$ each contain two different contributions. In addition, there are two new diagrams not displayed in Fig.~\ref{diagrams}, which we give in Fig.~\ref{Poly9}. They correspond to the diagrams $L_7$ and $L_9$ from Fig.~\ref{diagrams} with the sub-loop replaced by a tadpole, so we will include the contributions of the left and right diagram in Fig.~\ref{Poly9} in the following expressions for $L_7$ and $L_9$, respectively. Diagrams $L_8$, $L_{10}$, and $L_{12}$ do not contribute, because if the spatial gluons were of the scale $m_M$, then these would correspond to the one-loop diagram with a resummed spatial propagator, which we have already considered in the $\mathcal{O}\left(g^5\right)$ calculation. Diagram $L_{11}$ with one ghost propagator of the scale $m_M$ does not contribute, because from the gluon-ghost vertices there is a factor of the loop momentum squared in the numerator, so this diagram is of $\mathcal{O}\left(g^8\right)$.

We will do this calculation in Feynman gauge, because the expressions are somewhat shorter. We will label the momenta such that $k,p\sim m_D$ and $q\sim m_M$. Since we have to expand everything in $q/m_D$, we can just ignore $q$ in all other propagators at leading order. This simplifies the $q$ integration, which now contains only one propagator:
\begin{equation}
 \int_{q\sim m_M}D_{ij}(0,\q)=\frac{\delta_{ij}}{d}\int_{q\sim m_M}D_{kk}(0,\q)\,.
\end{equation}
The Kronecker delta can then be used to contract all indices in the $k$ and $p$ integrations, which can be carried out by the same methods as in the $\mathcal{O}\left(g^5\right)$ calculation.

The calculation of the different diagrams gives
\begin{align}
 L_1&=-C_RC_A^2g^6T\int\hspace{-18pt}\int_{k,\,p\sim m_D,\,q\sim q_M}\limits\hspace{-18pt}\int\left(\frac{4\left(\k^2\p^2-(\k\cdot\p)^2\right)}{\left(\p^2\right)^2P(\k+\p)\,P(\k)^3}+\frac{2\left(\k^2\p^2-(\k\cdot\p)^2\right)}{\left(\p^2\right)^2P(\k+\p)^2\,P(\k)^2}\right)\frac{D_{ii}(0,\q)}{d}\notag\\
 &=-\frac{2\pi}{3}\frac{C_RC_A^2\alpha_\mathrm{s}^3T}{m_D^2}\int_{q\sim m_M}\limits\,D_{ii}(0,\q)\,,\\
 L_2&=-C_RC_A^2g^6T\int\hspace{-18pt}\int_{k,\,p\sim m_D,\,q\sim q_M}\limits\hspace{-18pt}\int\frac{2(2\k+\p)^2\left(\k^2+\k\cdot\p\right)}{\p^2\,P(\k+\p)^2\,P(\k)^3}\frac{D_{ii}(0,\q)}{d}\notag\\
 &=-\frac{2\pi}{3}\frac{C_RC_A^2\alpha_\mathrm{s}^3T}{m_D^2}\int_{q\sim m_M}\limits\,D_{ii}(0,\q)\,,\\
 L_3&=-\frac{3}{2}C_RC_A^2g^6T\int\hspace{-18pt}\int_{k,\,p\sim m_D,\,q\sim q_M}\limits\hspace{-18pt}\int\frac{d}{\p^2\,P(\k+\p)\,P(\k)^2}\frac{D_{ii}(0,\q)}{d}\notag\\
 &=-\frac{3\pi}{2}\frac{C_RC_A^2\alpha_\mathrm{s}^3T}{m_D^2}\int_{q\sim m_M}\limits\,D_{ii}(0,\q)\,,\\
 L_4&=C_RC_A^2g^6T\int\hspace{-18pt}\int_{k,\,p\sim m_D,\,q\sim q_M}\limits\hspace{-18pt}\int\frac{3\left(2\k^2+\k\cdot\p\right)}{\p^2\,P(\k+\p)\,P(\k)^3}\frac{D_{ii}(0,\q)}{d}\notag\\
 &=\pi\frac{C_RC_A^2\alpha_\mathrm{s}^3T}{m_D^2}\int_{q\sim m_M}\limits\,D_{ii}(0,\q)\,,\\
 L_5&+L_6=C_RC_A^2g^6T\int\hspace{-18pt}\int_{k,\,p\sim m_D,\,q\sim q_M}\limits\hspace{-18pt}\int\left(\frac{3(2\k+\p)\cdot(\k+\p)}{\p^2\,P(\k+\p)^2\,P(\k)^2}+\frac{3\left(2\k^2+\k\cdot\p\right)}{\p^2\,P(\k+\p)\,P(\k)^3}\right)\frac{D_{ii}(0,\q)}{d}\notag\\
 &=2\pi\frac{C_RC_A^2\alpha_\mathrm{s}^3T}{m_D^2}\int_{q\sim m_M}\limits\,D_{ii}(0,\q)\,,\\
 L_7&=-\frac{1}{2}C_RC_A^2g^6T\int\hspace{-18pt}\int_{k,\,p\sim m_D,\,q\sim q_M}\limits\hspace{-18pt}\int\left(\frac{4(2\k+\p)^2(\k+\p)^2}{\p^2\,P(\k+\p)^3\,P(\k)^2}+\frac{4\k^2(2\k+\p)^2}{\p^2\,P(\k+\p)\,P(\k)^4}\right.\notag\\
 &\hspace{150pt}-\left.\frac{d(2\k+\p)^2}{\p^2\,P(\k+\p)^2\,P(\k)^2}\right)\frac{D_{ii}(0,\q)}{d}\notag\\
 &=-\frac{5\pi}{9}\frac{C_RC_A^2\alpha_\mathrm{s}^3T}{m_D^2}\int_{q\sim m_M}\limits\,D_{ii}(0,\q)\,,\\
 L_9&=-\frac{1}{2}C_RC_A^2g^6T\int\hspace{-22pt}\int_{k,\,p\sim m_D,\,q\sim q_M}\limits\hspace{-22pt}\int\biggl(\frac{20\k^2\p^2-4(6-d)(\k\cdot\p)^2-4(1-d)(\k\cdot\p)^2\p^2-(1-d)\left(\p^2\right)^2}{\left(\p^2\right)^3P(\k+\p)\,P(\k)^2}\notag\\
 &\hspace{150pt}-\frac{(d-1)(2\k+\p)^2}{\left(\p^2\right)^2P(\k+\p)\,P(\k)^2}\biggr)\frac{D_{ii}(0,\q)}{d}\notag\\
 &=\frac{\pi}{2}\frac{C_RC_A^2\alpha_\mathrm{s}^3T}{m_D^2}\int_{q\sim m_M}\limits\,D_{ii}(0,\q)\,.
\end{align}
From the square of the one-loop self-energy we have
\begin{align}
 &-\frac{C_Rg^2}{2T}\int_{k\sim m_D}\limits\frac{\Pi^{(1)}_{m_D}(0,k)\Pi^{(1)}_{m_M}(0,k)}{\left(k^2+m_D^2\right)^3}\notag\\
 &=-C_RC_A^2g^6T\int\hspace{-18pt}\int_{k,\,p\sim m_D,\,q\sim q_M}\limits\hspace{-18pt}\int\left(\frac{4\k^2(2\k+\p)^2}{\p^2\,P(\k+\p)\,P(\k)^4}-\frac{d\,(2\k+\p)^2}{\p^2\,P(\k+\p)\,P(\k)^3}\right)\frac{D_{ii}(0,\q)}{d}\notag\\
 &=-\frac{\pi}{9}\frac{C_RC_A^2\alpha_\mathrm{s}^3T}{m_D^2}\int_{q\sim m_M}\limits\,D_{ii}(0,\q)\,.
\end{align}
The sum of all these terms gives zero.

\section{Automatic reduction to master integrals}
\label{M3LI}

The method of how to solve the three-loop integrals appearing in this calculation has been described in~\cite{Broadhurst:1991fi}. Minimal modifications are required in order to account for the Euclidean metric. All integrals can be put in the two following forms:
\begin{equation}
 B_M(i_1,i_2,i_3,i_4,i_5,i_6)=\int_k\int_p\int_q\frac{1}{(\p^2)^{i_1}((\p-\q)^2)^{i_2}(\q^2)^{i_3}P(\k+\p)^{i_4}P(\k+\q)^{i_5}P(\k)^{i_6}}\,,
\end{equation}
\begin{equation}
 B_N(i_1,i_2,i_3,i_4,i_5,i_6)=\int_k\int_p\int_q\frac{1}{(\p^2)^{i_1}(\q^2)^{i_2}P(\k+\p)^{i_3}P(\k+\q)^{i_4}P(\k+\p+\q)^{i_5}P(\k)^{i_6}}\,,
\end{equation}
with $P(\k)=k^2+m_D^2$. In this framework, the exponents $i_1,\dots,i_6$ are integers.

By relabeling or shifting the integration variables $\k$, $\p$, and $\q$ several identities between the different $B_M$ and $B_N$ can be established:
\begin{align}
 &B_M(i_1,i_2,i_3,i_4,i_5,i_6)=B_M(i_2,i_1,i_3,i_4,i_6,i_5)\notag\\
 &=B_M(i_3,i_1,i_2,i_6,i_4,i_5)=B_M(i_3,i_2,i_1,i_5,i_4,i_6)\notag\\
 &=B_M(i_2,i_3,i_1,i_5,i_6,i_4)=B_M(i_1,i_3,i_2,i_6,i_5,i_4)\,,\\
 &B_N(i_1,i_2,i_3,i_4,i_5,i_6)=B_N(i_1,i_2,i_5,i_6,i_3,i_4)\notag\\
 &=B_N(i_1,i_2,i_4,i_3,i_6,i_5)=B_N(i_1,i_2,i_6,i_5,i_4,i_3)\notag\\
 &=B_N(i_2,i_1,i_4,i_3,i_5,i_6)=B_N(i_2,i_1,i_5,i_6,i_4,i_3)\notag\\
 &=B_N(i_2,i_1,i_3,i_4,i_6,i_5)=B_N(i_2,i_1,i_6,i_5,i_3,i_4)\,.
\end{align}
In addition, any $B_N$ with an index $i_3,\dots,i_6$ zero or negative can be turned into a $B_M$. The obvious relation is
\begin{equation}
 B_N(i_1,i_2,i_3,i_4,0,i_6)=B_M(i_1,0,i_2,i_3,i_4,i_6)\,.
\end{equation}
If $i_5$ is negative, one can expand the numerator after substituting
\begin{equation}
 (\k+\p+\q)^2+m_D^2=\p^2-(\p-\q)^2+\q^2+\left((\k+\p)^2+m_D^2\right)+\left((\k+\q)^2+m_D^2\right)-\left(\k^2+m_D^2\right)\,.
\end{equation}
All these terms appear to some power in the denominator, so they can be canceled to give proper $B_M$ integrals. If any of the other indices $i_3,\dots,i_6$ is zero or negative, then one can use the identities above to shift that to the fifth position and then use the relation for $i_5\leq0$.

Other identities can be found by acting with $\bm{\nabla}_i\cdot\k_j$ on the integrand, where $\k_i$ and $\k_j$ can be any of the three loop momenta. The total expression has to be zero, since it is an integral over a total derivative, but calculating the derivative explicitly gives a number of other $B_M$ and $B_N$ integrals. These new identities include integrals with changed indices $i_1,\dots,i_6$, while the identities above just shift them:
\begin{align}
 &\left(d-2i_1-i_2-i_4\right)B_M\left(i_1,i_2,i_3,i_4,i_5,i_6\right)\notag\\
 &=i_2B_M\left(i_1-1,i_2+1,i_3,i_4,i_5,i_6\right)-i_2B_M\left(i_1,i_2+1,i_3-1,i_4,i_5,i_6\right)\notag\\
 &\hspace{13pt}+i_4B_M\left(i_1-1,i_2,i_3,i_4+1,i_5,i_6\right)-i_4B_M\left(i_1,i_2,i_3,i_4+1,i_5,i_6-1\right)\,,\\
 &\left(d-i_1-i_3-2i_6\right)B_M\left(i_1,i_2,i_3,i_4,i_5,i_6\right)=-2i_6m_D^2B_M\left(i_1,i_2,i_3,i_4,i_5,i_6+1\right)\notag\\*
 &\hspace{13pt}+i_1B_M\left(i_1+1,i_2,i_3,i_4,i_5,i_6-1\right)-i_1B_M\left(i_1+1,i_2,i_3,i_4-1,i_5,i_6\right)\notag\\*
 &\hspace{13pt}+i_3B_M\left(i_1,i_2,i_3+1,i_4,i_5,i_6-1\right)-i_3B_M\left(i_1,i_2,i_3+1,i_4,i_5-1,i_6\right)\,,\\
 &\left(d-i_1-2i_4-i_6\right)B_N\left(i_1,i_2,i_3,i_4,i_5,i_6\right)\notag\\
 &=i_1B_N\left(i_1+1,i_2,i_3,i_4-1,i_5,i_6\right)-i_1B_N\left(i_1+1,i_2,i_3,i_4,i_5-1,i_6\right)\notag\\
 &\hspace{13pt}+i_6B_N\left(i_1,i_2,i_3,i_4-1,i_5,i_6+1\right)-i_6B_N\left(i_1,i_2-1,i_3,i_4,i_5,i_6+1\right)\notag\\
 &\hspace{13pt}-2i_4m_D^2B_N\left(i_1,i_2,i_3,i_4+1,i_5,i_6\right)-2i_6m_D^2B_N\left(i_1,i_2,i_3,i_4,i_5,i_6+1\right)\,,\\
 &\left(i_1+i_2+i_3+i_4+i_5+i_6-\frac{3d}{2}\right)B_N\left(i_1,i_2,i_3,i_4,i_5,i_6\right)\notag\\
 &=i_3m_D^2B_N\left(i_1,i_2,i_3+1,i_4,i_5,i_6\right)+i_4m_D^2B_N\left(i_1,i_2,i_3,i_4+1,i_5,i_6\right)\notag\\
 &\hspace{13pt}+i_5m_D^2B_N\left(i_1,i_2,i_3,i_4,i_5+1,i_6\right)+i_6m_D^2B_N\left(i_1,i_2,i_3,i_4,i_5,i_6+1\right)\,.
\end{align}
There are 16 further identities, which can be obtained from these four by combining them with the index shifts given above.

By repeated use of these identities every $B_N$ integral can be reduced to $B_N(0,0,1,1,1,1)$ plus a bunch of $B_M$ integrals. In the same way every $B_M$ integral can be reduced to $B_M(0,0,0,1,1,1)$ plus $B_M$ integrals where at least one of the indices $i_4,\dots,i_6$ is zero or negative, for which there exists a general solution. So all integrals appearing in our calculation can be put into the form of a few master integrals. The needed results for those can be found in~\cite{Broadhurst:1991fi,Gray:1990yh,Braaten:1995jr}.\footnote{As a check that our programs are running correctly we have calculated all the integrals given in the appendix of~\cite{Braaten:1995jr} and reproduced their results.}

\section{Calculation of the master integrals}

For the sake of completeness, we attach here how the master integrals, whose results are given in~\cite{Broadhurst:1991fi,Gray:1990yh,Braaten:1995jr}, can be calculated. The simplest one is $B_M(0,0,0,1,1,1)$, because in this case all three loop integrations decouple by shifting $\p\to\p-\k$ and $\q\to\q-\k$:
\begin{equation}
 B_M(0,0,0,1,1,1)=\left(\int_k\frac{1}{k^2+m_D^2}\right)^3=\frac{\Gamma\left(1-\frac{d}{2}\right)^3}{(4\pi)^{\frac{3d}{2}}}\,m_D^{3d-6}\,.
\end{equation}

For the other $B_M$ integrals instead of a closed expression we will rather give another algorithm for their solution. We will assume that the zero or negative index is $i_4$, because if it is $i_5$ or $i_6$ instead then one can exchange those with $i_4$ by one of the identities. After also performing the shift $\p\to\p-\k$ and $\q\to\q-\k$ we can integrate over $\p$ without problems, because it no longer appears in a massive denominator:
\begin{align}
 &\int_p\frac{\left(p^2+m_D^2\right)^{-i_4}}{\left((\p-\k)^2\right)^{i_1}\left((\p-\q)^2\right)^{i_2}}\notag\\
 &=\frac{\Gamma(i_1+i_2)}{\Gamma(i_1)\Gamma(i_2)}\int_0^1dx\int_p\frac{(p^2+m_D^2)^{-i_4}x^{i_1-1}(1-x)^{i_2-1}}{(p^2-2x\p\cdot\k-2(1-x)\p\cdot\q+xk^2+(1-x)q^2)^{i_1+i_2}}\notag\\
 &=\frac{\Gamma(i_1+i_2)}{\Gamma(i_1)\Gamma(i_2)}\int_0^1dx\int_p\frac{\left((\p+x\k+(1-x)\q)^2+m_D^2\right)^{-i_4}x^{i_1-1}(1-x)^{i_2-1}}{(p^2+x(1-x)(\k-\q)^2)^{i_1+i_2}}\,.
\end{align}
Now we have to expand the numerator, which we can do because $-i_4$ is a non-negative integer, and then use the identity
\begin{align}
 \int_p(\p\cdot\q)^nf\left(p^2\right)&=\frac{2}{(4\pi)^{\frac{d-1}{2}}\Gamma\left(\frac{d-1}{2}\right)}\int_{-\infty}^\infty\frac{dp_\parallel}{2\pi}\int_0^\infty dp_\perp^{\,}\,p_\perp^{d-2}(p_\parallel q)^nf\left(p_\parallel^2+p_\perp^2\right)\notag\\
 &=\frac{4}{(4\pi)^{\frac{d+1}{2}}\Gamma\left(\frac{d-1}{2}\right)}\int_0^\infty dp\int_{-1}^1dx\frac{p}{\sqrt{1-x^2}}\,\left(\sqrt{1-x^2}\,p\right)^{d-2}(xpq)^nf\left(p^2\right)\notag\\
 &=\frac{4}{(4\pi)^{\frac{d+1}{2}}\Gamma\left(\frac{d-1}{2}\right)}\int_0^\infty dp\int_0^1dx\,(1-x)^{\frac{d-3}{2}}x^{\frac{n-1}{2}}p^{d-1+n}q^nf\left(p^2\right)\notag\\
 &=\frac{\Gamma\left(\frac{n+1}{2}\right)\Gamma\left(\frac{d}{2}\right)}{\sqrt{\pi}\,\Gamma\left(\frac{d+n}{2}\right)}\int_p\,p^nq^nf\left(p^2\right)=\frac{\Gamma(n)\Gamma\left(\frac{d}{2}\right)}{2^{n-1}\Gamma\left(\frac{n}{2}\right)\Gamma\left(\frac{d+n}{2}\right)}\int_p\,p^nq^nf\left(p^2\right)\,,
\end{align}
if $n$ is even, or $0$ if it is odd. Then the expanded numerator consists only of a sum of powers of $p^2$, $m_D^2$, and $(x\k+(1-x)\q)^2$, the last of which can be re-expressed as:
\begin{equation}
 (x\k+(1-x)\q)^2=x(k^2+m_D^2)-x(1-x)(\k-\q)^2+(1-x)(q^2+m_D^2)-m_D^2\,.
\end{equation}
The $p$ and $x$ integrations now all have the form
\begin{align}
 &\int_0^1dx\int_p\frac{x^{\alpha-1}(1-x)^{\beta-1}p^{2\gamma}}{\left(p^2+x(1-x)(\k-\q)^2\right)^\delta}\notag\\
 &=\frac{\Gamma\left(\delta-\gamma-\frac{d}{2}\right)\Gamma\left(\frac{d}{2}+\gamma\right)}{(4\pi)^{\frac{d}{2}}\Gamma(\delta)\Gamma\left(\frac{d}{2}\right)}\int_0^1dx\,\frac{x^{\alpha+\gamma+\frac{d}{2}-\delta-1}(1-x)^{\beta+\gamma+\frac{d}{2}-\delta-1}}{\left((\k-\q)^2\right)^{\delta-\gamma-\frac{d}{2}}}\notag\\
 &=\frac{\Gamma\left(\delta-\gamma-\frac{d}{2}\right)\Gamma\left(\frac{d}{2}+\gamma\right)\Gamma\left(\alpha+\gamma+\frac{d}{2}-\delta\right)\Gamma\left(\beta+\gamma+\frac{d}{2}-\delta\right)\mu^{3-d}}{(4\pi)^{\frac{d}{2}}\Gamma(\delta)\Gamma\left(\frac{d}{2}\right)\Gamma(\alpha+\beta+2\gamma+d-2\delta)\left((\k-\q)^2\right)^{\delta-\gamma-\frac{d}{2}}}\,.
\end{align}

We see that the remaining loop momenta $\k$ and $\q$ appear only in the combination $(\k-\q)^2$ in the denominator, which can be combined with the term $\left((\k-\q)^2\right)^{i_3}$ from the original $B_M$ integral. The numerator has already been expressed through terms appearing in the denominator, which can also be combined so that we have a sum of integrals of the form

\begin{align}
 &\int_k\int_q\frac{1}{\left((\k-\q)^2\right)^\alpha(k^2+m_D)^\beta(q^2+m_D^2)^\gamma}\notag\\*
 &=\frac{\Gamma(\alpha+\beta)}{\Gamma(\alpha)\Gamma(\beta)}\int_0^1dx\int_k\int_q\frac{(1-x)^{\alpha-1}x^{\beta-1}}{(k^2+x(1-x)q^2+xm_D^2)^{\alpha+\beta}(q^2+m_D^2)^{\gamma}}\notag\\
 &=\frac{\Gamma\left(\alpha+\beta-\frac{d}{2}\right)}{(4\pi)^{\frac{d}{2}}\Gamma(\alpha)\Gamma(\beta)}\int_q\frac{(1-x)^{\alpha-1}x^{\frac{d}{2}-\alpha-1}\mu^{3-d}}{((1-x)q^2+m_D^2)^{\alpha+\beta-\frac{d}{2}}(q^2+m_D^2)^\gamma}\notag\\
 &=\frac{\Gamma\left(\alpha+\beta+\gamma-\frac{d}{2}\right)}{(4\pi)^{\frac{d}{2}}\Gamma(\alpha)\Gamma(\beta)\Gamma(\gamma)}\int_0^1dx\int_0^1dy\int_q\frac{(1-x)^{\alpha-1}x^{\frac{d}{2}-\alpha-1}(1-y)^{\gamma-1}y^{\alpha+\beta-\frac{d}{2}-1}\mu^{3-d}}{\left((1-xy)q^2+m_D^2\right)^{\alpha+\beta+\gamma-\frac{d}{2}}}\notag\\
 &=\frac{\Gamma(\alpha+\beta+\gamma-d)}{(4\pi)^d\Gamma(\alpha)\Gamma(\beta)\Gamma(\gamma)}\int_0^1dx\int_0^1dy\,\frac{(1-x)^{\alpha-1}x^{\frac{d}{2}-\alpha-1}(1-y)^{\gamma-1}y^{\alpha+\beta-\frac{d}{2}-1}\mu^{6-2d}}{(1-xy)^{\frac{d}{2}}m_D^{2\alpha+2\beta+2\gamma-2d}}\,.
\end{align}
If we now perform the substitution
\begin{equation}
 z=\frac{(1-y)x}{1-xy}\,,\hspace{20pt}1-z=\frac{1-x}{1-xy}\,,\hspace{20pt}dz=\frac{1-y}{(1-xy)^2}dx\,,
\end{equation}
where for $x$ from $0$ to $1$ also $z$ ranges from $0$ to $1$ independently of $y$, then the two Feynman parameter integrations decouple:
\begin{align}
 &\int_k\int_q\frac{1}{\left((\k-\q)^2\right)^\alpha(k^2+m_D)^\beta(q^2+m_D^2)^\gamma}\notag\\
 &=\frac{\Gamma(\alpha+\beta+\gamma-d)}{(4\pi)^d\Gamma(\alpha)\Gamma(\beta)\Gamma(\gamma)}\int_0^1dy\int_0^1dz\,\frac{(1-z)^{\alpha-1}z^{\frac{d}{2}-\alpha-1}(1-y)^{\alpha+\gamma-\frac{d}{2}-1}y^{\alpha+\beta-\frac{d}{2}-1}\mu^{6-2d}}{m_D^{2\alpha+2\beta+2\gamma-2d}}\notag\\
 &=\frac{\Gamma(\alpha+\beta+\gamma-d)\Gamma\left(\frac{d}{2}-\alpha\right)\Gamma\left(\alpha+\beta-\frac{d}{2}\right)\Gamma\left(\alpha+\gamma-\frac{d}{2}\right)\mu^{6-2d}}{(4\pi)^d\Gamma(\beta)\Gamma(\gamma)\Gamma\left(\frac{d}{2}\right)\Gamma(2\alpha+\beta+\gamma-d)m_D^{2\alpha+2\beta+2\gamma-2d}}\,.
\end{align}
In this way all $B_M$ integrals with a zero or negative index $i_4,\dots,i_6$ can be expressed through gamma functions.

The final missing integral $B_N(0,0,1,1,1,1)$ is more complicated and we are not aware of a solution for general $d$, so we will show how to calculate it to $\mathcal{O}(\epsilon^0)$. In fact, it is easier to calculate $B_N(0,0,2,1,1,2)$, because unlike $B_N(0,0,1,1,1,1)$ this integral is finite. Through the algorithm described above we get the relation
\begin{equation}
 B_N(0,0,2,1,1,2)=\frac{(3d-8)(3d-10)(d-3)}{64(d-4)m^4}B_N(0,0,1,1,1,1)+\frac{(d-2)^3\Gamma\left(1-\frac{d}{2}\right)^3\mu^{3d-9}}{32(d-4)(4\pi)^{\frac{3d}{2}}m_D^{3d-6}}\,.
\end{equation}
We see that, because of the coefficient $(d-3)$, in order to get $B_N(0,0,1,1,1,1)$ to $\mathcal{O}(\epsilon^0)$ we need to calculate $B_N(0,0,2,1,1,2)$ to $\mathcal{O}(\epsilon^1)$. After performing the shift $\p\to\p-\k$ the $\p$ and $\k$ integrations are identical:
\begin{align}
 &B_N(0,0,2,1,1,2)
 =\int_k\int_p\int_q\frac{1}{\left(p^2+m_D^2\right)^2\left((\k+\q)^2+m_D^2\right)\left((\p+\q)^2+m_D^2\right)\left(k^2+m_D^2\right)^2}\notag\\
 &=\int_q\left(\int_k\frac{1}{\left((\k+\q)^2+m_D^2\right)\left(k^2+m_D^2\right)^2}\right)^2\notag\\
 &=\int_q\left(\int_0^1dx\int_k\frac{2x}{\left(k^2+x(1-x)q^2+m_D^2\right)^3}\right)^2\notag\\
 &=\int_q\left(\frac{\Gamma\left(3-\frac{d}{2}\right)}{(4\pi)^{\frac{d}{2}}}\int_0^1dx\frac{x\,\mu^{3-d}}{\left(x(1-x)q^2+m_D^2\right)^{3-\frac{d}{2}}}\right)^2\notag\\
 &=\int_0^\infty dq\,\frac{2q^{d-1}\mu^{3-d}}{(4\pi)^{\frac{d}{2}}\Gamma\left(\frac{d}{2}\right)}\left(\frac{1}{8\pi m_D\left(q^2+4m_D^2\right)}\left(1-\gamma_E\,\epsilon+\ln\frac{\mu^2\pi}{m_D^2}\,\epsilon\right)+\frac{\tan^{-1}\frac{q}{2m_D}}{2\pi q\left(q^2+4m_D^2\right)}\,\epsilon\right)^2\notag\\
 &=\int_0^\infty dq\,\frac{q^2}{128\pi^4m_D^2\left(q^2+m_D^2\right)^2}\left(1+2\epsilon-3\gamma_E\,\epsilon+\ln\frac{\mu^6\pi^3}{m_D^4q^2}\,\epsilon+\frac{8m_D}{q}\tan^{-1}\frac{q}{2m_D}\,\epsilon\right)\notag\\
 &=\frac{1}{16(4\pi)^3m_D^3}\left(1+2\epsilon-3\gamma_E\,\epsilon-2\ln2\,\epsilon+3\ln\frac{\mu^2\pi}{m_D^2}\,\epsilon\right)+\mathcal{O}\left(\epsilon^2\right)\,.
\end{align}
From these two results we obtain
\begin{equation}
 B_N(0,0,1,1,1,1)=\frac{m_D}{(4\pi)^3}\left(-\frac{1}{\epsilon}-8+3\gamma_E+4\ln2-3\ln\frac{\mu^2\pi}{m_D^2}\right)+\mathcal{O}(\epsilon)\,.
\end{equation}

\section{Color coefficients of the unconnected three-gluon diagrams}
\label{Coefficients}

\begin{figure}[t]
 \includegraphics[width=\linewidth]{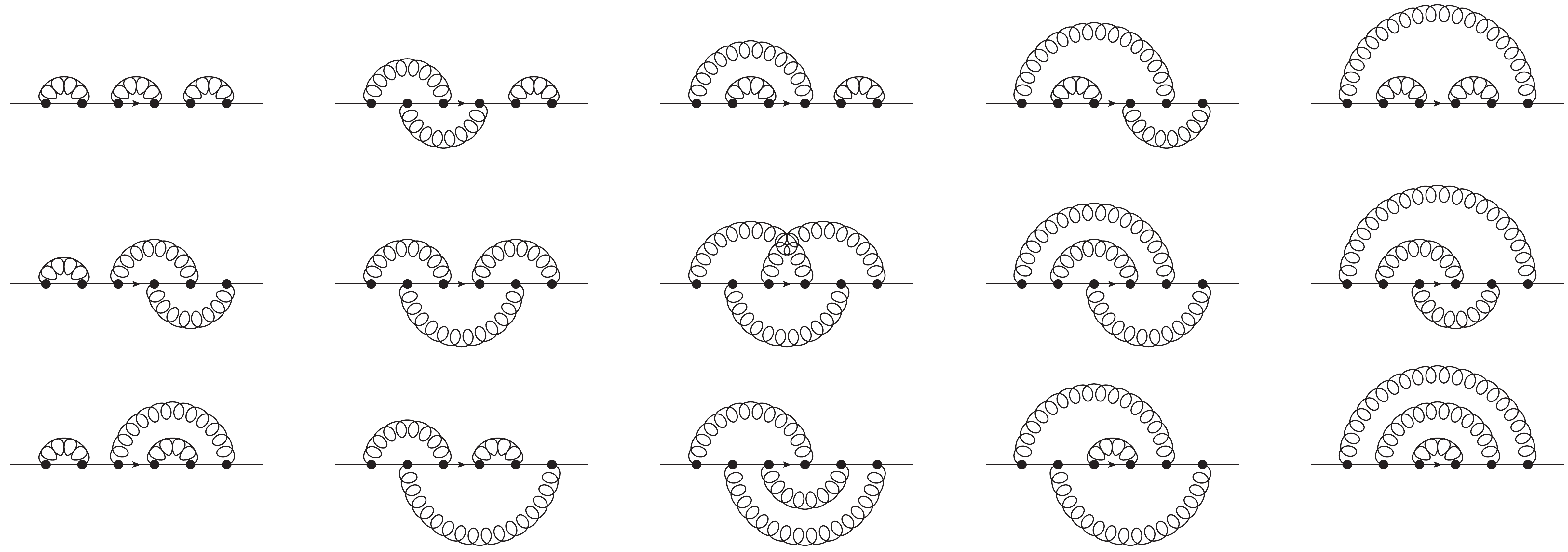}
 \caption{All unconnected three-gluon diagrams. The corresponding color coefficients are labeled $C_{ij}$, where $i$ denotes the row and $j$ the column in which the diagram is listed.}
 \label{Poly7}
\end{figure}

All unconnected three-gluon diagrams are given in Fig.~\ref{Poly7}. The standard color coefficients are labeled $C_{ij}$ according to the caption, while the coefficients that appear in the logarithm are called $\widetilde{C}_{ij}$.

The most straightforward prescription to calculate the coefficients in the logarithm comes from the replica trick~\cite{Gardi:2010rn,*Gardi:2013ita}. First one attaches an index from $1$ to $n$ to each gluon, where $n$ is some integer, then rearranges each diagram such that gluons with a higher index are moved along the Polyakov loop contour to the right of gluons with a lower index, while gluons with the same index keep their current configuration. After summing over all combinations of indices one expands in $n$ and takes the coefficient of the linear term.

Here we have three different possibilities, either all three gluons have a different index, two have the same but the third index is different, or all three indices are the same. It is then only a matter of combinatorics to count the number of possible index combinations. For three different indices there are $n(n-1)(n-2)$ possibilities, while when all three are the same there are $n$. When only two are the same there are $n(n-1)$ index combinations and $3$ ways to choose the one gluon that has a different index. Rearranging the gluons according to their index number always gives $C_{11}=C_R^3$ for three different indices and the standard (i.e., QCD) color coefficient when all indices are the same. When only two are the same, then in half of the index combinations the single index will be smaller then the double index and larger for the other half, but in both cases the color coefficient is the same, so we do not have to differentiate between them. The $3$ different ways to choose the single index gluon may or may not give different color coefficients after rearranging the gluons according to their indices.

The standard color factors are
\begin{align}
 C_{11}&=C_R^3\,, & C_{21}&=C_R^2\left(C_R-\frac{1}{2}C_A\right)\,, & C_{31}&=C_R^3\,,\notag\\
 C_{12}&=C_R^2\left(C_R-\frac{1}{2}C_A\right)\,, & C_{22}&=C_R\left(C_R-\frac{1}{2}C_A\right)^2, & C_{32}&=C_R^2\left(C_R-\frac{1}{2}C_A\right)\,,\notag\\
 C_{13}&=C_R^3\,, & C_{23}&=C_R\left(C_R-\frac{1}{2}C_A\right)\left(C_R-C_A\right)\,, & C_{33}&=C_R\left(C_R-\frac{1}{2}C_A\right)^2,\notag\\
 C_{14}&=C_R^2\left(C_R-\frac{1}{2}C_A\right)\,, & C_{24}&=C_R\left(C_R-\frac{1}{2}C_A\right)^2, & C_{34}&=C_R^2\left(C_R-\frac{1}{2}C_A\right)\,,\notag\\
 C_{15}&=C_R^3\,, & C_{25}&=C_R^2\left(C_R-\frac{1}{2}C_A\right)\,, & C_{35}&=C_R^3\,.
\end{align}

Then we can calculate the coefficients in the logarithm:
\begin{align}
 \widetilde{C}_{11}&=nC_{11}+3n(n-1)C_{11}+n(n-1)(n-2)C_{11}\Bigr|_{\mathcal{O}(n)}=0\,,\\
 \widetilde{C}_{12}&=nC_{12}+2n(n-1)C_{11}+n(n-1)C_{12}+n(n-1)(n-2)C_{11}\Bigr|_{\mathcal{O}(n)}=0\,,\\
 \widetilde{C}_{13}&=nC_{13}+2n(n-1)C_{11}+n(n-1)C_{13}+n(n-1)(n-2)C_{11}\Bigr|_{\mathcal{O}(n)}=0\,,\\
 \widetilde{C}_{14}&=nC_{14}+n(n-1)C_{11}+n(n-1)C_{12}+n(n-1)C_{13}\notag\\
 &\hspace{13pt}+n(n-1)(n-2)C_{11}\Bigr|_{\mathcal{O}(n)}=C_{11}-C_{12}-C_{13}+C_{14}=0\,,\\
 \widetilde{C}_{15}&=nC_{15}+n(n-1)C_{11}+2n(n-1)C_{13}+n(n-1)(n-2)C_{11}\Bigr|_{\mathcal{O}(n)}\notag\\
 &=C_{11}-2C_{13}+C_{15}=0\,,\\
 \widetilde{C}_{21}&=nC_{21}+2n(n-1)C_{11}+n(n-1)C_{21}+n(n-1)(n-2)C_{11}\Bigr|_{\mathcal{O}(n)}=0\,,\\
 \widetilde{C}_{22}&=nC_{22}+2n(n-1)C_{21}+n(n-1)C_{11}+n(n-1)(n-2)C_{11}\Bigr|_{\mathcal{O}(n)}\notag\\
 &=C_{22}-2C_{21}+C_{11}=\frac{1}{4}C_RC_A^2\,,\\
 \widetilde{C}_{23}&=nC_{23}+3n(n-1)C_{21}+n(n-1)(n-2)C_{11}\Bigr|_{\mathcal{O}(n)}\notag\\
 &=C_{23}-3C_{21}+2C_{11}=\frac{1}{2}C_RC_A^2\,,\\
 \widetilde{C}_{24}&=nC_{24}+2n(n-1)C_{21}+n(n-1)C_{31}+n(n-1)(n-2)C_{11}\Bigr|_{\mathcal{O}(n)}\notag\\
 &=C_{24}-2C_{21}-C_{31}+2C_{11}=\frac{1}{4}C_RC_A^2\,,\\
 \widetilde{C}_{25}&=nC_{25}+n(n-1)C_{21}+2n(n-1)C_{31}+n(n-1)(n-2)C_{11}\Bigr|_{\mathcal{O}(n)}\notag\\
 &=C_{25}-C_{21}-2C_{31}+2C_{11}=0\,,\\
 \widetilde{C}_{31}&=nC_{31}+2n(n-1)C_{11}+n(n-1)C_{31}+n(n-1)(n-2)C_{11}\Bigr|_{\mathcal{O}(n)}=0\,,\\
 \widetilde{C}_{32}&=nC_{32}+n(n-1)C_{11}+n(n-1)C_{21}+n(n-1)C_{31}+n(n-1)(n-2)C_{11}\Bigr|_{\mathcal{O}(n)}\notag\\
 &=C_{32}-C_{21}-C_{31}+C_{11}=0\,,\\
 \widetilde{C}_{33}&=nC_{33}+2n(n-1)C_{21}+n(n-1)C_{31}+n(n-1)(n-2)C_{11}\Bigr|_{\mathcal{O}(n)}\notag\\
 &=C_{33}-2C_{21}-C_{31}+2C_{11}=\frac{1}{4}C_RC_A^2\,,\\
 \widetilde{C}_{34}&=nC_{34}+n(n-1)C_{21}+2n(n-1)C_{31}+n(n-1)(n-2)C_{11}\Bigr|_{\mathcal{O}(n)}\notag\\
 &=C_{34}-C_{21}-2C_{31}+2C_{11}=0\,,\\
 \widetilde{C}_{35}&=nC_{35}+3n(n-1)C_{31}+n(n-1)(n-2)C_{11}\Bigr|_{\mathcal{O}(n)}\notag\\
 &=C_{35}-3C_{31}+2C_{11}=0\,.
\end{align}

Here we see the general property confirmed that only two-particle irreducible diagrams appear in the logarithm. This means that the color coefficient in the logarithm vanishes for any diagram where one can cut the (closed) Polyakov loop contour in two points such that there are no gluons connecting from one segment of the contour to the other. These are the so-called two-particle reducible diagrams, the diagrams where this is not possible are called two-particle irreducible. Here the considerable reduction in the number of diagrams is even more apparent than in the two-gluon diagrams: out of fifteen unconnected three-gluon diagrams only four survive in the logarithm.

We also see that all higher power terms of $C_R$ are canceled, only the linear term remains and only the two-particle irreducible diagrams have a linear term. This is in accordance with the theorem shown in~\cite{Gardi:2013ita} that the color coefficients in the logarithm all correspond to those of fully connected diagrams. The coefficients of fully connected diagrams depend only linearly on $C_R$ or the $C_R^{(n)}$. So the only terms that can break Casimir scaling come from the $C_R^{(n)}$, see Eq.~\eqref{Casin}.

\bibliography{ref}

\begin{thebibliography}{62}%
\makeatletter
\providecommand \@ifxundefined [1]{%
 \@ifx{#1\undefined}
}%
\providecommand \@ifnum [1]{%
 \ifnum #1\expandafter \@firstoftwo
 \else \expandafter \@secondoftwo
 \fi
}%
\providecommand \@ifx [1]{%
 \ifx #1\expandafter \@firstoftwo
 \else \expandafter \@secondoftwo
 \fi
}%
\providecommand \natexlab [1]{#1}%
\providecommand \enquote  [1]{``#1''}%
\providecommand \bibnamefont  [1]{#1}%
\providecommand \bibfnamefont [1]{#1}%
\providecommand \citenamefont [1]{#1}%
\providecommand \href@noop [0]{\@secondoftwo}%
\providecommand \href [0]{\begingroup \@sanitize@url \@href}%
\providecommand \@href[1]{\@@startlink{#1}\@@href}%
\providecommand \@@href[1]{\endgroup#1\@@endlink}%
\providecommand \@sanitize@url [0]{\catcode `\\12\catcode `\$12\catcode
  `\&12\catcode `\#12\catcode `\^12\catcode `\_12\catcode `\%12\relax}%
\providecommand \@@startlink[1]{}%
\providecommand \@@endlink[0]{}%
\providecommand \url  [0]{\begingroup\@sanitize@url \@url }%
\providecommand \@url [1]{\endgroup\@href {#1}{\urlprefix }}%
\providecommand \urlprefix  [0]{URL }%
\providecommand \Eprint [0]{\href }%
\providecommand \doibase [0]{http://dx.doi.org/}%
\providecommand \selectlanguage [0]{\@gobble}%
\providecommand \bibinfo  [0]{\@secondoftwo}%
\providecommand \bibfield  [0]{\@secondoftwo}%
\providecommand \translation [1]{[#1]}%
\providecommand \BibitemOpen [0]{}%
\providecommand \bibitemStop [0]{}%
\providecommand \bibitemNoStop [0]{.\EOS\space}%
\providecommand \EOS [0]{\spacefactor3000\relax}%
\providecommand \BibitemShut  [1]{\csname bibitem#1\endcsname}%
\let\auto@bib@innerbib\@empty
\bibitem [{\citenamefont {Polyakov}(1978)}]{Polyakov:1978vu}%
  \BibitemOpen
  \bibfield  {author} {\bibinfo {author} {\bibfnamefont {A.~M.}\ \bibnamefont
  {Polyakov}},\ }\href {\doibase 10.1016/0370-2693(78)90737-2} {\bibfield
  {journal} {\bibinfo  {journal} {Phys. Lett.}\ }\textbf {\bibinfo {volume}
  {B72}},\ \bibinfo {pages} {477} (\bibinfo {year} {1978})}\BibitemShut
  {NoStop}%
\bibitem [{\citenamefont {Kuti}\ \emph {et~al.}(1981)\citenamefont {Kuti},
  \citenamefont {Polonyi},\ and\ \citenamefont {Szlachanyi}}]{Kuti:1980gh}%
  \BibitemOpen
  \bibfield  {author} {\bibinfo {author} {\bibfnamefont {J.}~\bibnamefont
  {Kuti}}, \bibinfo {author} {\bibfnamefont {J.}~\bibnamefont {Polonyi}}, \
  and\ \bibinfo {author} {\bibfnamefont {K.}~\bibnamefont {Szlachanyi}},\
  }\href {\doibase 10.1016/0370-2693(81)90987-4} {\bibfield  {journal}
  {\bibinfo  {journal} {Phys. Lett.}\ }\textbf {\bibinfo {volume} {B98}},\
  \bibinfo {pages} {199} (\bibinfo {year} {1981})}\BibitemShut {NoStop}%
\bibitem [{\citenamefont {McLerran}\ and\ \citenamefont
  {Svetitsky}(1981)}]{McLerran:1981pb}%
  \BibitemOpen
  \bibfield  {author} {\bibinfo {author} {\bibfnamefont {L.~D.}\ \bibnamefont
  {McLerran}}\ and\ \bibinfo {author} {\bibfnamefont {B.}~\bibnamefont
  {Svetitsky}},\ }\href {\doibase 10.1103/PhysRevD.24.450} {\bibfield
  {journal} {\bibinfo  {journal} {Phys. Rev.}\ }\textbf {\bibinfo {volume}
  {D24}},\ \bibinfo {pages} {450} (\bibinfo {year} {1981})}\BibitemShut
  {NoStop}%
\bibitem [{\citenamefont {Fukushima}(2003)}]{Fukushima:2002bk}%
  \BibitemOpen
  \bibfield  {author} {\bibinfo {author} {\bibfnamefont {K.}~\bibnamefont
  {Fukushima}},\ }\href {\doibase 10.1016/S0003-4916(03)00012-5} {\bibfield
  {journal} {\bibinfo  {journal} {Annals Phys.}\ }\textbf {\bibinfo {volume}
  {304}},\ \bibinfo {pages} {72} (\bibinfo {year} {2003})},\ \Eprint
  {http://arxiv.org/abs/hep-ph/0204302} {arXiv:hep-ph/0204302 [hep-ph]}
  \BibitemShut {NoStop}%
\bibitem [{\citenamefont {Gava}\ and\ \citenamefont
  {Jengo}(1981)}]{Gava:1981qd}%
  \BibitemOpen
  \bibfield  {author} {\bibinfo {author} {\bibfnamefont {E.}~\bibnamefont
  {Gava}}\ and\ \bibinfo {author} {\bibfnamefont {R.}~\bibnamefont {Jengo}},\
  }\href {\doibase 10.1016/0370-2693(81)90890-X} {\bibfield  {journal}
  {\bibinfo  {journal} {Phys. Lett.}\ }\textbf {\bibinfo {volume} {B105}},\
  \bibinfo {pages} {285} (\bibinfo {year} {1981})}\BibitemShut {NoStop}%
\bibitem [{\citenamefont {Burnier}\ \emph {et~al.}(2010)\citenamefont
  {Burnier}, \citenamefont {Laine},\ and\ \citenamefont
  {{Veps\"al\"ainen}}}]{Burnier:2009bk}%
  \BibitemOpen
  \bibfield  {author} {\bibinfo {author} {\bibfnamefont {Y.}~\bibnamefont
  {Burnier}}, \bibinfo {author} {\bibfnamefont {M.}~\bibnamefont {Laine}}, \
  and\ \bibinfo {author} {\bibfnamefont {M.}~\bibnamefont
  {{Veps\"al\"ainen}}},\ }\href {\doibase 10.1007/JHEP01(2010)054,
  10.1007/JHEP01(2013)180} {\bibfield  {journal} {\bibinfo  {journal} {JHEP}\
  }\textbf {\bibinfo {volume} {01}},\ \bibinfo {pages} {054} (\bibinfo {year}
  {2010})},\ \bibinfo {note} {[Erratum: JHEP {\bf 01}, 180 (2013)]},\ \Eprint
  {http://arxiv.org/abs/0911.3480} {arXiv:0911.3480 [hep-ph]} \BibitemShut
  {NoStop}%
\bibitem [{\citenamefont {Brambilla}\ \emph
  {et~al.}(2010{\natexlab{a}})\citenamefont {Brambilla}, \citenamefont
  {Ghiglieri}, \citenamefont {Petreczky},\ and\ \citenamefont
  {Vairo}}]{Brambilla:2010xn}%
  \BibitemOpen
  \bibfield  {author} {\bibinfo {author} {\bibfnamefont {N.}~\bibnamefont
  {Brambilla}}, \bibinfo {author} {\bibfnamefont {J.}~\bibnamefont
  {Ghiglieri}}, \bibinfo {author} {\bibfnamefont {P.}~\bibnamefont
  {Petreczky}}, \ and\ \bibinfo {author} {\bibfnamefont {A.}~\bibnamefont
  {Vairo}},\ }\href {\doibase 10.1103/PhysRevD.82.074019} {\bibfield  {journal}
  {\bibinfo  {journal} {Phys. Rev.}\ }\textbf {\bibinfo {volume} {D82}},\
  \bibinfo {pages} {074019} (\bibinfo {year} {2010}{\natexlab{a}})},\ \Eprint
  {http://arxiv.org/abs/1007.5172} {arXiv:1007.5172 [hep-ph]} \BibitemShut
  {NoStop}%
\bibitem [{\citenamefont {Kaczmarek}\ \emph {et~al.}(2002)\citenamefont
  {Kaczmarek}, \citenamefont {Karsch}, \citenamefont {Petreczky},\ and\
  \citenamefont {Zantow}}]{Kaczmarek:2002mc}%
  \BibitemOpen
  \bibfield  {author} {\bibinfo {author} {\bibfnamefont {O.}~\bibnamefont
  {Kaczmarek}}, \bibinfo {author} {\bibfnamefont {F.}~\bibnamefont {Karsch}},
  \bibinfo {author} {\bibfnamefont {P.}~\bibnamefont {Petreczky}}, \ and\
  \bibinfo {author} {\bibfnamefont {F.}~\bibnamefont {Zantow}},\ }\href
  {\doibase 10.1016/S0370-2693(02)02415-2} {\bibfield  {journal} {\bibinfo
  {journal} {Phys. Lett.}\ }\textbf {\bibinfo {volume} {B543}},\ \bibinfo
  {pages} {41} (\bibinfo {year} {2002})},\ \Eprint
  {http://arxiv.org/abs/hep-lat/0207002} {arXiv:hep-lat/0207002 [hep-lat]}
  \BibitemShut {NoStop}%
\bibitem [{\citenamefont {Kaczmarek}\ \emph {et~al.}(2004)\citenamefont
  {Kaczmarek}, \citenamefont {Karsch}, \citenamefont {Zantow},\ and\
  \citenamefont {Petreczky}}]{Kaczmarek:2004gv}%
  \BibitemOpen
  \bibfield  {author} {\bibinfo {author} {\bibfnamefont {O.}~\bibnamefont
  {Kaczmarek}}, \bibinfo {author} {\bibfnamefont {F.}~\bibnamefont {Karsch}},
  \bibinfo {author} {\bibfnamefont {F.}~\bibnamefont {Zantow}}, \ and\ \bibinfo
  {author} {\bibfnamefont {P.}~\bibnamefont {Petreczky}},\ }\href {\doibase
  10.1103/PhysRevD.70.074505, 10.1103/PhysRevD.72.059903} {\bibfield  {journal}
  {\bibinfo  {journal} {Phys. Rev.}\ }\textbf {\bibinfo {volume} {D70}},\
  \bibinfo {pages} {074505} (\bibinfo {year} {2004})},\ \bibinfo {note}
  {[Erratum: Phys. Rev. {\bf D72}, 059903 (2005)]},\ \Eprint
  {http://arxiv.org/abs/hep-lat/0406036} {arXiv:hep-lat/0406036 [hep-lat]}
  \BibitemShut {NoStop}%
\bibitem [{\citenamefont {Digal}\ \emph {et~al.}(2003)\citenamefont {Digal},
  \citenamefont {Fortunato},\ and\ \citenamefont {Petreczky}}]{Digal:2003jc}%
  \BibitemOpen
  \bibfield  {author} {\bibinfo {author} {\bibfnamefont {S.}~\bibnamefont
  {Digal}}, \bibinfo {author} {\bibfnamefont {S.}~\bibnamefont {Fortunato}}, \
  and\ \bibinfo {author} {\bibfnamefont {P.}~\bibnamefont {Petreczky}},\ }\href
  {\doibase 10.1103/PhysRevD.68.034008} {\bibfield  {journal} {\bibinfo
  {journal} {Phys. Rev.}\ }\textbf {\bibinfo {volume} {D68}},\ \bibinfo {pages}
  {034008} (\bibinfo {year} {2003})},\ \Eprint
  {http://arxiv.org/abs/hep-lat/0304017} {arXiv:hep-lat/0304017 [hep-lat]}
  \BibitemShut {NoStop}%
\bibitem [{\citenamefont {Mykkanen}\ \emph {et~al.}(2012)\citenamefont
  {Mykkanen}, \citenamefont {Panero},\ and\ \citenamefont
  {Rummukainen}}]{Mykkanen:2012ri}%
  \BibitemOpen
  \bibfield  {author} {\bibinfo {author} {\bibfnamefont {A.}~\bibnamefont
  {Mykkanen}}, \bibinfo {author} {\bibfnamefont {M.}~\bibnamefont {Panero}}, \
  and\ \bibinfo {author} {\bibfnamefont {K.}~\bibnamefont {Rummukainen}},\
  }\href {\doibase 10.1007/JHEP05(2012)069} {\bibfield  {journal} {\bibinfo
  {journal} {JHEP}\ }\textbf {\bibinfo {volume} {05}},\ \bibinfo {pages} {069}
  (\bibinfo {year} {2012})},\ \Eprint {http://arxiv.org/abs/1202.2762}
  {arXiv:1202.2762 [hep-lat]} \BibitemShut {NoStop}%
\bibitem [{\citenamefont {Aoki}\ \emph {et~al.}(2006)\citenamefont {Aoki},
  \citenamefont {Fodor}, \citenamefont {Katz},\ and\ \citenamefont
  {Szabo}}]{Aoki:2006br}%
  \BibitemOpen
  \bibfield  {author} {\bibinfo {author} {\bibfnamefont {Y.}~\bibnamefont
  {Aoki}}, \bibinfo {author} {\bibfnamefont {Z.}~\bibnamefont {Fodor}},
  \bibinfo {author} {\bibfnamefont {S.~D.}\ \bibnamefont {Katz}}, \ and\
  \bibinfo {author} {\bibfnamefont {K.~K.}\ \bibnamefont {Szabo}},\ }\href
  {\doibase 10.1016/j.physletb.2006.10.021} {\bibfield  {journal} {\bibinfo
  {journal} {Phys. Lett.}\ }\textbf {\bibinfo {volume} {B643}},\ \bibinfo
  {pages} {46} (\bibinfo {year} {2006})},\ \Eprint
  {http://arxiv.org/abs/hep-lat/0609068} {arXiv:hep-lat/0609068 [hep-lat]}
  \BibitemShut {NoStop}%
\bibitem [{\citenamefont {Cheng}\ \emph {et~al.}(2008)\citenamefont {Cheng}
  \emph {et~al.}}]{Cheng:2007jq}%
  \BibitemOpen
  \bibfield  {author} {\bibinfo {author} {\bibfnamefont {M.}~\bibnamefont
  {Cheng}} \emph {et~al.},\ }\href {\doibase 10.1103/PhysRevD.77.014511}
  {\bibfield  {journal} {\bibinfo  {journal} {Phys. Rev.}\ }\textbf {\bibinfo
  {volume} {D77}},\ \bibinfo {pages} {014511} (\bibinfo {year} {2008})},\
  \Eprint {http://arxiv.org/abs/0710.0354} {arXiv:0710.0354 [hep-lat]}
  \BibitemShut {NoStop}%
\bibitem [{\citenamefont {Aoki}\ \emph {et~al.}(2009)\citenamefont {Aoki},
  \citenamefont {{Bors\'anyi}}, \citenamefont {Durr}, \citenamefont {Fodor},
  \citenamefont {Katz}, \citenamefont {Krieg},\ and\ \citenamefont
  {{Szab\'o}}}]{Aoki:2009sc}%
  \BibitemOpen
  \bibfield  {author} {\bibinfo {author} {\bibfnamefont {Y.}~\bibnamefont
  {Aoki}}, \bibinfo {author} {\bibfnamefont {S.}~\bibnamefont {{Bors\'anyi}}},
  \bibinfo {author} {\bibfnamefont {S.}~\bibnamefont {Durr}}, \bibinfo {author}
  {\bibfnamefont {Z.}~\bibnamefont {Fodor}}, \bibinfo {author} {\bibfnamefont
  {S.~D.}\ \bibnamefont {Katz}}, \bibinfo {author} {\bibfnamefont
  {S.}~\bibnamefont {Krieg}}, \ and\ \bibinfo {author} {\bibfnamefont {K.~K.}\
  \bibnamefont {{Szab\'o}}},\ }\href {\doibase 10.1088/1126-6708/2009/06/088}
  {\bibfield  {journal} {\bibinfo  {journal} {JHEP}\ }\textbf {\bibinfo
  {volume} {06}},\ \bibinfo {pages} {088} (\bibinfo {year} {2009})},\ \Eprint
  {http://arxiv.org/abs/0903.4155} {arXiv:0903.4155 [hep-lat]} \BibitemShut
  {NoStop}%
\bibitem [{\citenamefont {Bazavov}\ \emph {et~al.}(2009)\citenamefont {Bazavov}
  \emph {et~al.}}]{Bazavov:2009zn}%
  \BibitemOpen
  \bibfield  {author} {\bibinfo {author} {\bibfnamefont {A.}~\bibnamefont
  {Bazavov}} \emph {et~al.},\ }\href {\doibase 10.1103/PhysRevD.80.014504}
  {\bibfield  {journal} {\bibinfo  {journal} {Phys. Rev.}\ }\textbf {\bibinfo
  {volume} {D80}},\ \bibinfo {pages} {014504} (\bibinfo {year} {2009})},\
  \Eprint {http://arxiv.org/abs/0903.4379} {arXiv:0903.4379 [hep-lat]}
  \BibitemShut {NoStop}%
\bibitem [{\citenamefont {{Bors\'anyi}}\ \emph {et~al.}(2010)\citenamefont
  {{Bors\'anyi}}, \citenamefont {Fodor}, \citenamefont {Hoelbling},
  \citenamefont {Katz}, \citenamefont {Krieg}, \citenamefont {Ratti},\ and\
  \citenamefont {Szabo}}]{Borsanyi:2010bp}%
  \BibitemOpen
  \bibfield  {author} {\bibinfo {author} {\bibfnamefont {S.}~\bibnamefont
  {{Bors\'anyi}}}, \bibinfo {author} {\bibfnamefont {Z.}~\bibnamefont {Fodor}},
  \bibinfo {author} {\bibfnamefont {C.}~\bibnamefont {Hoelbling}}, \bibinfo
  {author} {\bibfnamefont {S.~D.}\ \bibnamefont {Katz}}, \bibinfo {author}
  {\bibfnamefont {S.}~\bibnamefont {Krieg}}, \bibinfo {author} {\bibfnamefont
  {C.}~\bibnamefont {Ratti}}, \ and\ \bibinfo {author} {\bibfnamefont {K.~K.}\
  \bibnamefont {Szabo}} (\bibinfo {collaboration} {Wuppertal-Budapest}),\
  }\href {\doibase 10.1007/JHEP09(2010)073} {\bibfield  {journal} {\bibinfo
  {journal} {JHEP}\ }\textbf {\bibinfo {volume} {09}},\ \bibinfo {pages} {073}
  (\bibinfo {year} {2010})},\ \Eprint {http://arxiv.org/abs/1005.3508}
  {arXiv:1005.3508 [hep-lat]} \BibitemShut {NoStop}%
\bibitem [{\citenamefont {Bazavov}\ \emph
  {et~al.}(2012{\natexlab{a}})\citenamefont {Bazavov} \emph
  {et~al.}}]{Bazavov:2011nk}%
  \BibitemOpen
  \bibfield  {author} {\bibinfo {author} {\bibfnamefont {A.}~\bibnamefont
  {Bazavov}} \emph {et~al.},\ }\href {\doibase 10.1103/PhysRevD.85.054503}
  {\bibfield  {journal} {\bibinfo  {journal} {Phys. Rev.}\ }\textbf {\bibinfo
  {volume} {D85}},\ \bibinfo {pages} {054503} (\bibinfo {year}
  {2012}{\natexlab{a}})},\ \Eprint {http://arxiv.org/abs/1111.1710}
  {arXiv:1111.1710 [hep-lat]} \BibitemShut {NoStop}%
\bibitem [{\citenamefont {Bazavov}\ and\ \citenamefont
  {Petreczky}(2013)}]{Bazavov:2013yv}%
  \BibitemOpen
  \bibfield  {author} {\bibinfo {author} {\bibfnamefont {A.}~\bibnamefont
  {Bazavov}}\ and\ \bibinfo {author} {\bibfnamefont {P.}~\bibnamefont
  {Petreczky}},\ }\href {\doibase 10.1103/PhysRevD.87.094505} {\bibfield
  {journal} {\bibinfo  {journal} {Phys.Rev.}\ }\textbf {\bibinfo {volume}
  {D87}},\ \bibinfo {pages} {094505} (\bibinfo {year} {2013})},\ \Eprint
  {http://arxiv.org/abs/1301.3943} {arXiv:1301.3943 [hep-lat]} \BibitemShut
  {NoStop}%
\bibitem [{\citenamefont {{Bors\'anyi}}\ \emph {et~al.}(2015)\citenamefont
  {{Bors\'anyi}}, \citenamefont {Fodor}, \citenamefont {Katz}, \citenamefont
  {{P\'asztor}}, \citenamefont {{Szab\'o}},\ and\ \citenamefont
  {{T\"or\"ok}}}]{Borsanyi:2015yka}%
  \BibitemOpen
  \bibfield  {author} {\bibinfo {author} {\bibfnamefont {S.}~\bibnamefont
  {{Bors\'anyi}}}, \bibinfo {author} {\bibfnamefont {Z.}~\bibnamefont {Fodor}},
  \bibinfo {author} {\bibfnamefont {S.~D.}\ \bibnamefont {Katz}}, \bibinfo
  {author} {\bibfnamefont {A.}~\bibnamefont {{P\'asztor}}}, \bibinfo {author}
  {\bibfnamefont {K.~K.}\ \bibnamefont {{Szab\'o}}}, \ and\ \bibinfo {author}
  {\bibfnamefont {C.}~\bibnamefont {{T\"or\"ok}}},\ }\href {\doibase
  10.1007/JHEP04(2015)138} {\bibfield  {journal} {\bibinfo  {journal} {JHEP}\
  }\textbf {\bibinfo {volume} {04}},\ \bibinfo {pages} {138} (\bibinfo {year}
  {2015})},\ \Eprint {http://arxiv.org/abs/1501.02173} {arXiv:1501.02173
  [hep-lat]} \BibitemShut {NoStop}%
\bibitem [{\citenamefont {Gupta}\ \emph {et~al.}(2008)\citenamefont {Gupta},
  \citenamefont {Huebner},\ and\ \citenamefont {Kaczmarek}}]{Gupta:2007ax}%
  \BibitemOpen
  \bibfield  {author} {\bibinfo {author} {\bibfnamefont {S.}~\bibnamefont
  {Gupta}}, \bibinfo {author} {\bibfnamefont {K.}~\bibnamefont {Huebner}}, \
  and\ \bibinfo {author} {\bibfnamefont {O.}~\bibnamefont {Kaczmarek}},\ }\href
  {\doibase 10.1103/PhysRevD.77.034503} {\bibfield  {journal} {\bibinfo
  {journal} {Phys. Rev.}\ }\textbf {\bibinfo {volume} {D77}},\ \bibinfo {pages}
  {034503} (\bibinfo {year} {2008})},\ \Eprint {http://arxiv.org/abs/0711.2251}
  {arXiv:0711.2251 [hep-lat]} \BibitemShut {NoStop}%
\bibitem [{\citenamefont {Petreczky}\ and\ \citenamefont
  {Schadler}(2015)}]{Petreczky:2015yta}%
  \BibitemOpen
  \bibfield  {author} {\bibinfo {author} {\bibfnamefont {P.}~\bibnamefont
  {Petreczky}}\ and\ \bibinfo {author} {\bibfnamefont {H.~P.}\ \bibnamefont
  {Schadler}},\ }\href {\doibase 10.1103/PhysRevD.92.094517} {\bibfield
  {journal} {\bibinfo  {journal} {Phys. Rev.}\ }\textbf {\bibinfo {volume}
  {D92}},\ \bibinfo {pages} {094517} (\bibinfo {year} {2015})},\ \Eprint
  {http://arxiv.org/abs/1509.07874} {arXiv:1509.07874 [hep-lat]} \BibitemShut
  {NoStop}%
\bibitem [{\citenamefont {Gatheral}(1983)}]{Gatheral:1983cz}%
  \BibitemOpen
  \bibfield  {author} {\bibinfo {author} {\bibfnamefont {J.~G.~M.}\
  \bibnamefont {Gatheral}},\ }\href {\doibase 10.1016/0370-2693(83)90112-0}
  {\bibfield  {journal} {\bibinfo  {journal} {Phys. Lett.}\ }\textbf {\bibinfo
  {volume} {B133}},\ \bibinfo {pages} {90} (\bibinfo {year}
  {1983})}\BibitemShut {NoStop}%
\bibitem [{\citenamefont {Frenkel}\ and\ \citenamefont
  {Taylor}(1984)}]{Frenkel:1984pz}%
  \BibitemOpen
  \bibfield  {author} {\bibinfo {author} {\bibfnamefont {J.}~\bibnamefont
  {Frenkel}}\ and\ \bibinfo {author} {\bibfnamefont {J.~C.}\ \bibnamefont
  {Taylor}},\ }\href {\doibase 10.1016/0550-3213(84)90294-3} {\bibfield
  {journal} {\bibinfo  {journal} {Nucl. Phys.}\ }\textbf {\bibinfo {volume}
  {B246}},\ \bibinfo {pages} {231} (\bibinfo {year} {1984})}\BibitemShut
  {NoStop}%
\bibitem [{\citenamefont {Gardi}\ \emph {et~al.}(2010)\citenamefont {Gardi},
  \citenamefont {Laenen}, \citenamefont {Stavenga},\ and\ \citenamefont
  {White}}]{Gardi:2010rn}%
  \BibitemOpen
  \bibfield  {author} {\bibinfo {author} {\bibfnamefont {E.}~\bibnamefont
  {Gardi}}, \bibinfo {author} {\bibfnamefont {E.}~\bibnamefont {Laenen}},
  \bibinfo {author} {\bibfnamefont {G.}~\bibnamefont {Stavenga}}, \ and\
  \bibinfo {author} {\bibfnamefont {C.~D.}\ \bibnamefont {White}},\ }\href
  {\doibase 10.1007/JHEP11(2010)155} {\bibfield  {journal} {\bibinfo  {journal}
  {JHEP}\ }\textbf {\bibinfo {volume} {11}},\ \bibinfo {pages} {155} (\bibinfo
  {year} {2010})},\ \Eprint {http://arxiv.org/abs/1008.0098} {arXiv:1008.0098
  [hep-ph]} \BibitemShut {NoStop}%
\bibitem [{\citenamefont {Gardi}\ \emph {et~al.}(2013)\citenamefont {Gardi},
  \citenamefont {Smillie},\ and\ \citenamefont {White}}]{Gardi:2013ita}%
  \BibitemOpen
  \bibfield  {author} {\bibinfo {author} {\bibfnamefont {E.}~\bibnamefont
  {Gardi}}, \bibinfo {author} {\bibfnamefont {J.~M.}\ \bibnamefont {Smillie}},
  \ and\ \bibinfo {author} {\bibfnamefont {C.~D.}\ \bibnamefont {White}},\
  }\href {\doibase 10.1007/JHEP06(2013)088} {\bibfield  {journal} {\bibinfo
  {journal} {JHEP}\ }\textbf {\bibinfo {volume} {1306}},\ \bibinfo {pages}
  {088} (\bibinfo {year} {2013})},\ \Eprint {http://arxiv.org/abs/1304.7040}
  {arXiv:1304.7040 [hep-ph]} \BibitemShut {NoStop}%
\bibitem [{\citenamefont {Berwein}\ \emph {et~al.}(2013)\citenamefont
  {Berwein}, \citenamefont {Brambilla}, \citenamefont {Ghiglieri},\ and\
  \citenamefont {Vairo}}]{Berwein:2012mw}%
  \BibitemOpen
  \bibfield  {author} {\bibinfo {author} {\bibfnamefont {M.}~\bibnamefont
  {Berwein}}, \bibinfo {author} {\bibfnamefont {N.}~\bibnamefont {Brambilla}},
  \bibinfo {author} {\bibfnamefont {J.}~\bibnamefont {Ghiglieri}}, \ and\
  \bibinfo {author} {\bibfnamefont {A.}~\bibnamefont {Vairo}},\ }\href
  {\doibase 10.1007/JHEP03(2013)069} {\bibfield  {journal} {\bibinfo  {journal}
  {JHEP}\ }\textbf {\bibinfo {volume} {03}},\ \bibinfo {pages} {069} (\bibinfo
  {year} {2013})},\ \Eprint {http://arxiv.org/abs/1212.4413} {arXiv:1212.4413
  [hep-th]} \BibitemShut {NoStop}%
\bibitem [{\citenamefont {Berwein}\ \emph {et~al.}(2014)\citenamefont
  {Berwein}, \citenamefont {Brambilla},\ and\ \citenamefont
  {Vairo}}]{Berwein:2013xza}%
  \BibitemOpen
  \bibfield  {author} {\bibinfo {author} {\bibfnamefont {M.}~\bibnamefont
  {Berwein}}, \bibinfo {author} {\bibfnamefont {N.}~\bibnamefont {Brambilla}},
  \ and\ \bibinfo {author} {\bibfnamefont {A.}~\bibnamefont {Vairo}},\ }\href
  {\doibase 10.1134/S1063779614040029} {\bibfield  {journal} {\bibinfo
  {journal} {Phys. Part. Nucl.}\ }\textbf {\bibinfo {volume} {45}},\ \bibinfo
  {pages} {656} (\bibinfo {year} {2014})},\ \Eprint
  {http://arxiv.org/abs/1312.6651} {arXiv:1312.6651 [hep-th]} \BibitemShut
  {NoStop}%
\bibitem [{\citenamefont {Braaten}\ and\ \citenamefont
  {Nieto}(1996)}]{Braaten:1995jr}%
  \BibitemOpen
  \bibfield  {author} {\bibinfo {author} {\bibfnamefont {E.}~\bibnamefont
  {Braaten}}\ and\ \bibinfo {author} {\bibfnamefont {A.}~\bibnamefont
  {Nieto}},\ }\href {\doibase 10.1103/PhysRevD.53.3421} {\bibfield  {journal}
  {\bibinfo  {journal} {Phys. Rev.}\ }\textbf {\bibinfo {volume} {D53}},\
  \bibinfo {pages} {3421} (\bibinfo {year} {1996})},\ \Eprint
  {http://arxiv.org/abs/hep-ph/9510408} {arXiv:hep-ph/9510408 [hep-ph]}
  \BibitemShut {NoStop}%
\bibitem [{\citenamefont {Kajantie}\ \emph
  {et~al.}(1997{\natexlab{a}})\citenamefont {Kajantie}, \citenamefont {Laine},
  \citenamefont {Rummukainen},\ and\ \citenamefont
  {Shaposhnikov}}]{Kajantie:1997tt}%
  \BibitemOpen
  \bibfield  {author} {\bibinfo {author} {\bibfnamefont {K.}~\bibnamefont
  {Kajantie}}, \bibinfo {author} {\bibfnamefont {M.}~\bibnamefont {Laine}},
  \bibinfo {author} {\bibfnamefont {K.}~\bibnamefont {Rummukainen}}, \ and\
  \bibinfo {author} {\bibfnamefont {M.~E.}\ \bibnamefont {Shaposhnikov}},\
  }\href {\doibase 10.1016/S0550-3213(97)00425-2} {\bibfield  {journal}
  {\bibinfo  {journal} {Nucl. Phys.}\ }\textbf {\bibinfo {volume} {B503}},\
  \bibinfo {pages} {357} (\bibinfo {year} {1997}{\natexlab{a}})},\ \Eprint
  {http://arxiv.org/abs/hep-ph/9704416} {arXiv:hep-ph/9704416 [hep-ph]}
  \BibitemShut {NoStop}%
\bibitem [{\citenamefont {Ginsparg}(1980)}]{Ginsparg:1980ef}%
  \BibitemOpen
  \bibfield  {author} {\bibinfo {author} {\bibfnamefont {P.~H.}\ \bibnamefont
  {Ginsparg}},\ }\href {\doibase 10.1016/0550-3213(80)90418-6} {\bibfield
  {journal} {\bibinfo  {journal} {Nucl. Phys.}\ }\textbf {\bibinfo {volume}
  {B170}},\ \bibinfo {pages} {388} (\bibinfo {year} {1980})}\BibitemShut
  {NoStop}%
\bibitem [{\citenamefont {Appelquist}\ and\ \citenamefont
  {Pisarski}(1981)}]{Appelquist:1981vg}%
  \BibitemOpen
  \bibfield  {author} {\bibinfo {author} {\bibfnamefont {T.}~\bibnamefont
  {Appelquist}}\ and\ \bibinfo {author} {\bibfnamefont {R.~D.}\ \bibnamefont
  {Pisarski}},\ }\href {\doibase 10.1103/PhysRevD.23.2305} {\bibfield
  {journal} {\bibinfo  {journal} {Phys. Rev.}\ }\textbf {\bibinfo {volume}
  {D23}},\ \bibinfo {pages} {2305} (\bibinfo {year} {1981})}\BibitemShut
  {NoStop}%
\bibitem [{\citenamefont {D'Hoker}(1982)}]{D'Hoker:1981us}%
  \BibitemOpen
  \bibfield  {author} {\bibinfo {author} {\bibfnamefont {E.}~\bibnamefont
  {D'Hoker}},\ }\href {\doibase 10.1016/0550-3213(82)90441-2} {\bibfield
  {journal} {\bibinfo  {journal} {Nucl. Phys.}\ }\textbf {\bibinfo {volume}
  {B201}},\ \bibinfo {pages} {401} (\bibinfo {year} {1982})}\BibitemShut
  {NoStop}%
\bibitem [{\citenamefont {Nadkarni}(1983)}]{Nadkarni:1982kb}%
  \BibitemOpen
  \bibfield  {author} {\bibinfo {author} {\bibfnamefont {S.}~\bibnamefont
  {Nadkarni}},\ }\href {\doibase 10.1103/PhysRevD.27.917} {\bibfield  {journal}
  {\bibinfo  {journal} {Phys. Rev.}\ }\textbf {\bibinfo {volume} {D27}},\
  \bibinfo {pages} {917} (\bibinfo {year} {1983})}\BibitemShut {NoStop}%
\bibitem [{\citenamefont {Landsman}(1989)}]{Landsman:1989be}%
  \BibitemOpen
  \bibfield  {author} {\bibinfo {author} {\bibfnamefont {N.~P.}\ \bibnamefont
  {Landsman}},\ }\href {\doibase 10.1016/0550-3213(89)90424-0} {\bibfield
  {journal} {\bibinfo  {journal} {Nucl. Phys.}\ }\textbf {\bibinfo {volume}
  {B322}},\ \bibinfo {pages} {498} (\bibinfo {year} {1989})}\BibitemShut
  {NoStop}%
\bibitem [{\citenamefont {Reisz}(1992)}]{Reisz:1991er}%
  \BibitemOpen
  \bibfield  {author} {\bibinfo {author} {\bibfnamefont {T.}~\bibnamefont
  {Reisz}},\ }\href {\doibase 10.1007/BF01483886} {\bibfield  {journal}
  {\bibinfo  {journal} {Z. Phys.}\ }\textbf {\bibinfo {volume} {C53}},\
  \bibinfo {pages} {169} (\bibinfo {year} {1992})}\BibitemShut {NoStop}%
\bibitem [{\citenamefont {LaCock}\ \emph {et~al.}(1992)\citenamefont {LaCock},
  \citenamefont {Miller},\ and\ \citenamefont {Reisz}}]{LaCock:1991hh}%
  \BibitemOpen
  \bibfield  {author} {\bibinfo {author} {\bibfnamefont {P.}~\bibnamefont
  {LaCock}}, \bibinfo {author} {\bibfnamefont {D.~E.}\ \bibnamefont {Miller}},
  \ and\ \bibinfo {author} {\bibfnamefont {T.}~\bibnamefont {Reisz}},\ }\href
  {\doibase 10.1016/0550-3213(92)90396-S} {\bibfield  {journal} {\bibinfo
  {journal} {Nucl. Phys.}\ }\textbf {\bibinfo {volume} {B369}},\ \bibinfo
  {pages} {501} (\bibinfo {year} {1992})}\BibitemShut {NoStop}%
\bibitem [{\citenamefont {Karkkainen}\ \emph {et~al.}(1993)\citenamefont
  {Karkkainen}, \citenamefont {Lacock}, \citenamefont {Petersson},\ and\
  \citenamefont {Reisz}}]{Karkkainen:1992wu}%
  \BibitemOpen
  \bibfield  {author} {\bibinfo {author} {\bibfnamefont {L.}~\bibnamefont
  {Karkkainen}}, \bibinfo {author} {\bibfnamefont {P.}~\bibnamefont {Lacock}},
  \bibinfo {author} {\bibfnamefont {B.}~\bibnamefont {Petersson}}, \ and\
  \bibinfo {author} {\bibfnamefont {T.}~\bibnamefont {Reisz}},\ }\href
  {\doibase 10.1016/0550-3213(93)90055-T} {\bibfield  {journal} {\bibinfo
  {journal} {Nucl. Phys.}\ }\textbf {\bibinfo {volume} {B395}},\ \bibinfo
  {pages} {733} (\bibinfo {year} {1993})}\BibitemShut {NoStop}%
\bibitem [{\citenamefont {Karkkainen}\ \emph {et~al.}(1994)\citenamefont
  {Karkkainen}, \citenamefont {Lacock}, \citenamefont {Miller}, \citenamefont
  {Petersson},\ and\ \citenamefont {Reisz}}]{Karkkainen:1993bu}%
  \BibitemOpen
  \bibfield  {author} {\bibinfo {author} {\bibfnamefont {L.}~\bibnamefont
  {Karkkainen}}, \bibinfo {author} {\bibfnamefont {P.}~\bibnamefont {Lacock}},
  \bibinfo {author} {\bibfnamefont {D.~E.}\ \bibnamefont {Miller}}, \bibinfo
  {author} {\bibfnamefont {B.}~\bibnamefont {Petersson}}, \ and\ \bibinfo
  {author} {\bibfnamefont {T.}~\bibnamefont {Reisz}},\ }\href {\doibase
  10.1016/0550-3213(94)90235-6} {\bibfield  {journal} {\bibinfo  {journal}
  {Nucl. Phys.}\ }\textbf {\bibinfo {volume} {B418}},\ \bibinfo {pages} {3}
  (\bibinfo {year} {1994})},\ \Eprint {http://arxiv.org/abs/hep-lat/9310014}
  {arXiv:hep-lat/9310014 [hep-lat]} \BibitemShut {NoStop}%
\bibitem [{\citenamefont {Patkos}\ \emph {et~al.}(1998)\citenamefont {Patkos},
  \citenamefont {Petreczky},\ and\ \citenamefont {Szep}}]{Patkos:1997cw}%
  \BibitemOpen
  \bibfield  {author} {\bibinfo {author} {\bibfnamefont {A.}~\bibnamefont
  {Patkos}}, \bibinfo {author} {\bibfnamefont {P.}~\bibnamefont {Petreczky}}, \
  and\ \bibinfo {author} {\bibfnamefont {Z.}~\bibnamefont {Szep}},\ }\href
  {\doibase 10.1007/s100520050277} {\bibfield  {journal} {\bibinfo  {journal}
  {Eur. Phys. J.}\ }\textbf {\bibinfo {volume} {C5}},\ \bibinfo {pages} {337}
  (\bibinfo {year} {1998})},\ \Eprint {http://arxiv.org/abs/hep-ph/9711263}
  {arXiv:hep-ph/9711263 [hep-ph]} \BibitemShut {NoStop}%
\bibitem [{\citenamefont {Karsch}\ \emph {et~al.}(1998)\citenamefont {Karsch},
  \citenamefont {Oevers},\ and\ \citenamefont {Petreczky}}]{Karsch:1998tx}%
  \BibitemOpen
  \bibfield  {author} {\bibinfo {author} {\bibfnamefont {F.}~\bibnamefont
  {Karsch}}, \bibinfo {author} {\bibfnamefont {M.}~\bibnamefont {Oevers}}, \
  and\ \bibinfo {author} {\bibfnamefont {P.}~\bibnamefont {Petreczky}},\ }\href
  {\doibase 10.1016/S0370-2693(98)01248-9} {\bibfield  {journal} {\bibinfo
  {journal} {Phys. Lett.}\ }\textbf {\bibinfo {volume} {B442}},\ \bibinfo
  {pages} {291} (\bibinfo {year} {1998})},\ \Eprint
  {http://arxiv.org/abs/hep-lat/9807035} {arXiv:hep-lat/9807035 [hep-lat]}
  \BibitemShut {NoStop}%
\bibitem [{\citenamefont {Kajantie}\ \emph
  {et~al.}(1997{\natexlab{b}})\citenamefont {Kajantie}, \citenamefont {Laine},
  \citenamefont {Peisa}, \citenamefont {Rajantie}, \citenamefont
  {Rummukainen},\ and\ \citenamefont {Shaposhnikov}}]{Kajantie:1997pd}%
  \BibitemOpen
  \bibfield  {author} {\bibinfo {author} {\bibfnamefont {K.}~\bibnamefont
  {Kajantie}}, \bibinfo {author} {\bibfnamefont {M.}~\bibnamefont {Laine}},
  \bibinfo {author} {\bibfnamefont {J.}~\bibnamefont {Peisa}}, \bibinfo
  {author} {\bibfnamefont {A.}~\bibnamefont {Rajantie}}, \bibinfo {author}
  {\bibfnamefont {K.}~\bibnamefont {Rummukainen}}, \ and\ \bibinfo {author}
  {\bibfnamefont {M.~E.}\ \bibnamefont {Shaposhnikov}},\ }\href {\doibase
  10.1103/PhysRevLett.79.3130} {\bibfield  {journal} {\bibinfo  {journal}
  {Phys. Rev. Lett.}\ }\textbf {\bibinfo {volume} {79}},\ \bibinfo {pages}
  {3130} (\bibinfo {year} {1997}{\natexlab{b}})},\ \Eprint
  {http://arxiv.org/abs/hep-ph/9708207} {arXiv:hep-ph/9708207 [hep-ph]}
  \BibitemShut {NoStop}%
\bibitem [{\citenamefont {Laine}\ and\ \citenamefont
  {Philipsen}(1999)}]{Laine:1999hh}%
  \BibitemOpen
  \bibfield  {author} {\bibinfo {author} {\bibfnamefont {M.}~\bibnamefont
  {Laine}}\ and\ \bibinfo {author} {\bibfnamefont {O.}~\bibnamefont
  {Philipsen}},\ }\href {\doibase 10.1016/S0370-2693(99)00641-3} {\bibfield
  {journal} {\bibinfo  {journal} {Phys. Lett.}\ }\textbf {\bibinfo {volume}
  {B459}},\ \bibinfo {pages} {259} (\bibinfo {year} {1999})},\ \Eprint
  {http://arxiv.org/abs/hep-lat/9905004} {arXiv:hep-lat/9905004 [hep-lat]}
  \BibitemShut {NoStop}%
\bibitem [{\citenamefont {Hart}\ and\ \citenamefont
  {Philipsen}(2000)}]{Hart:1999dj}%
  \BibitemOpen
  \bibfield  {author} {\bibinfo {author} {\bibfnamefont {A.}~\bibnamefont
  {Hart}}\ and\ \bibinfo {author} {\bibfnamefont {O.}~\bibnamefont
  {Philipsen}},\ }\href {\doibase 10.1016/S0550-3213(99)00742-7} {\bibfield
  {journal} {\bibinfo  {journal} {Nucl. Phys.}\ }\textbf {\bibinfo {volume}
  {B572}},\ \bibinfo {pages} {243} (\bibinfo {year} {2000})},\ \Eprint
  {http://arxiv.org/abs/hep-lat/9908041} {arXiv:hep-lat/9908041 [hep-lat]}
  \BibitemShut {NoStop}%
\bibitem [{\citenamefont {Hart}\ \emph {et~al.}(2000)\citenamefont {Hart},
  \citenamefont {Laine},\ and\ \citenamefont {Philipsen}}]{Hart:2000ha}%
  \BibitemOpen
  \bibfield  {author} {\bibinfo {author} {\bibfnamefont {A.}~\bibnamefont
  {Hart}}, \bibinfo {author} {\bibfnamefont {M.}~\bibnamefont {Laine}}, \ and\
  \bibinfo {author} {\bibfnamefont {O.}~\bibnamefont {Philipsen}},\ }\href
  {\doibase 10.1016/S0550-3213(00)00418-1} {\bibfield  {journal} {\bibinfo
  {journal} {Nucl. Phys.}\ }\textbf {\bibinfo {volume} {B586}},\ \bibinfo
  {pages} {443} (\bibinfo {year} {2000})},\ \Eprint
  {http://arxiv.org/abs/hep-ph/0004060} {arXiv:hep-ph/0004060 [hep-ph]}
  \BibitemShut {NoStop}%
\bibitem [{\citenamefont {Cucchieri}\ \emph
  {et~al.}(2001{\natexlab{a}})\citenamefont {Cucchieri}, \citenamefont
  {Karsch},\ and\ \citenamefont {Petreczky}}]{Cucchieri:2000cy}%
  \BibitemOpen
  \bibfield  {author} {\bibinfo {author} {\bibfnamefont {A.}~\bibnamefont
  {Cucchieri}}, \bibinfo {author} {\bibfnamefont {F.}~\bibnamefont {Karsch}}, \
  and\ \bibinfo {author} {\bibfnamefont {P.}~\bibnamefont {Petreczky}},\ }\href
  {\doibase 10.1016/S0370-2693(00)01331-9} {\bibfield  {journal} {\bibinfo
  {journal} {Phys. Lett.}\ }\textbf {\bibinfo {volume} {B497}},\ \bibinfo
  {pages} {80} (\bibinfo {year} {2001}{\natexlab{a}})},\ \Eprint
  {http://arxiv.org/abs/hep-lat/0004027} {arXiv:hep-lat/0004027 [hep-lat]}
  \BibitemShut {NoStop}%
\bibitem [{\citenamefont {Cucchieri}\ \emph
  {et~al.}(2001{\natexlab{b}})\citenamefont {Cucchieri}, \citenamefont
  {Karsch},\ and\ \citenamefont {Petreczky}}]{Cucchieri:2001tw}%
  \BibitemOpen
  \bibfield  {author} {\bibinfo {author} {\bibfnamefont {A.}~\bibnamefont
  {Cucchieri}}, \bibinfo {author} {\bibfnamefont {F.}~\bibnamefont {Karsch}}, \
  and\ \bibinfo {author} {\bibfnamefont {P.}~\bibnamefont {Petreczky}},\ }\href
  {\doibase 10.1103/PhysRevD.64.036001} {\bibfield  {journal} {\bibinfo
  {journal} {Phys. Rev.}\ }\textbf {\bibinfo {volume} {D64}},\ \bibinfo {pages}
  {036001} (\bibinfo {year} {2001}{\natexlab{b}})},\ \Eprint
  {http://arxiv.org/abs/hep-lat/0103009} {arXiv:hep-lat/0103009 [hep-lat]}
  \BibitemShut {NoStop}%
\bibitem [{\citenamefont {Linde}(1980)}]{Linde:1980ts}%
  \BibitemOpen
  \bibfield  {author} {\bibinfo {author} {\bibfnamefont {A.~D.}\ \bibnamefont
  {Linde}},\ }\href {\doibase 10.1016/0370-2693(80)90769-8} {\bibfield
  {journal} {\bibinfo  {journal} {Phys. Lett.}\ }\textbf {\bibinfo {volume}
  {B96}},\ \bibinfo {pages} {289} (\bibinfo {year} {1980})}\BibitemShut
  {NoStop}%
\bibitem [{\citenamefont {Laine}\ and\ \citenamefont
  {{Schr\"oder}}(2005)}]{Laine:2005ai}%
  \BibitemOpen
  \bibfield  {author} {\bibinfo {author} {\bibfnamefont {M.}~\bibnamefont
  {Laine}}\ and\ \bibinfo {author} {\bibfnamefont {Y.}~\bibnamefont
  {{Schr\"oder}}},\ }\href {\doibase 10.1088/1126-6708/2005/03/067} {\bibfield
  {journal} {\bibinfo  {journal} {JHEP}\ }\textbf {\bibinfo {volume} {03}},\
  \bibinfo {pages} {067} (\bibinfo {year} {2005})},\ \Eprint
  {http://arxiv.org/abs/hep-ph/0503061} {arXiv:hep-ph/0503061 [hep-ph]}
  \BibitemShut {NoStop}%
\bibitem [{\citenamefont {Braaten}\ and\ \citenamefont
  {Nieto}(1995)}]{Braaten:1994qx}%
  \BibitemOpen
  \bibfield  {author} {\bibinfo {author} {\bibfnamefont {E.}~\bibnamefont
  {Braaten}}\ and\ \bibinfo {author} {\bibfnamefont {A.}~\bibnamefont
  {Nieto}},\ }\href {\doibase 10.1103/PhysRevLett.74.3530} {\bibfield
  {journal} {\bibinfo  {journal} {Phys. Rev. Lett.}\ }\textbf {\bibinfo
  {volume} {74}},\ \bibinfo {pages} {3530} (\bibinfo {year} {1995})},\ \Eprint
  {http://arxiv.org/abs/hep-ph/9410218} {arXiv:hep-ph/9410218 [hep-ph]}
  \BibitemShut {NoStop}%
\bibitem [{\citenamefont {Hietanen}\ \emph {et~al.}(2009)\citenamefont
  {Hietanen}, \citenamefont {Kajantie}, \citenamefont {Laine}, \citenamefont
  {Rummukainen},\ and\ \citenamefont {{Schr\"oder}}}]{Hietanen:2008tv}%
  \BibitemOpen
  \bibfield  {author} {\bibinfo {author} {\bibfnamefont {A.}~\bibnamefont
  {Hietanen}}, \bibinfo {author} {\bibfnamefont {K.}~\bibnamefont {Kajantie}},
  \bibinfo {author} {\bibfnamefont {M.}~\bibnamefont {Laine}}, \bibinfo
  {author} {\bibfnamefont {K.}~\bibnamefont {Rummukainen}}, \ and\ \bibinfo
  {author} {\bibfnamefont {Y.}~\bibnamefont {{Schr\"oder}}},\ }\href {\doibase
  10.1103/PhysRevD.79.045018} {\bibfield  {journal} {\bibinfo  {journal} {Phys.
  Rev.}\ }\textbf {\bibinfo {volume} {D79}},\ \bibinfo {pages} {045018}
  (\bibinfo {year} {2009})},\ \Eprint {http://arxiv.org/abs/0811.4664}
  {arXiv:0811.4664 [hep-lat]} \BibitemShut {NoStop}%
\bibitem [{\citenamefont {Kajantie}\ \emph {et~al.}(2003)\citenamefont
  {Kajantie}, \citenamefont {Laine}, \citenamefont {Rummukainen},\ and\
  \citenamefont {{Schr\"oder}}}]{Kajantie:2003ax}%
  \BibitemOpen
  \bibfield  {author} {\bibinfo {author} {\bibfnamefont {K.}~\bibnamefont
  {Kajantie}}, \bibinfo {author} {\bibfnamefont {M.}~\bibnamefont {Laine}},
  \bibinfo {author} {\bibfnamefont {K.}~\bibnamefont {Rummukainen}}, \ and\
  \bibinfo {author} {\bibfnamefont {Y.}~\bibnamefont {{Schr\"oder}}},\ }\href
  {\doibase 10.1088/1126-6708/2003/04/036} {\bibfield  {journal} {\bibinfo
  {journal} {JHEP}\ }\textbf {\bibinfo {volume} {04}},\ \bibinfo {pages} {036}
  (\bibinfo {year} {2003})},\ \Eprint {http://arxiv.org/abs/hep-ph/0304048}
  {arXiv:hep-ph/0304048 [hep-ph]} \BibitemShut {NoStop}%
\bibitem [{\citenamefont {Necco}(2004)}]{Necco:2003vh}%
  \BibitemOpen
  \bibfield  {author} {\bibinfo {author} {\bibfnamefont {S.}~\bibnamefont
  {Necco}},\ }\href {\doibase 10.1016/j.nuclphysb.2004.01.032} {\bibfield
  {journal} {\bibinfo  {journal} {Nucl. Phys.}\ }\textbf {\bibinfo {volume}
  {B683}},\ \bibinfo {pages} {137} (\bibinfo {year} {2004})},\ \Eprint
  {http://arxiv.org/abs/hep-lat/0309017} {arXiv:hep-lat/0309017 [hep-lat]}
  \BibitemShut {NoStop}%
\bibitem [{\citenamefont {Sommer}(1994)}]{Sommer:1993ce}%
  \BibitemOpen
  \bibfield  {author} {\bibinfo {author} {\bibfnamefont {R.}~\bibnamefont
  {Sommer}},\ }\href {\doibase 10.1016/0550-3213(94)90473-1} {\bibfield
  {journal} {\bibinfo  {journal} {Nucl. Phys.}\ }\textbf {\bibinfo {volume}
  {B411}},\ \bibinfo {pages} {839} (\bibinfo {year} {1994})},\ \Eprint
  {http://arxiv.org/abs/hep-lat/9310022} {arXiv:hep-lat/9310022 [hep-lat]}
  \BibitemShut {NoStop}%
\bibitem [{\citenamefont {Brambilla}\ \emph
  {et~al.}(2010{\natexlab{b}})\citenamefont {Brambilla}, \citenamefont
  {Garcia~i Tormo}, \citenamefont {Soto},\ and\ \citenamefont
  {Vairo}}]{Brambilla:2010pp}%
  \BibitemOpen
  \bibfield  {author} {\bibinfo {author} {\bibfnamefont {N.}~\bibnamefont
  {Brambilla}}, \bibinfo {author} {\bibfnamefont {X.}~\bibnamefont {Garcia~i
  Tormo}}, \bibinfo {author} {\bibfnamefont {J.}~\bibnamefont {Soto}}, \ and\
  \bibinfo {author} {\bibfnamefont {A.}~\bibnamefont {Vairo}},\ }\href
  {\doibase 10.1103/PhysRevLett.105.212001, 10.1103/PhysRevLett.108.269903}
  {\bibfield  {journal} {\bibinfo  {journal} {Phys. Rev. Lett.}\ }\textbf
  {\bibinfo {volume} {105}},\ \bibinfo {pages} {212001} (\bibinfo {year}
  {2010}{\natexlab{b}})},\ \bibinfo {note} {[Erratum: Phys. Rev. Lett. {\bf
  108}, 269903 (2012)]},\ \Eprint {http://arxiv.org/abs/1006.2066}
  {arXiv:1006.2066 [hep-ph]} \BibitemShut {NoStop}%
\bibitem [{\citenamefont {Bazavov}\ \emph
  {et~al.}(2012{\natexlab{b}})\citenamefont {Bazavov}, \citenamefont
  {Brambilla}, \citenamefont {Garcia~i Tormo}, \citenamefont {Petreczky},
  \citenamefont {Soto},\ and\ \citenamefont {Vairo}}]{Bazavov:2012ka}%
  \BibitemOpen
  \bibfield  {author} {\bibinfo {author} {\bibfnamefont {A.}~\bibnamefont
  {Bazavov}}, \bibinfo {author} {\bibfnamefont {N.}~\bibnamefont {Brambilla}},
  \bibinfo {author} {\bibfnamefont {X.}~\bibnamefont {Garcia~i Tormo}},
  \bibinfo {author} {\bibfnamefont {P.}~\bibnamefont {Petreczky}}, \bibinfo
  {author} {\bibfnamefont {J.}~\bibnamefont {Soto}}, \ and\ \bibinfo {author}
  {\bibfnamefont {A.}~\bibnamefont {Vairo}},\ }\href {\doibase
  10.1103/PhysRevD.86.114031} {\bibfield  {journal} {\bibinfo  {journal} {Phys.
  Rev.}\ }\textbf {\bibinfo {volume} {D86}},\ \bibinfo {pages} {114031}
  (\bibinfo {year} {2012}{\natexlab{b}})},\ \Eprint
  {http://arxiv.org/abs/1205.6155} {arXiv:1205.6155 [hep-ph]} \BibitemShut
  {NoStop}%
\bibitem [{\citenamefont {Bazavov}\ \emph {et~al.}(2014)\citenamefont
  {Bazavov}, \citenamefont {Brambilla}, \citenamefont {Garcia~i Tormo},
  \citenamefont {Petreczky}, \citenamefont {Soto},\ and\ \citenamefont
  {Vairo}}]{Bazavov:2014soa}%
  \BibitemOpen
  \bibfield  {author} {\bibinfo {author} {\bibfnamefont {A.}~\bibnamefont
  {Bazavov}}, \bibinfo {author} {\bibfnamefont {N.}~\bibnamefont {Brambilla}},
  \bibinfo {author} {\bibfnamefont {X.}~\bibnamefont {Garcia~i Tormo}},
  \bibinfo {author} {\bibfnamefont {P.}~\bibnamefont {Petreczky}}, \bibinfo
  {author} {\bibfnamefont {J.}~\bibnamefont {Soto}}, \ and\ \bibinfo {author}
  {\bibfnamefont {A.}~\bibnamefont {Vairo}},\ }\href {\doibase
  10.1103/PhysRevD.90.074038} {\bibfield  {journal} {\bibinfo  {journal} {Phys.
  Rev.}\ }\textbf {\bibinfo {volume} {D90}},\ \bibinfo {pages} {074038}
  (\bibinfo {year} {2014})},\ \Eprint {http://arxiv.org/abs/1407.8437}
  {arXiv:1407.8437 [hep-ph]} \BibitemShut {NoStop}%
\bibitem [{Rpa()}]{Rpackage}%
  \BibitemOpen
  \href@noop {} {}\bibinfo {note} {\url{http://www.r-project.org/}}\BibitemShut
  {NoStop}%
\bibitem [{\citenamefont {Schwinger}(1962)}]{Schwinger:1962wd}%
  \BibitemOpen
  \bibfield  {author} {\bibinfo {author} {\bibfnamefont {J.}~\bibnamefont
  {Schwinger}},\ }\href {\doibase 10.1103/PhysRev.127.324} {\bibfield
  {journal} {\bibinfo  {journal} {Phys. Rev.}\ }\textbf {\bibinfo {volume}
  {127}},\ \bibinfo {pages} {324} (\bibinfo {year} {1962})}\BibitemShut
  {NoStop}%
\bibitem [{\citenamefont {Christ}\ and\ \citenamefont
  {Lee}(1980)}]{Christ:1980ku}%
  \BibitemOpen
  \bibfield  {author} {\bibinfo {author} {\bibfnamefont {N.~H.}\ \bibnamefont
  {Christ}}\ and\ \bibinfo {author} {\bibfnamefont {T.~D.}\ \bibnamefont
  {Lee}},\ }\href {\doibase 10.1103/PhysRevD.22.939} {\bibfield  {journal}
  {\bibinfo  {journal} {Phys. Rev.}\ }\textbf {\bibinfo {volume} {D22}},\
  \bibinfo {pages} {939} (\bibinfo {year} {1980})}\BibitemShut {NoStop}%
\bibitem [{\citenamefont {Andrasi}(2004)}]{Andrasi:2003zf}%
  \BibitemOpen
  \bibfield  {author} {\bibinfo {author} {\bibfnamefont {A.}~\bibnamefont
  {Andrasi}},\ }\href {\doibase 10.1140/epjc/s2004-02005-2} {\bibfield
  {journal} {\bibinfo  {journal} {Eur. Phys. J.}\ }\textbf {\bibinfo {volume}
  {C37}},\ \bibinfo {pages} {307} (\bibinfo {year} {2004})},\ \Eprint
  {http://arxiv.org/abs/hep-th/0311118} {arXiv:hep-th/0311118 [hep-th]}
  \BibitemShut {NoStop}%
\bibitem [{\citenamefont {Broadhurst}(1992)}]{Broadhurst:1991fi}%
  \BibitemOpen
  \bibfield  {author} {\bibinfo {author} {\bibfnamefont {D.~J.}\ \bibnamefont
  {Broadhurst}},\ }\href {\doibase 10.1007/BF01559486} {\bibfield  {journal}
  {\bibinfo  {journal} {Z. Phys.}\ }\textbf {\bibinfo {volume} {C54}},\
  \bibinfo {pages} {599} (\bibinfo {year} {1992})}\BibitemShut {NoStop}%
\bibitem [{\citenamefont {Gray}\ \emph {et~al.}(1990)\citenamefont {Gray},
  \citenamefont {Broadhurst}, \citenamefont {Grafe},\ and\ \citenamefont
  {Schilcher}}]{Gray:1990yh}%
  \BibitemOpen
  \bibfield  {author} {\bibinfo {author} {\bibfnamefont {N.}~\bibnamefont
  {Gray}}, \bibinfo {author} {\bibfnamefont {D.~J.}\ \bibnamefont
  {Broadhurst}}, \bibinfo {author} {\bibfnamefont {W.}~\bibnamefont {Grafe}}, \
  and\ \bibinfo {author} {\bibfnamefont {K.}~\bibnamefont {Schilcher}},\ }\href
  {\doibase 10.1007/BF01614703} {\bibfield  {journal} {\bibinfo  {journal} {Z.
  Phys.}\ }\textbf {\bibinfo {volume} {C48}},\ \bibinfo {pages} {673} (\bibinfo
  {year} {1990})}\BibitemShut {NoStop}%
\end{thebibliography}%
\end{document}